\documentclass[twoside,11pt]{article}

\usepackage[accepted]{melba}

\usepackage[english]{babel}
\usepackage[utf8]{inputenc}

\usepackage{amsmath,amssymb,amsfonts}
\usepackage{textcomp}
\usepackage{caption}
\usepackage{subcaption}
\usepackage{dsfont}
\usepackage{mathtools}
\usepackage{algorithm}
\usepackage{algpseudocode}
\usepackage{footnote}
\usepackage{bbm}

\usepackage{hyperref}
\usepackage{cleveref}
\usepackage{bm}
\usepackage{bbold}
\usepackage{breqn}
\usepackage{graphics,graphicx,wrapfig}
\usepackage{booktabs}
\usepackage{stmaryrd}
\usepackage{array}
\usepackage{tikz}
\usepackage{pgfplots}
\usepackage{multirow}
\usepackage{adjustbox}
\usepackage{makecell}
\usepackage{ulem}
\usepackage{rotating}

\usetikzlibrary{arrows}

\tikzset{
    cross/.pic = {
    \draw[rotate = 45] (-#1,0) -- (#1,0);
    \draw[rotate = 45] (0,-#1) -- (0, #1);
    }
}

\DeclareMathOperator*{\argmin}{arg\,min}

\def\R{\mathbb{R}}
\def\C{\mathbb{C}}

\def\E{\mathbb{E}}

\def\dist{\mathrm{dist}}

\def\Sc{\mathcal{S}}
\def\Pc{\mathcal{P}}
\def\Cc{\mathcal{C}}
\def\Lc{\mathcal{L}}
\def\Dc{\mathcal{D}}
\def\Zc{\mathcal{Z}}
\def\f12{\frac{1}{2}}

\def\Id{\mathrm{Id}}
\def\eqdef{\stackrel{\mathrm{def}}{=}}
\def\one{\mathds{1}}
\newcommand{\rev}[1]{#1}

\usepackage{pifont}
\definecolor{orange}{rgb}{0.8,0.5,0}
\definecolor{lightblue}{rgb}{0.075,0.541,0.855}

\makeatletter
\newcommand*{\shifttext}[2]{%
  \settowidth{\@tempdima}{#2}%
  \makebox[\@tempdima]{\hspace*{#1}#2}%
}
\makeatother

\makeatletter
\newcommand*{\plottraj}[8]{%
    \begin{subfigure}[t]{.4\linewidth}
        \centering
        \begin{tikzpicture}
          \node[anchor=south west,inner sep=0] (fig) at (0,0) {\includegraphics[trim={0.5cm 0 0 0},clip,width=\textwidth]{#1}};
          \node[rectangle] (box) at (2.88,3.03) [draw,black,thick,minimum width=0.6cm,minimum height=0.6cm] {};
          \draw [-stealth, thick](box.north east) -- (6.1,4.5);
        \end{tikzpicture}%
        \caption{#3}%
        \ifthenelse{\equal{#7}{0}}{}{\label{#7}}
  \end{subfigure}%
  \raisebox{0.2\linewidth}{%
      \begin{subfigure}[t]{.2\linewidth}\centering
          \vspace{-3.24cm}%
          \includegraphics[trim={0.6cm 0 0 0},clip,width=\linewidth]{#2}%
          \vspace{-0.2cm}%
          \ifthenelse{\equal{#5}{0}}{}{ \includegraphics[trim={0.6cm 0 0 0},clip,width=\linewidth]{#5} }
      \end{subfigure}%
  }%
  \ifthenelse{\equal{#4}{0}}{}{%
    \begin{subfigure}[t]{.4\linewidth}
        \vspace{-6.4cm}%
        \centering
        \begin{tikzpicture}
          \node[anchor=south west,inner sep=0] (fig) at (0,0) {\includegraphics[trim={0.5cm 0 0 0},clip,width=\textwidth]{#4}};
          \node[rectangle] (box) at (2.88,2.95) [draw,black,thick,minimum width=0.6cm,minimum height=0.6cm] {};
          \draw [-stealth, thick](box.south west) -- (0,1.6);
        \end{tikzpicture}
        \caption{#6}
        \ifthenelse{\equal{#8}{0}}{}{\label{#8}}
    \end{subfigure}
  }
}
\makeatother

\usepackage{orcidlink}

\melbaid{2023:009}  
\doi{https://doi.org/10.59275/j.melba.2023-8172}
\melbaauthors{Gossard, de Gournay and Weiss}  
\volume{2}
\firstpageno{253}  
\melbayear{2023}  
\datesubmitted{09/2022}  
\datepublished{06/2023}  

\ShortHeadings{Bayesian Optimization of Sampling Densities in MRI}{Gossard, de Gournay and Weiss}

\title{Bayesian Optimization of Sampling Densities in MRI}

\author{\name Alban Gossard \orcidlink{0000-0001-6782-5080} \email alban.paul.gossard@gmail.com \\
  \addr \hspace*{1em} Institut de Mathématiques de Toulouse (IMT); UMR5219; Université de Toulouse; CNRS, France \\
  \addr \hspace*{1em} Université Paul Sabatier, F-31062 Toulouse Cedex 9, France
  \AND
  \name Frédéric de Gournay \orcidlink{0000-0003-4721-3137} \email degourna@insa-toulouse.fr \\
  \addr \hspace*{1em} Institut de Mathématiques de Toulouse (IMT); UMR5219; Université de Toulouse; CNRS, France \\
  \addr \hspace*{1em} INSA de Toulouse; F-31077 Toulouse, France
  \AND
  \name Pierre Weiss \orcidlink{0000-0001-9083-214X} \email pierre.weiss@cnrs.fr \\
  \addr \hspace*{1em} Institut de Mathématiques de Toulouse (IMT); UMR5219; Université de Toulouse; CNRS, France \\
  \addr \hspace*{1em} Centre de Biologie Intégrative (CBI), Laboratoire MCD, F-31062 Toulouse Cedex, France
}

\begin{document}
\maketitle

\begin{abstract}
Data-driven optimization of sampling patterns in MRI has recently received a significant attention.
Following recent observations on the combinatorial number of minimizers in off-the-grid optimization, we propose a framework to globally optimize the sampling densities using Bayesian optimization. Using a dimension reduction technique, we optimize the sampling trajectories more than 20 times faster than conventional off-the-grid methods, with a restricted number of training samples. This method -- among other benefits -- discards the need of automatic differentiation.
Its performance is slightly worse than state-of-the-art learned trajectories since it reduces the space of admissible trajectories, but comes with significant computational advantages.
Other contributions include: i) a careful evaluation of the distance in probability space to generate trajectories ii) a specific training procedure on families of operators for unrolled reconstruction networks and iii) a gradient projection based scheme for trajectory optimization.

\end{abstract}

\begin{keywords}
Sampling theory, compressed sensing, MRI, Fourier transform, data-driven optimization, globalization, Bayesian optimization
\end{keywords}


\section{Introduction}
\label{sec:introduction}

The quest for efficient acquisition and reconstruction mechanisms in Magnetic Resonance Imaging (MRI) has been ongoing since its invention in the 1970's. 
This led to a few major breakthrough, which comprise the design of efficient pulse sequences \cite{bernstein2004handbook}, the use of parallel imaging \cite{roemer1990nmr,blaimer2004smash}, the theory and application of compressed sensing \cite{lustig2008compressed} and its recent improvements thanks to the progresses in learning and GPU computing \cite{knoll2020advancing}. While the first attempts to use neural networks in this field were primarily focused on the efficient design of reconstruction algorithms \cite{jacob2020computational}, some recent works began investigating the design of efficient sampling schemes or joint sampling/reconstruction schemes. The aim of this paper is to make progress in the numerical analysis of this nascent and challenging field.

\subsection{Some sampling theory}\label{sec:sampling_theory}

In a simplified way, an MRI scanner measures values of the Fourier transform of the image to reconstruct at different locations $(\xi_m)_{1\leq m \leq M}$ in the so-called k-space. 
The locations $(\xi_m)$ are obtained by sampling a continuous trajectory defined through a gradient sequence. 
The problem we tackle in this paper is: how to choose the points $(\xi_m)$ or the underlying trajectories in an efficient or optimal way?

\paragraph{Shannon-Nyquist}
This question was first addressed using Shannon-Nyquist theorem, which certifies that sampling the k-space on a sufficiently fine Euclidean grid provides exact reconstructions using linear reconstructors.
This motivated the design of many trajectories, such as the ones in echo-planar imaging (EPI) \cite{schmitt2012echo}.
Progresses on non-uniform sampling theory \cite{feichtinger1994theory} then provided guidelines to produce efficient sampling/reconstruction schemes for linear reconstructors. This theory is now mature for the reconstruction of bandlimited functions.
In a nutshell, it advocates the use of a sampling set which covers the k-space sufficiently densely with well spread samples.

\paragraph{Compressed sensing theory}
Shannon-Nyquist theory requires sampling the k-space densely, resulting in long scanning times. 
It was observed in the 1980's that subsampling the high frequencies using variable density radial patterns did not compromise the image quality too much \cite{ahn1986high,jackson1992twisting}.
The first theoretical elements justifying this evidence were provided by the theory of compressed sensing, when using nonlinear reconstructors.
This seminal theory is based on concepts such as the restricted isometry property (RIP) or the incoherence between the measurements \cite{candes2006robust,lustig2005faster}. 
However it soon became evident that these concepts were not suited to the practice of MRI and a refined theory based on local coherence appeared in \cite{adcock2017breaking,boyer2019compressed}. 
The main teaching is that a good sampling scheme for $\ell^1$-based reconstruction methods must have a variable density that depends on the sparsity basis and on the sparsity pattern of the images. 
To the best of our knowledge, this theory is currently the one that provides the best explanation of the success of sub-sampling. 
In particular, analytical expressions of the optimal densities \cite{adcockoracle} can be derived and fit relatively well with the best empirical ones.

\paragraph{The main teachings}
To date, there is still a significant discrepancy between the theory and practice of sampling in MRI. 
A mix between theory and common sense however provides the following main insights. 
A good sampling scheme should \cite{boyer2016generation}:
\begin{itemize}
	\item have a variable density, decaying with the distance to the center of the k-space,
	\item have a sufficiently high density in the center to comply with the Shannon-Nyquist criterion, and sufficiently low to avoid dense clusters which would not bring additional information,
	\item have a locally uniform coverage of the k-space. In particular, nearby samples are detrimental to the reconstruction since they are highly correlated and increase the condition number of partial Fourier matrices.
\end{itemize}
These considerations are all satisfied when using Poisson disk sampling with an adequate density \cite{vasanawala2011practical} for pointwise sampling. 
They also led to the development of the Sparkling trajectories \cite{chauffert2017projection,lazarus2019sparkling}, which incorporate additional trajectory constraints in the design. 

\paragraph{What can still be optimized?}
Given the previous remarks, an important question remains open: how to choose the sampling density? 
An axiomatic approach leads to choosing radial densities with a plateau (constant value) at the center. 
The radial character ensures rotation invariance, which seems natural to image organs in arbitrary orientations. 
The plateau enforces Nyquist rate at the center. 
However, it may still be possible to improve the results for specific datasets.

\subsection{Data-driven sampling schemes}

The first attempts to learn a sampling density \cite{knoll2011adapted,zhang2014energy} were based on the average energy of the k-space coefficients on a collection of reference images. While this principle is valid for linear reconstructions, it is not supported by a theoretical background when using nonlinear reconstructors. Motivated by the recent breakthroughs of learning and deep learning, many authors recently proposed to learn either the reconstructor \rev{\cite{hammernik2018learning,korkmaz2022unsupervised,muckley2021results}}, the sampling pattern \cite{baldassarre2016learning,gozcu2018learning,zibetti2021fast,sherry2019learning}, or both \rev{\cite{jin2019self,bahadir2020deep,weiss2021pilot,aggarwal2020j,zibetti2022alternating,wang2022b}}. Data-driven optimization has emerged as a promising approach to tailor the sampling schemes with respect to the reconstructor and to the image structure.
In \cite{baldassarre2016learning,gozcu2018learning,sherry2019learning,sanchez2020scalable,zibetti2021fast}, the authors look for an optimal subset of a fixed set of k-space positions. The initial algorithms are based on simple greedy approaches that generated a sampling pattern by iteratively selecting a discrete horizontal line that minimizes the residual error of the reconstructed image. This approach is limited to low dimensional sets of parallel lines. 
Some efforts have been spent on finding better and more scalable solutions to this hard combinatorial problem using stochastic greedy algorithms \cite{sanchez2020scalable}, $\ell^1$-relaxation and bi-level programming \cite{sherry2019learning} or bias-accelerated subset selection \cite{zibetti2021fast}. This method is reported to provide results over 3D images and seems to have solved some of the scalability issues.

To the best of our knowledge, the first work investigating the joint optimization of a sampling pattern and a reconstruction algorithm was proposed in \cite{jin2019self}. In this work, the authors use a Monte Carlo Tree Search which allows them to optimize a policy that determines the positions to sample.
This sampling relies on lines and the reconstruction process is an image to image domain with an inverse Fourier transform performed on the data before the denoising step.
In the same spirit, \cite{bahadir2020deep} proposes to learn MRI trajectories by optimizing a binary mask over a Cartesian grid with some sparsity constraint. The reconstruction is decomposed into two steps: a regridding using an inverse Fourier transform and a U-NET for de-aliasing. Finally, a new class of reconstruction methods called \textit{algorithm unrolling}, mimicking classical variational approaches have emerged. 
These approaches improve the interpretability of deep learning based methods. 
Optimizing the weights of a CNN that plays the role of a denoiser in a conjugate gradient descent has been investigated in \cite{aggarwal2020j}. 
The authors jointly optimize the sampling pattern and a denoising network based on an unrolled conjugate gradient scheme. 
The sampling scheme is expressed as the tensor product of 1D sampling patterns which significantly restricts the possible sampling schemes.

Overall, the previous works suffer from some limitations: the sampling points are required to live on a Cartesian grid, which may be non physical and lead to combinatorial problems; the methods cannot incorporate advanced constraints on the sampling trajectory and therefore focus on ``rigid'' constraints such as selecting a subset of horizontal lines.

To address these issues, some recent works propose to optimize points that can move freely in a continuous domain \cite{weiss2021pilot,wang2022b}. This approach allows handling real kinematic constraints.
In \cite{weiss2021pilot}, the authors propose to reconstruct an image using a rough inversion of the partial Fourier transform, followed by a U-NET to eliminate the residual artifacts. They optimize jointly the weights of the U-NET together with the k-space positions using a stochastic gradient method. 
The physical kinematic constraints are handled using two different ingredients.
First, the k-space points are regularly ordered by solving a traveling salesman problem, ensuring a low distance between consecutive points.
Second, the constraints are promoted using a penalization function.
This re-ordering step was then abandoned in \cite{wang2022b}, where the authors use a B-spline parameterization of the trajectories with a penalization over the constraints in the cost function.
Instead of using a rough inversion with a U-NET, the authors opted for an unrolled ADMM reconstructor where the proximal operator is replaced by a DIDN CNN \cite{yu2019deep}.
The k-space locations and the CNN weights are optimized jointly.
In both works, long computation times and memory requirements are reported.
We also observed significant convergence issues related to the existence of spurious minimizers \cite{gossard2022spurious}.

\subsection{Our contribution}

The purpose of this work is to improve the process of optimizing sampling schemes from a methodological perspective.
We propose a framework that optimizes the sampling density using Bayesian Optimization (BO).
Our method has a few advantages compared to recent learning based approaches: i) it globalizes the convergence by reducing the dimensionality of the optimization problem, ii) it reduces the computing times drastically, iii) it requires only a small number of reference images and iv) it works off-the-grid and handles arbitrary physical constraints.
The first three features are essential to make sampling scheme optimization tractable in a wide range a different MRI scanners. The last one allows more versatility in the sampling patterns that can take advantage of all the degrees of freedom offered by an MRI scanner.

\section{The proposed approach}
\label{sec:approach}

In this section, we describe the main ideas of this work after having introduced the notation. 

\subsection{Preliminaries}\label{subsec:preliminaries}
\paragraph{Images} Let $\mathcal{X}$ denote the set of $K$ training images $\mathcal{X}=\{x_1,\hdots,x_K \}$. A $D$-dimensional image is a vector of $\C^N$, where $N=N_1\hdots N_D$ and $N_d\in 2\mathbb{N}$ denotes the number of pixels in the $d$-th direction. In this work, each index $n\in \llbracket 1, N\rrbracket$, is associated with  a position $p_n\in \left\llbracket -\frac{N_1}{2},\frac {N_1}{2}-1\right\rrbracket \times \hdots \times \left\llbracket-\frac{N_D}{2},\frac{N_D}{2}-1\right\rrbracket$ on Euclidean grid. It describes the location of the $n$-th pixel in the k-space. 
With a slight abuse of notation, we associate to each discrete image $x_k\in\C^N$, a function still denoted $x_k$, defined by
\begin{equation*}
x_k=\left(\sum_{n=1}^N x_k[n] \delta_{p_n}\right)\star \psi,
\end{equation*}
where $\star$ denotes the convolution-product and where $\psi$ is an interpolation function. For instance, we can set $\psi$ as the indicator of a grid element to generate piece-wise constant images.

\paragraph{Image quality} To measure the reconstruction quality, we consider an image quality metric $\eta:\R^N\times \R^N\to \R_+$. The experiments in this work are conducted using the squared $\ell^2$ distance $\eta(\tilde x,x)=\frac{1}{2}\|\tilde x - x\|_2^2$. Any other metric could be used instead with the proposed approach.

\paragraph{The Non-Uniform Fourier Transform} Throughout the paper, we let $\xi =(\xi_1, \hdots, \xi_M)\in (\R^D)^M$ denote a set of locations in the k-space (or Fourier domain). Let $A(\xi)\in \C^{M\times N}$ denote the forward non-uniform Fourier transform defined for all $m\in \llbracket 1,M\rrbracket$ and $x\in \C^N$ by
\begin{align}
 [A(\xi)(x)]_m &= \int_{t\in \R^D} \exp(- i \langle t,\xi_m\rangle) x(t) \,dt\nonumber \\
  &=\Psi(\xi_m) \cdot\sum_{n=1}^N x[n] \exp(-i \langle p_n, \xi_m\rangle),\label{eq:def_NUFFT}
\end{align}
where $\Psi$ is the Fourier transform of the interpolation function $\psi$.

\paragraph{Image reconstruction} We let $R:\C^M\times (\R^D)^M \times \R^J\to \C^N$ denote an image reconstruction mapping.
For a measurement vector $y\in \C^M$, a sampling scheme $\xi\in (\R^D)^M$, and a parameter $\lambda\in \R^J$, we let $\tilde x = R(\xi, y,\lambda)$ denote the reconstructed image.
In this paper, we will consider two different reconstructors:
\begin{itemize}
    \item A Total Variation (TV) reconstructor \cite{lustig2008compressed}, which is a standard baseline:
    \begin{equation}\label{eq:reconstructor_TV}
    R_1(\xi, y,\lambda)=\argmin_{x\in \C^N} \frac{1}{2}\|A(\xi)x - y\|_2^2 + \lambda \|\nabla x\|_1,
    \end{equation}
where $\lambda\geq 0$ is a regularization parameter. The approximate solution of this problem is obtained with an iterative algorithm run for a fixed number of iterations.  We refer the reader to Appendix~\ref{sec:reconstruction_algorithm_TV} for the algorithmic details. This allows us to use the automatic differentiation of PyTorch as described in \cite{ochs2015bilevel}.

    \item An unrolled neural network $R_2(\xi,y,\lambda)$, where $\lambda$ denotes the weights of the neural network.
There is now a multitude of such reconstructors available in the literature \cite{muckley2021results}.
They draw their inspiration from classical model-based reconstructors with hand-crafted priors.
The details are provided in Appendix~\ref{sec:appendix_neural_net_definition}.
\end{itemize}

\paragraph{Constraints on the sampling scheme} As mentioned in the introduction, the sampling positions $\xi=(\xi_1,\hdots,\xi_M)$ correspond to the discretization of a k-space trajectory subject to kinematic constraints. Throughout the paper, we  let $\Xi \subset (\R^D)^M$ denote the constraint set for $\xi$. A sampling set consists of $N_s \in \mathbb{N}$ trajectories (shots) with $P$ measurements per shot. We consider realistic kinematic constraints on these trajectories. Let $\alpha$ denote the maximal speed of a discrete trajectory and $\beta$ denote its maximal acceleration (the slew rate). We let
\begin{equation}\label{eq:constraint_sets}
  Q_P^{\alpha,\beta} = \left\{ \xi \in ([-\pi,\pi]^D)^P, \|\dot \xi \|_\infty\leq\alpha, \|\ddot \xi \|_\infty\leq\beta, C\xi=b \right\},
\end{equation}
where 
\begin{align*}
\|\dot \xi \|_\infty &= \max_{1\leq p \leq P-1} \|\xi_{p+1}-\xi_p\|_2 \\
\|\ddot \xi \|_\infty &= \max_{2\leq p \leq P-1} \|\xi_{p+1}+\xi_{p-1} - 2 \xi_p\|_2,
\end{align*}
where $b$ is a vector and $C$ a matrix encoding some position constraints. For instance, we enforce the first point of each trajectory to start at the origin. Since the sampling schemes consists of $N_s$ trajectories, the constraint set on the sampling is $\Xi=(Q_P^{\alpha,\beta})^{N_s}$. The total number of measurements $M$ is equal to $M=N_s\cdot P$.
We refer the reader to \cite{chauffert2016projection} for more details on these constraints.

\subsection{The challenges of sampling scheme optimization}
\label{sec:first-section}

In this paper, we consider the optimization of a sampling scheme for a fixed reconstruction mapping $R$. A good sampling scheme should reconstruct the images in the training set $\mathcal{X}$ efficiently on average. Hence, a natural  optimization criterion is 
\begin{equation}\label{eq:objective_xi}
    \min_{\xi \in \Xi} \E \left(\frac{1}{K}\sum_{k=1}^K \eta\left( R(\xi,A(\xi)x_k + n,\lambda), x_k) \right)\right).
\end{equation}
The term $A(\xi)x_k$ corresponds to the measurements of the image $x_k$ associated with the sampling scheme $\xi$. The expectation is taken with respect to the term $n\in \C^N$ which models noise on the measurements.
More elaborate forward models can be designed to account for sensibility matrices in multi-coil imaging or for trajectory errors.
We will not consider these extensions in this paper. Their integration is straightforward -- at least at an abstract level.

Even if problem \eqref{eq:objective_xi} is simple to state (and very similar to \cite{weiss2021pilot, wang2022b}), the practical optimization looks extremely challenging for the following reasons:
\begin{itemize}
	\item The computation of the cost function is very costly.
	\item Computing the derivative of the cost function using backward differentiation requires differentiating a Non-uniform Fast Fourier Transform (NFFT). It also requires a consequent quantity of memory that limits the complexity of the reconstruction mapping.
	\item The energetic landscape of the functional is usually full of spurious minimizers \cite{gossard2022spurious}.
	\item The minimization of an expectation calls for the use of stochastic gradient descent, but the additional presence of a constraint set $\Xi$ reduces the number of solvers available.
\end{itemize}
Hence, the design of efficient computational solutions is a major issue. It will be the main focus of this paper.
The following sections are dedicated to the simplification of \eqref{eq:objective_xi} and to the design of a lightweight solver.
We also propose a home-made solver that attacks \eqref{eq:objective_xi} directly. Since similar ideas were proposed in \cite{wang2022b}, we describe the main ideas and differences in Appendix \ref{sec:solving_xi} only.

\subsection{Regularization and dimensionality reduction}\label{subsec:parametrization}

The non-convexity of \eqref{eq:objective_xi}  is a major hurdle inducing spurious minimizers \cite{gossard2022spurious}.  We discuss the existing solutions to mitigate this problem and give our solution of choice.

\subsubsection{Existing strategies and their limitation}
In \cite{weiss2021pilot,wang2022b}, the authors propose to avoid local minima by using a multi-scale optimization approach starting from a trajectory described through a small number of control points and progressively getting more complex through the iterations.
The use of the stochastic Adam optimizer can also allow escaping from narrow basins of attraction. In addition, Adam optimizer can be seen as a preconditioning technique, which can accelerate the convergence, especially for the high frequencies \cite{gossard2022spurious}.
This optimizer together with a multi-scale approach can yield sampling schemes with improved reconstruction quality at the cost of a long training process.
However, despite heuristic approaches to globalize the convergence, we experienced significant difficulties in getting reproducible results.

\begin{figure}
    \centering
    \begin{subfigure}[t]{.24\linewidth}
        \centering
        \includegraphics[width=1.0\textwidth]{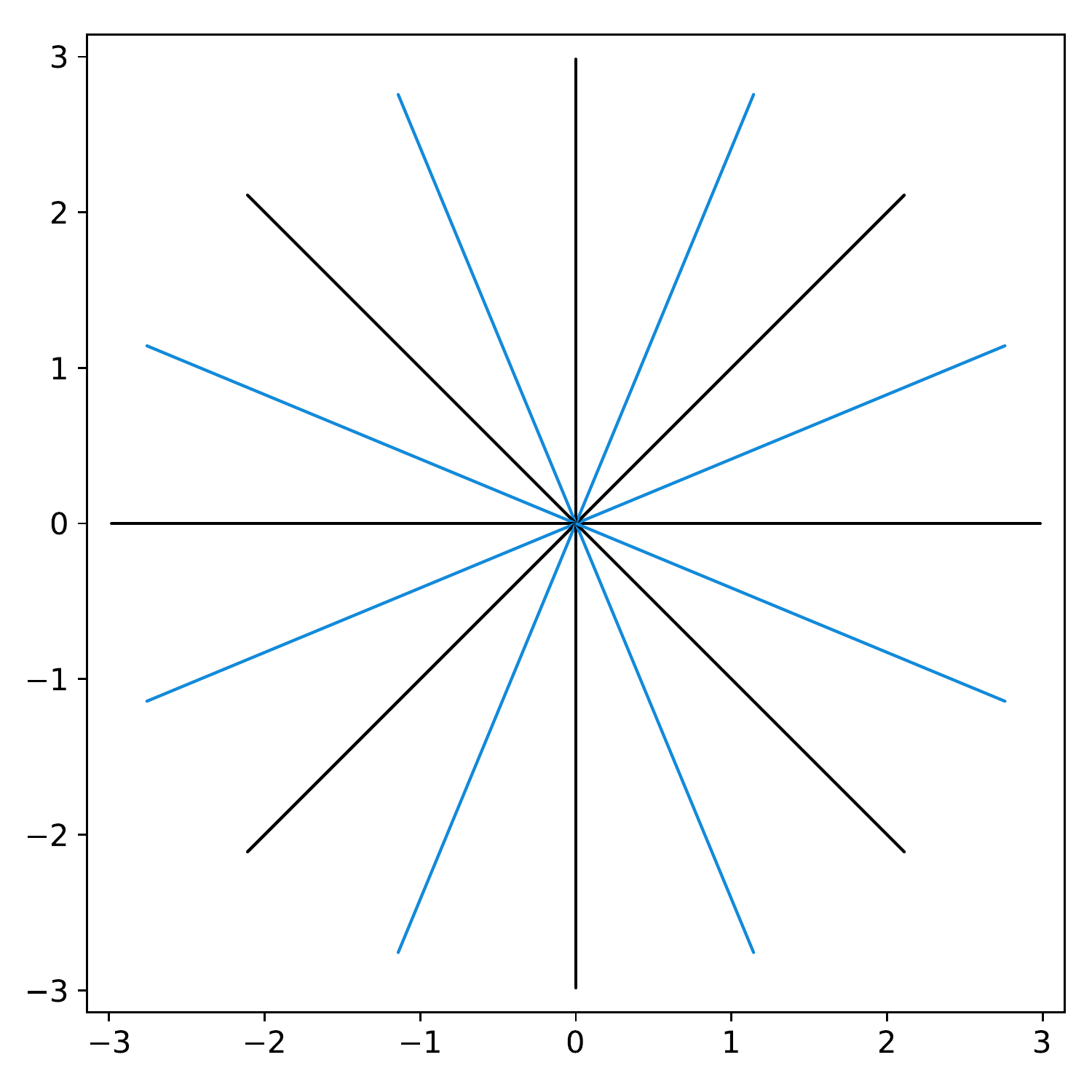}
        \caption{Radial init.}
    \end{subfigure}
    \begin{subfigure}[t]{.24\linewidth}
        \centering
        \includegraphics[width=1.0\textwidth]{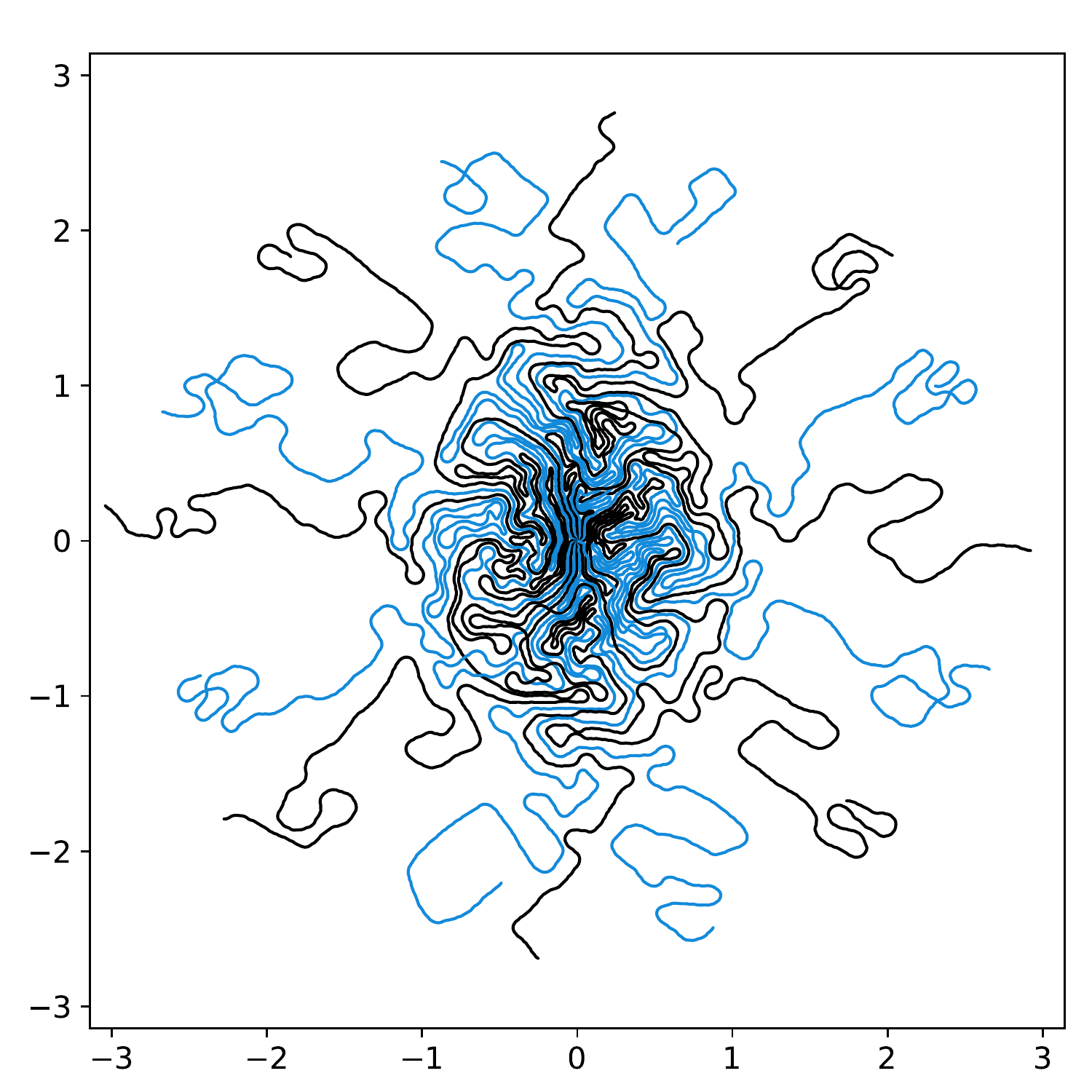}
        \caption{Optimized scheme\\$33.48$dB}
    \end{subfigure}
    \begin{subfigure}[t]{.24\linewidth}
        \centering
        \includegraphics[width=1.0\textwidth]{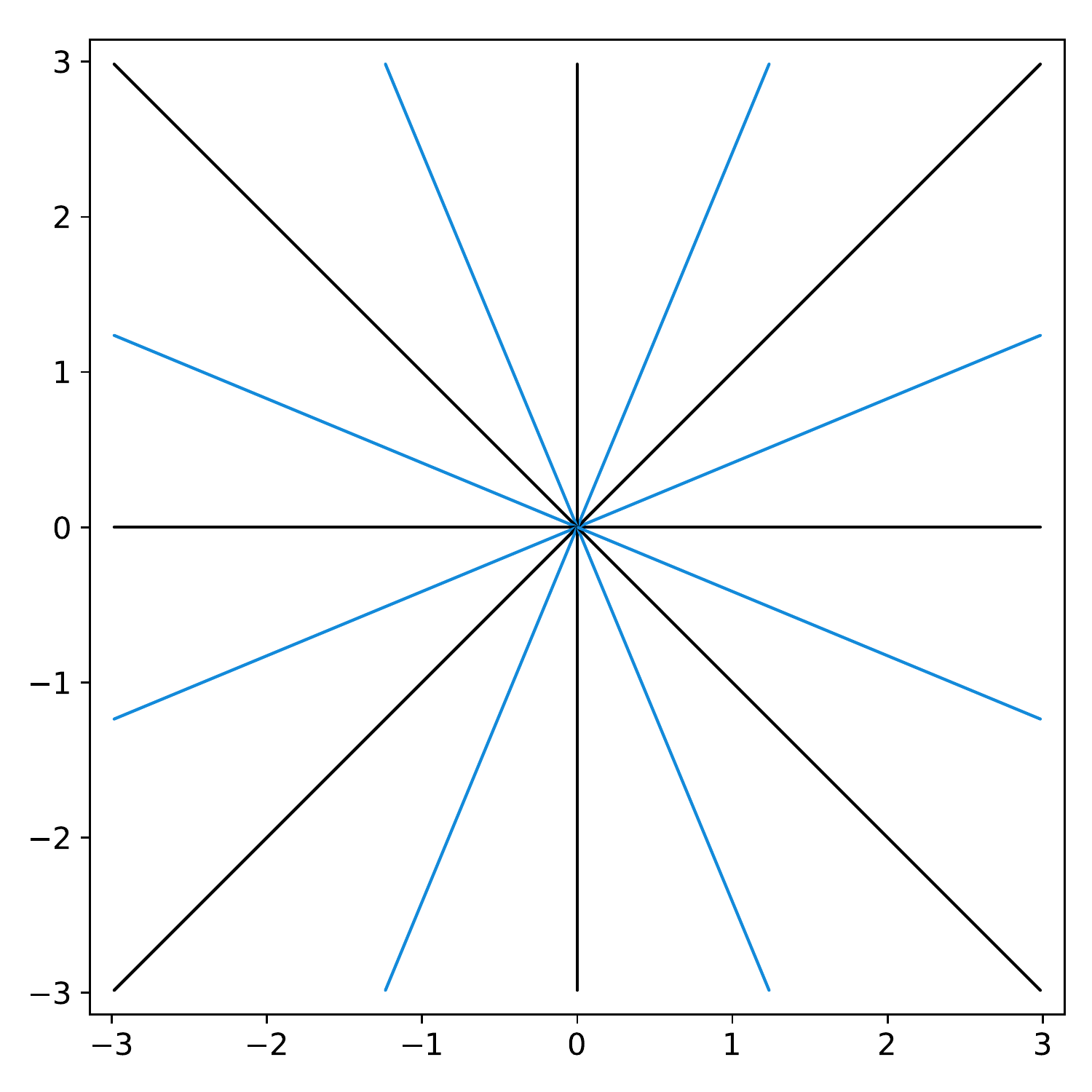}
        \caption{Box init.}
    \end{subfigure}
    \begin{subfigure}[t]{.24\linewidth}
        \centering
        \includegraphics[width=1.0\textwidth]{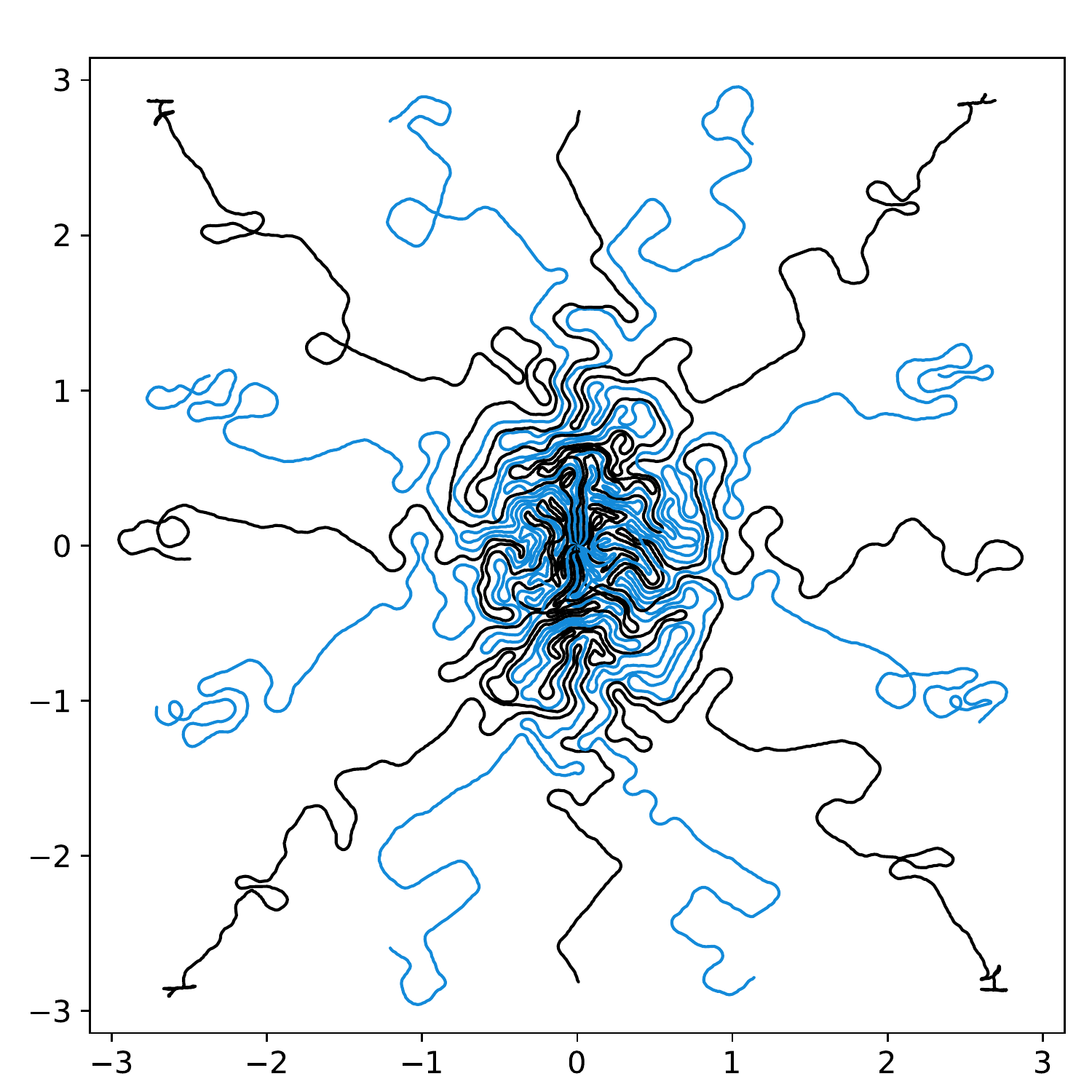}
        \caption{Optimized scheme\\ $33.17$dB}
    \end{subfigure}
    \caption{The globalization issue: optimizing a scheme with an advanced multi-scale approach yields different average PSNR when starting from different point configurations. In this experiment, we used a total variation reconstruction algorithm and $10\%$ undersampling.}
    \label{fig:optimmultiTVinit}
\end{figure}

To illustrate this fact, we conducted a simple experiment in Fig.~\ref{fig:optimmultiTVinit}.
Starting from two similar initial sampling trajectories, we let a multi-scale solver run for 14 epochs and 85 hours on the fastMRI knee database.
We then evaluate the average reconstruction peak signal-to-noise ratio (PSNR) on the validation set.
As can be seen, the final point configuration \rev{and the average performance varies by $0.3$dB, which is significant.}
\rev{This suggests that the algorithm was trapped in a spurious local minimizer and illustrates the difficulty to globalize the convergence.}

\subsubsection{Optimizing a sampling density}\label{subsubsec:reparam}

The key idea in this paper is to regularize the problem by optimizing a sampling density rather than the point positions directly. 
To formalize this principle we need to introduce two additional ingredients:
\begin{enumerate}
    \item A \emph{probability density generator} $\rho:\R^L\to \mathcal{P}$, where $\mathcal{P}$ is the set of probability distributions on $\R^D$. In this paper, $\rho$ will be defined as a simple affine mapping, but we could also consider more advanced generators such as Generative Adversarial Networks.
    \item A \emph{trajectory sampler} $\Sc_M:\Pc \to (\R^D)^M$, which maps a density $\rho$ to a point configuration $\Sc_M(\rho) \in (\R^D)^M$. Various possibilities could be considered such as Poisson point sampling, Poisson disk sampling. In this paper, we will use discrepancy based methods \cite{boyer2016generation}.
\end{enumerate}

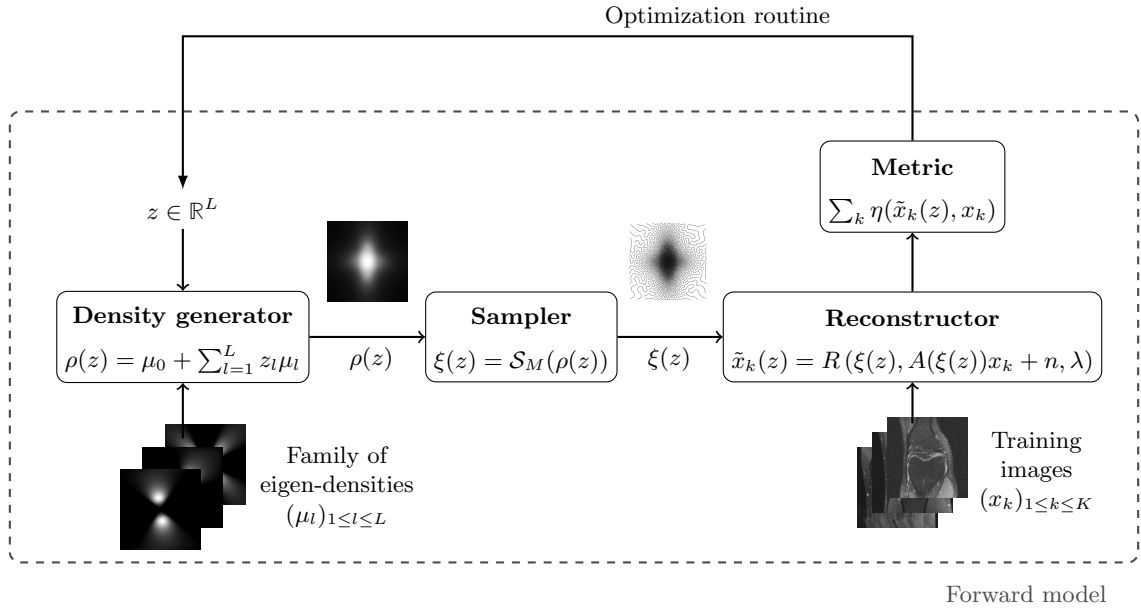
\begin{figure*}
    \begin{center}
      \tikzset{transition/.style = {rectangle, rounded corners, draw=black!70, dashed, thick, minimum width=15cm, minimum height=6cm},}
      \begin{tikzpicture}
        \footnotesize
        \node[align=center] (input) at (0.5,2.7) {$z\in\R^L$};
        \node[draw, rounded corners, align=center, minimum height=0.9cm, text height=0.9cm] (denformula) at (0.5,1) {$\rho(z)=\mu_0+\sum_{l=1}^L z_l\mu_l$};
        \node[align=center] (density_title) at ([yshift=-10pt]denformula.north) {\textbf{Density generator}};

        \node[align=center] (den1) at (0.8,-0.7)     {\includegraphics[width=0.07\textwidth]{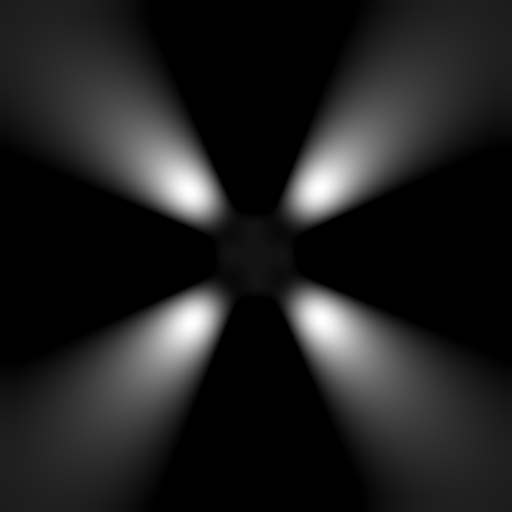}};
        \node[align=center] (den2) at (0.5,-1) {\includegraphics[width=0.07\textwidth]{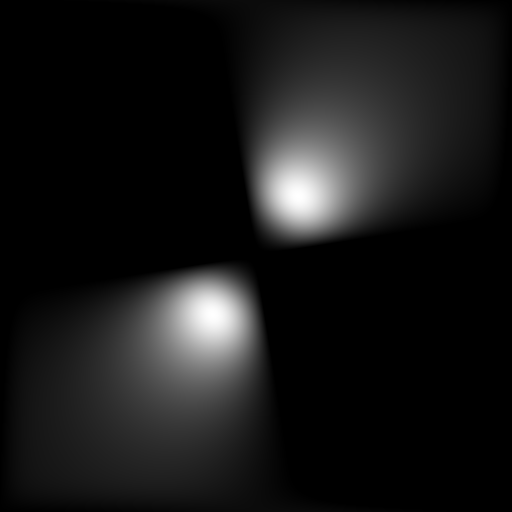}};
        \node[align=center] (den3) at (0.2,-1.3) {\includegraphics[width=0.07\textwidth]{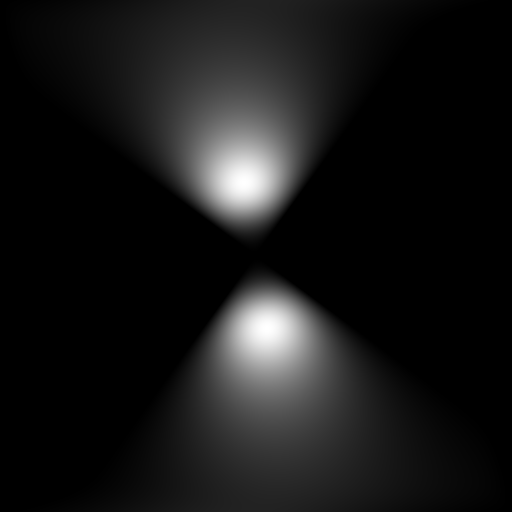}};
        \node[align=center] (text_family_den) at ([xshift=1.4cm]den2.east) {Family of\\ eigen-densities\\ $(\mu_l)_{1 \leq l\leq L}$};

        \node[align=center] (denTV) at (2.95,2)  {\includegraphics[width=0.07\textwidth]{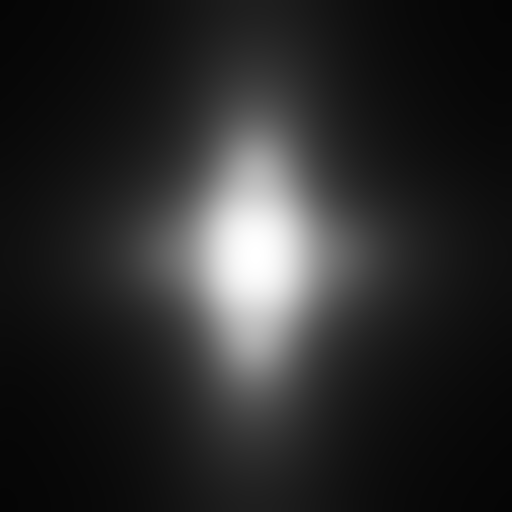}};
        \node[align=center] (xiTV)  at (6.95,2)  {\includegraphics[width=0.07\textwidth]{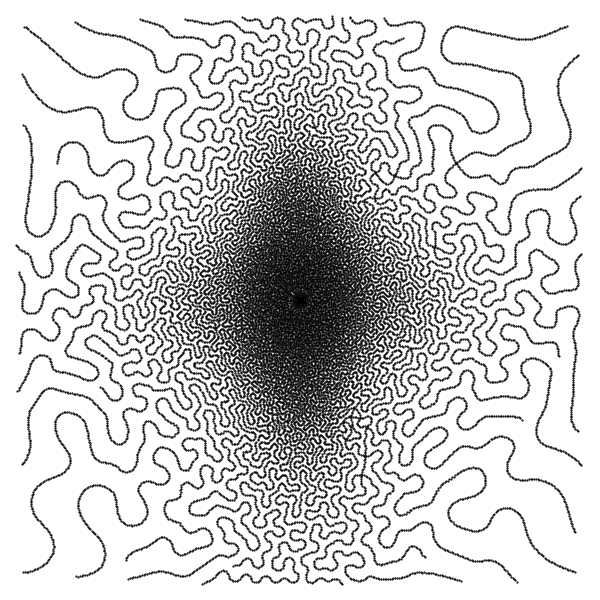}};

        \node[draw, rounded corners, align=center, minimum height=0.9cm, text height=0.9cm] (sampler) at (5,1) {$\xi(z) = \Sc_M(\rho(z))$};
        \node[align=center] (density_title) at ([yshift=-10pt]sampler.north) {\textbf{Sampler}};

        \node[align=center] (img3) at (10,-1) {\includegraphics[width=0.07\textwidth]{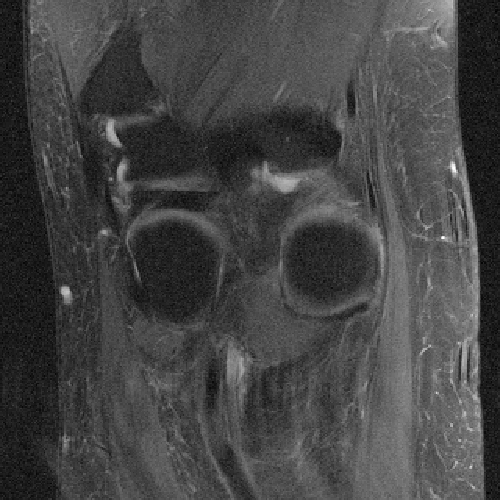}};
        \node[align=center] (img2) at (10.2,-0.8) {\includegraphics[width=0.07\textwidth]{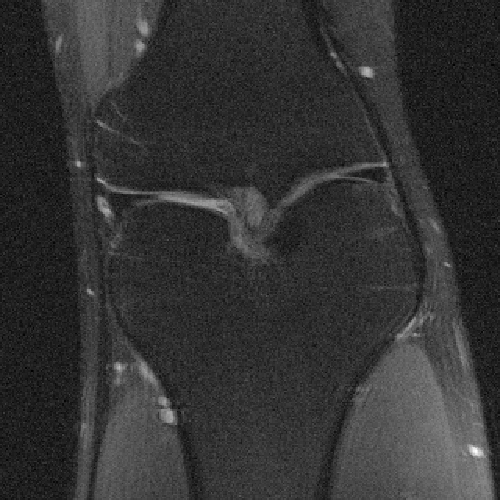}};
        \node[align=center] (img1) at (10.4,-0.6) {\includegraphics[width=0.07\textwidth]{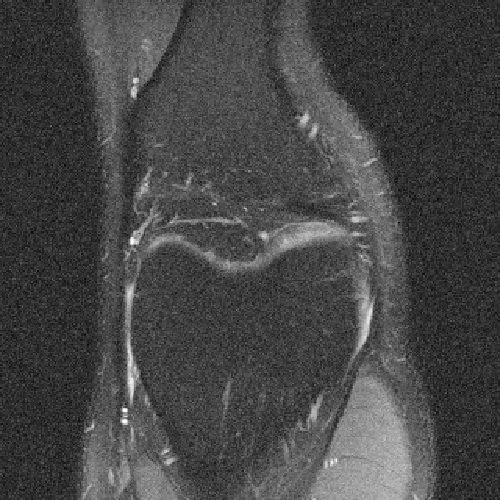}};
        \node[align=center] (text_training_img) at ([xshift=1cm]img2.east) {Training\\ images \\ $(x_k)_{1 \leq k\leq K}$};

        \node[draw, transition] (forward) at (5.7,1) {};
        \node[align=center, black!70] (text_forward) at ([yshift=-0.4cm, xshift=-1.5cm]forward.south east) {Forward model};

        \node[draw, rounded corners, align=center, minimum height=0.9cm, text height=0.9cm] (recon) at (10.2,1) {$\tilde x_k(z) = R\left(\xi(z),A(\xi(z))x_k+n,\lambda\right)$};
        \node[align=center] (recon_title) at ([yshift=-10pt]recon.north) {\textbf{Reconstructor}};
        \node[draw, rounded corners, align=center, minimum width=1.5cm, minimum height=0.9cm, text height=0.9cm] (metric) at (10.2,3) {$\sum_k \eta(\tilde x_k(z),x_k)$};
        \node[align=center] (metric_title) at ([yshift=-10pt]metric.north) {\textbf{Metric}};
        \draw[->,draw,thick,black] (input.south) to (denformula.north);
        \draw[->,draw,thick,black] (denformula.east) to (sampler.west) node [left,xshift=-0.3cm,yshift=-0.3cm] {$\rho(z)$};
        \draw[->,draw,thick,black] (sampler.east) to (recon.west) node [left,xshift=-0.3cm,yshift=-0.3cm] {$\xi(z)$};
        \draw[->,draw,thick,black] (recon.north) to (metric.south);
        \draw[->,draw,thick,black] (den2.north) to (denformula.south);
        \draw[->,draw,thick,black] (img2.north) to (recon.south);

        \path[draw,thick,-latex] (metric) -- ++(0,2) -| (input) node [pos=0.05,left,yshift=0.25cm] {Optimization routine};
      \end{tikzpicture}
    \end{center}
    \caption{A schematic illustration of the proposed algorithm. We generate a sampling density $\rho(z)$ through an affine combination of eigen-elements $(\mu_l)$. The density is then used in a sampling pattern generator $\Sc_M$ which yields a sampling trajectory $\xi(z)$. A set of training images are then reconstructed using this scheme. This allows computation of the (batch) average error. A zero-th, or first order (automatic differentiation) optimization routine optimizes the sampling density iteratively. \label{fig:optimizationdensity}}
\end{figure*}

Instead of minimizing \eqref{eq:objective_xi}, we propose to work directly with the density. 
Letting $\xi:\R^L\to (\R^D)^M$ denote the mapping defined by
\begin{equation}
\xi(z) \eqdef \Sc_M(\rho(z)),
\end{equation}
we propose to minimize:
\begin{equation}\label{eq:objective_rho}
F(z) \eqdef \min_{z \in \Cc\subset \R^L} \frac{1}{K}\sum_{k=1}^K \E\left(\eta\left[ R( \xi(z) ,A(\xi(z))x_k + n,\lambda), x_k \right]\right),
\end{equation}
where the expectation is taken with respect to the noise term $n$. A schematic illustration of this approach is proposed in Fig.~\ref{fig:optimizationdensity}.

\subsubsection{The density generator}\label{subsubsec:dengenerator}

Various approaches could be used to define a density generator $\rho$.
In this work, we simply define $\rho(z)$ as an affine mapping, i.e.
\begin{equation}
    \rho(z) \eqdef \mu_0 + \sum_{l=1}^L z_l \mu_l,
\end{equation}
where $z$ belongs to a properly defined convex set $\Cc$. 
We describe hereafter how the eigen-elements $(\mu_l)_l$ and the set $\Cc$ are constructed.

\paragraph{A candidate space of densities}

The general idea of our construction is to define a family of elementary densities and to enrich it by taking convex combinations of its elements. 

Let $\theta\in [0,\pi[$ denote a rotation angle, $\sigma_x,\sigma_y$ denote lengths, $r>0$ denote a density at the center and $\gamma>0$ a decay rate. 
For $(x,y)\in \R^2$, let $x_\theta = x\cos(\theta)+y\sin(\theta)$, $y_\theta=-\sin(\theta)x + \cos(\theta)y$. We define
\begin{equation}\label{eq:family_of_densities}
    g(x,y;\sigma_x,\sigma_y,\theta, r, \gamma) = \frac{1}{c} \min\left(r, \frac{1}{ \left(\left(\frac{x_\theta}{\sigma_x}\right)^2+\left(\frac{y_\theta}{\sigma_y}\right)^2+\epsilon \right)^\gamma}\right),
\end{equation}
where $c$ is a normalizing constant such that $\int_{\R^2} g = 1$. We then smooth the function $g$ by convolving it with a Gaussian function $G_\kappa$ of standard deviation $\kappa>0$:
\begin{equation}
\pi = G_\kappa \star g.
\end{equation}

\begin{figure*}
    \centering
    \begin{subfigure}[t]{0.49\textwidth}
        \centering
        \includegraphics[width=0.2\textwidth]{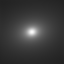}%
        \includegraphics[width=0.2\textwidth]{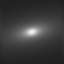}%
        \includegraphics[width=0.2\textwidth]{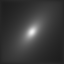}%
        \includegraphics[width=0.2\textwidth]{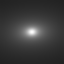}%
        \includegraphics[width=0.2\textwidth]{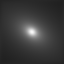}%
        \caption{Examples of $\pi_i$ \label{fig:exampledensities1}}
    \end{subfigure}\hfill
    \begin{subfigure}[t]{0.49\textwidth}
        \centering
        \includegraphics[width=0.2\textwidth]{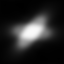}%
        \includegraphics[width=0.2\textwidth]{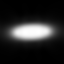}%
        \includegraphics[width=0.2\textwidth]{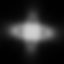}%
        \includegraphics[width=0.2\textwidth]{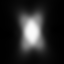}%
        \includegraphics[width=0.2\textwidth]{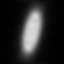}%
        \caption{Examples of $\rho(z)$ \label{fig:exampledensities2}}
    \end{subfigure}\\
    \begin{subfigure}[t]{\textwidth}
        \centering
        \includegraphics[width=0.1\textwidth]{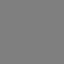}%
        \includegraphics[width=0.1\textwidth]{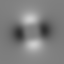}%
        \includegraphics[width=0.1\textwidth]{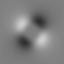}%
        \includegraphics[width=0.1\textwidth]{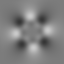}%
        \includegraphics[width=0.1\textwidth]{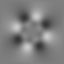}%
        \includegraphics[width=0.1\textwidth]{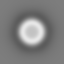}%
        \includegraphics[width=0.1\textwidth]{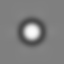}%
        \includegraphics[width=0.1\textwidth]{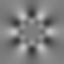}%
        \includegraphics[width=0.1\textwidth]{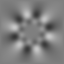}%
        \includegraphics[width=0.1\textwidth]{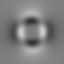}%
        \caption{$\mu_l$ for $0\leq l\leq 9$ \label{fig:exampledensities3}}
    \end{subfigure}
    \caption{Examples of densities using the proposed parameterization.}
    \label{fig:exampledensities}
\end{figure*}

The elements in this family are good candidates for sampling densities: i) they are nearly constant and approximately equal to $r$ at the center of the k-space, ii) they can be anisotropic to accommodate for specific image orientations and iii) they have various decay rates, allowing sampling the high frequencies more or less densely. Some examples of such densities are displayed in Fig.~\ref{fig:exampledensities1}. 
However, the family of densities generated by this procedure is quite poor. For instance, it is impossible to sample densely both the $x$ and $y$ axes simultaneously. 
In order to enrich it, we propose to consider the set of convex combinations of these elementary densities.
This allows us to construct more general multi-modal densities, see  Fig.~\ref{fig:exampledensities2} for examples of such convex combinations.

\paragraph{Dimensionality reduction}

In order to construct the family $(\mu_0,\hdots,\mu_L)$, we first draw a large family of $I\gg L$ densities $(\pi_i)_{1\leq i \leq I}$.
They are generated at random by uniform draws of the parameters $(\sigma_x,\sigma_y,\theta, t, \gamma)$ inside a box. 
We then perform a principal component analysis (PCA) on this family to generate some eigen-elements $(\nu_l)_{0\leq l \leq L}$.
We set $\mu_0 = \nu_0/\langle \nu_0, \one\rangle$. 
Since probability densities must sum to $1$, we orthogonalize the family $(\nu_l)$ with respect to the vector $\mu_0$. 
Thereby, we obtain a second family $(\mu_l)_{0\leq l\leq L}$ that satisfies $\langle \mu_0 , \one\rangle = 1$ and $\langle \mu_l, \one\rangle = 0$ for all $1\leq l\leq L$. This procedure discards one dimension. The resulting PCA basis is illustrated in Fig.~\ref{fig:exampledensities3}. 

Let $\mathcal E$ denote the intersection of the span of $(\mu_l)_{l\le L}$ with the probability densities and $\Pi_{\mathcal E}$ the orthogonal projection on $\mathcal E$. The space of densities is the convex hull of the family $(\pi_i)_i$ projected on $\mathcal E$:
\begin{equation}\label{eq:convex_hull}
\Cc \eqdef \text{Conv}\left(\Pi_{\mathcal E}\left(\pi_i\right), 1\leq i \leq I \right).
\end{equation}
As illustrated in Fig.~\ref{fig:exampledensities2}, this process overall provides a rather rich and natural family.
\rev{In practice, we select a value $L=20$, so that the $\ell_2$ tail of the singular values contains less than $1\%$ of the energy.
This value is also a compromise between numerical complexity (the higher $L$, the more complex) and the richness of the family of densities that can be generated.}

\subsubsection{The sampler}
\label{sec:sampler}

The sampler $\Sc_M:\Pc \to (\R^D)^M$ is based on discrepancy minimization \cite{schmaltz2010electrostatic,graf2012quadrature,chauffert2017projection}. It is defined as an approximate solution of
\begin{equation}\label{eq:discrepancy}
    \Sc_M(\rho) = \argmin_{\xi \in \Xi} \mathrm{dist}\left(\frac{1}{n}\sum_{m=1}^M \delta_{\xi[m]} , \rho\right),
\end{equation}
where \rev{$\delta_{\xi[m]}$ indicates the Dirac delta function, $\Xi\subset(\R^D)^M$ is the set of feasible discrete trajectories} and $\mathrm{dist}$ is a discrepancy defined by
\begin{equation*}
    \mathrm{dist}(\mu,\nu) = \sqrt{\langle h\star (\mu-\nu), (\mu-\nu)\rangle_{L^2(\R^D)}},
\end{equation*}
where $h$ is a positive definite kernel (i.e. a function with a real positive Fourier transform). 
Other metrics on the set of probability distributions could be used such as the transportation distance \cite{lebrat2019optimal}.
The formulation \eqref{eq:discrepancy} has already been proposed in \cite{chauffert2017projection} and it is at the core of the Sparkling scheme generation \cite{lazarus2019sparkling}. We will discuss the choice of the kernel $h$ in the numerical experiments: it turns out to play a critical role.

In practice \eqref{eq:discrepancy} is not solved exactly: an iterative solver \cite{chauffert2016projection} is ran for a fixed number of iterations. This allows the use of automatic differentiation in order to compute the Jacobian of $\xi$ w.r.t. $z$.
Technical details about the implementation of this sampler are provided in Appendix~\ref{subsec:computational_details}.

\subsubsection{The pros and cons of this strategy}

The optimization problem \eqref{eq:objective_rho} presents significant advantages compared to the original one \eqref{eq:objective_xi}:
\begin{itemize}
    \item The number of optimization variables is considerably reduced: instead of working with $D\cdot M$ variables, we now only work with $L \ll D\cdot M$ variables defining a continuous density. In this paper we set $L=20$ which is considerably smaller in comparison to the $M=25 801$ 2D sampling points for the formulation of \eqref{eq:objective_xi} with $25\%$ undersampling on $320\times 320$ images. This allows resorting to global optimization routines. Hereafter, we will describe a Bayesian optimization approach.
    \item The point configurations generated by this algorithm are always locally uniform since they correspond to the minimizers of a discrepancy. Clusters are therefore naturally discarded, which can be seen as a natural regularization scheme.
    \item As discussed in the numerical experiments, the regularization effect allows optimizing the sampling density with a small dataset with a similar performance. Optimizing the function with as little as $32$ reference images yields a near optimal density.
    This aspect might be critical for small databases.
\end{itemize}

On the negative side, notice that we considerably constrained the family of achievable trajectories, thereby reducing the maximal achievable gain. We 
will show later that the trajectories obtained by minimizing \eqref{eq:objective_rho} are indeed slightly less efficient than those obtained with \eqref{eq:objective_xi}. This price might be affordable if we compare it to the advantages of having a significantly faster and more robust solver requiring only a fraction of the data needed for solving \eqref{eq:objective_xi}.

\subsection{The optimization routine}\label{subsec:bayesian_optim}

In this section, we describe an algorithmic approach to attack the problem \eqref{eq:objective_rho}.

\subsubsection{The non informativeness of the gradient}\label{subsec:bayesian_optim:gradient}

\begin{figure*}[htbp]
    \centering
        \begin{subfigure}[b]{0.65\textwidth}
            \centering
            \includegraphics[width=1\textwidth]{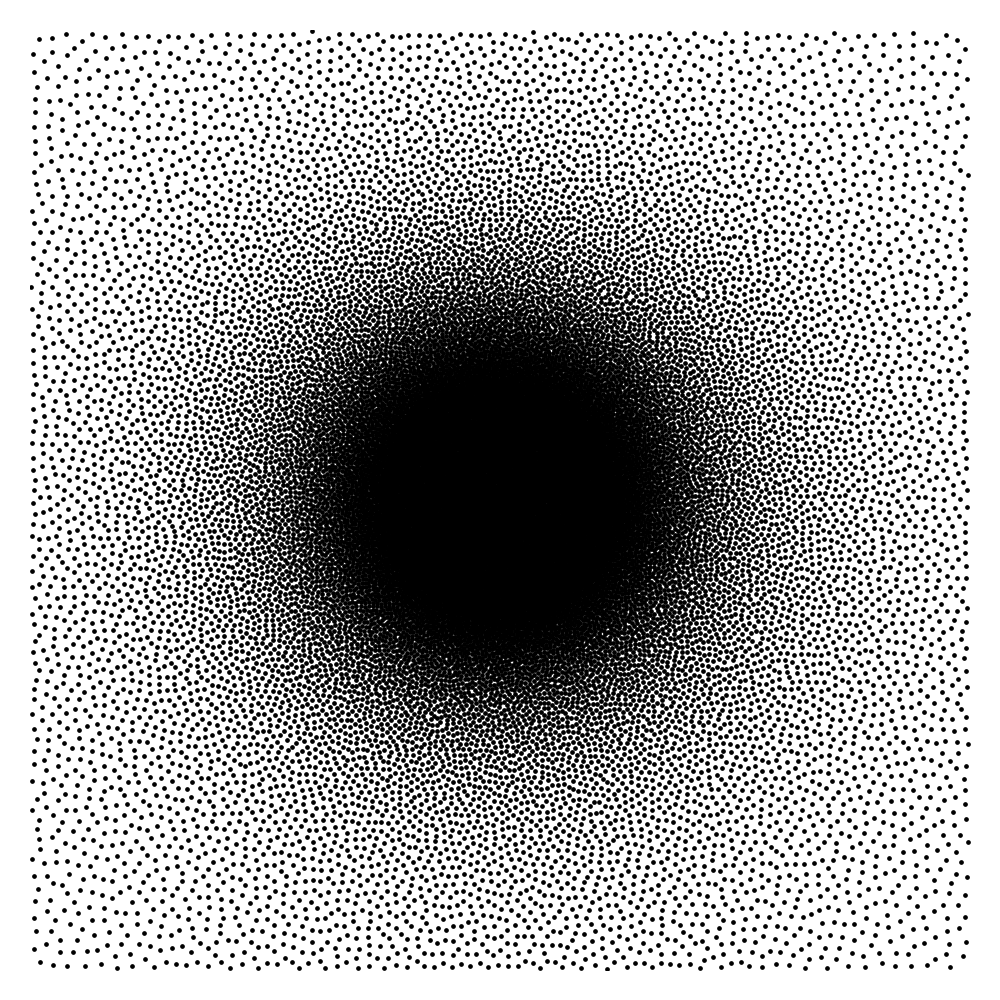}
            \caption{Sampling scheme \\ \ \label{fig:oscillations_TV_left}}
        \end{subfigure} 
        \begin{subfigure}[b]{0.34\textwidth}
            \begin{center}
                \begin{tikzpicture}
                    \node[align=center] at (0,0){\includegraphics[height=3.3cm]{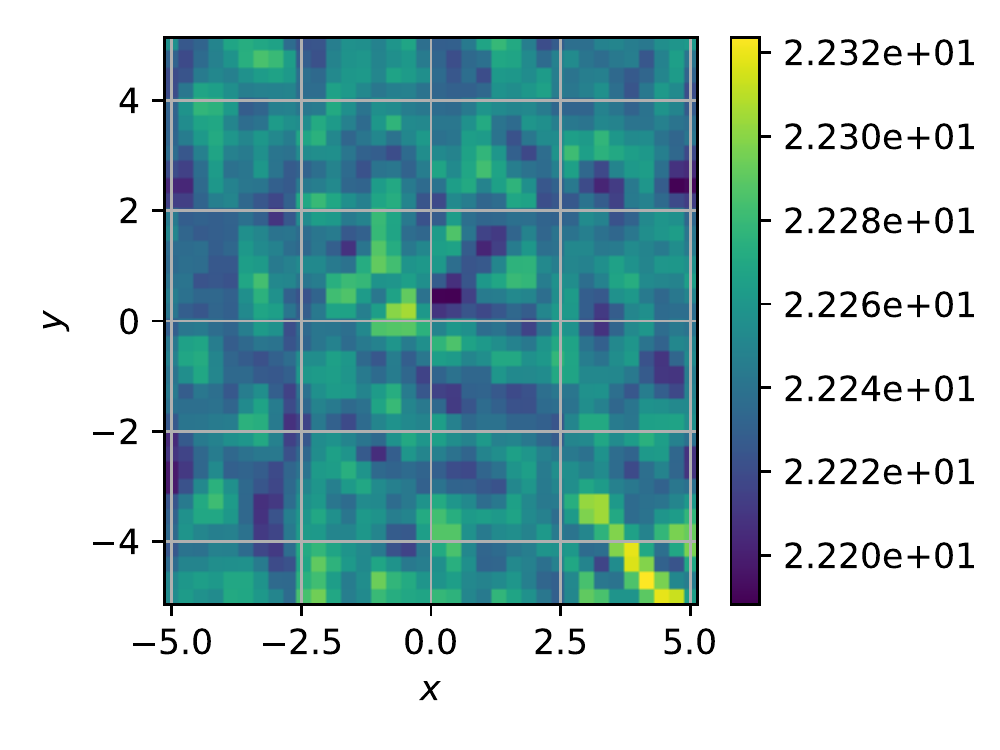}};
                    \node[align=center] at (0,3.3){\includegraphics[height=3.3cm]{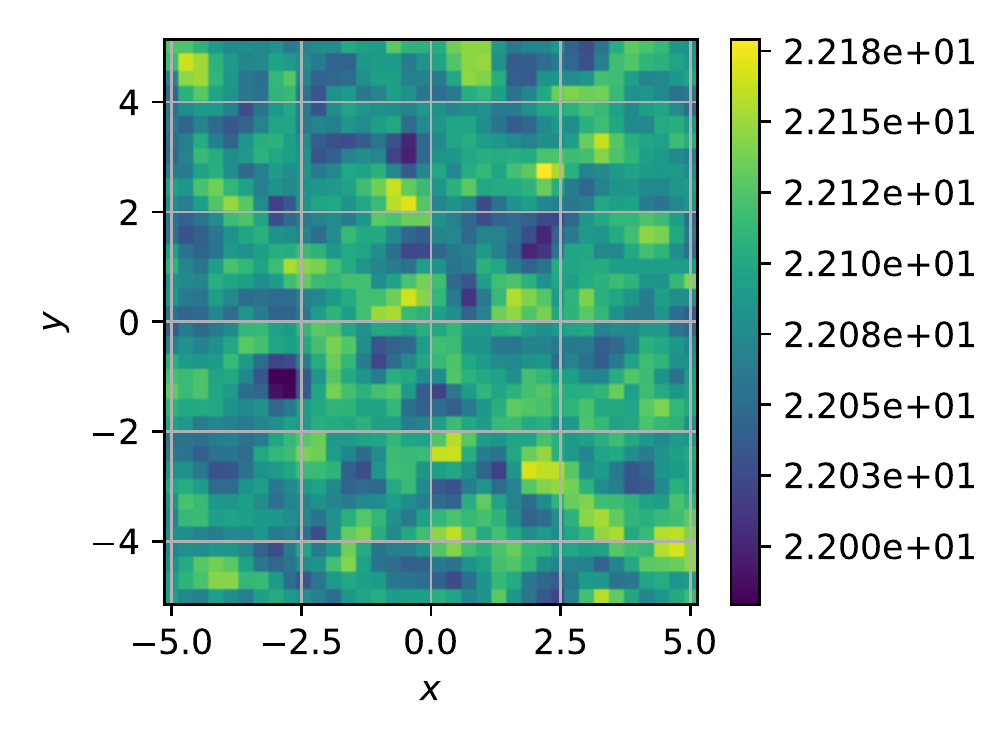}};
                    \node[align=center] at (0,6.6){\includegraphics[height=3.3cm]{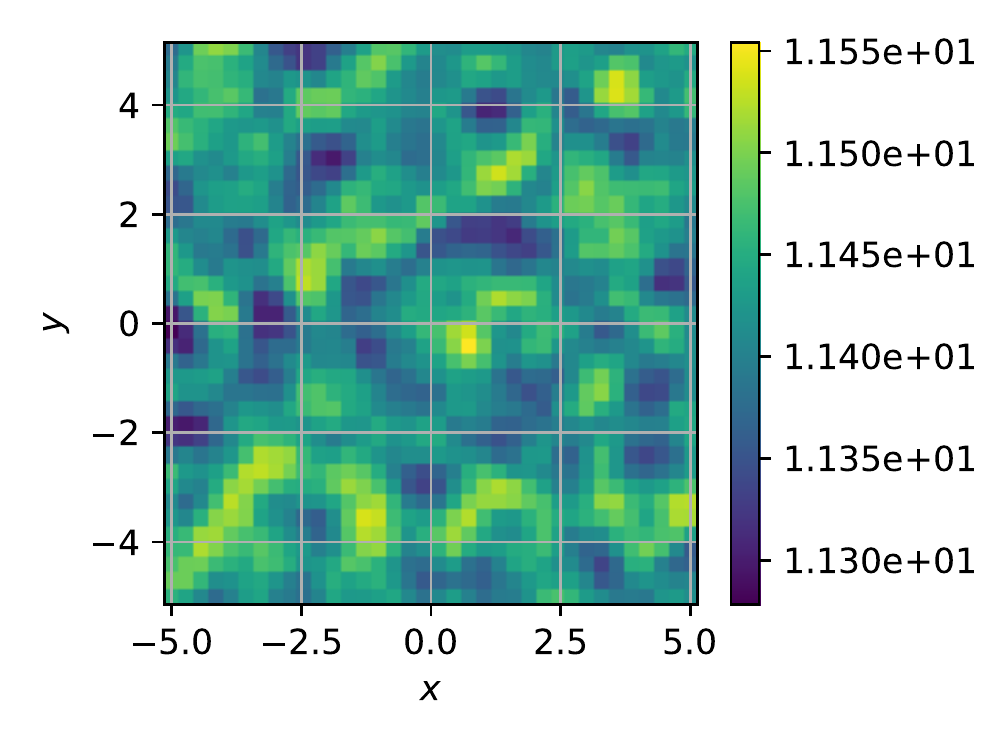}};
                \end{tikzpicture}
            \end{center}
            \vspace{-2em}
            \caption{Energy profiles for \\ $K\in \{1, 10, 100\}$ images. \label{fig:energyprofiles}}
        \end{subfigure}
    \caption{\rev{Spurious minimizers through the shift experiment. Here, the sampling scheme on the left \ref{fig:oscillations_TV_left} is continuously shifted on the x and y axes. For each $(x,y)$-shift position, we reconstruct a set of $K$ images, and evaluate the mean square reconstruction error. This way, we can visualize a slice of the energy profile. The results are show in Fig.~\ref{fig:energyprofiles} for $K=1,10,100$ images from top to bottom. In this Figure, the gray grid indicates the Shannon sampling distance.  Here, we consider a total variation reconstructor and $25\%$ undersampling. Observe that the oscillation amplitude decays with the number of images, but spurious minimizers are present whatever the number of images.\label{fig:oscillations_TV}}}
\end{figure*}

\begin{figure*}[htbp]
    \newcommand{\fle}{3.3cm}
    \newcommand{\fli}{2.8cm}
    \centering
    \begin{subfigure}[b]{0.65\textwidth}
        \begin{center}
            \begin{tikzpicture}
                \footnotesize
                \node[align=center] (im02) at (0,2*\fle) {\includegraphics[width=\fli]{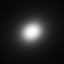}};
                \node[align=center] (im12) at (\fle,2*\fle) {\includegraphics[width=\fli]{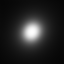}};
                \node[align=center] (im22) at (2*\fle,2*\fle) {\includegraphics[width=\fli]{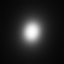}};
                \node[align=center] (im01) at (0,1*\fle) {\includegraphics[width=\fli]{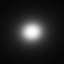}};
                \node[align=center] (im11) at (\fle,\fle) {\includegraphics[width=\fli]{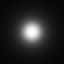}};
                \node[align=center] (im21) at (2*\fle,\fle) {\includegraphics[width=\fli]{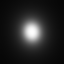}};
                \node[align=center] (im00) at (0,0) {\includegraphics[width=\fli]{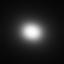}};
                \node[align=center] (im10) at (\fle,0) {\includegraphics[width=\fli]{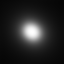}};
                \node[align=center] (im20) at (2*\fle,0) {\includegraphics[width=\fli]{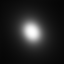}};
                \draw[->,thick] ([yshift=-0.2cm]im00.south west) -- ([yshift=-0.2cm]im20.south east) node [pos=0.5,below,font=\footnotesize] {$x$};
                \draw[->,thick] ([xshift=-0.2cm]im00.south west) -- ([xshift=-0.2cm]im02.north west) node [pos=0.5,left,font=\footnotesize] {$y$};
            \end{tikzpicture}
        \end{center}
        \caption{Sampling densities \label{fig:oscillations_TV2:1}}
    \end{subfigure}
    \begin{subfigure}[b]{0.34\textwidth}
        \begin{center}
            \begin{tikzpicture}
                \node[align=center] at (0,0){\includegraphics[height=3.6cm]{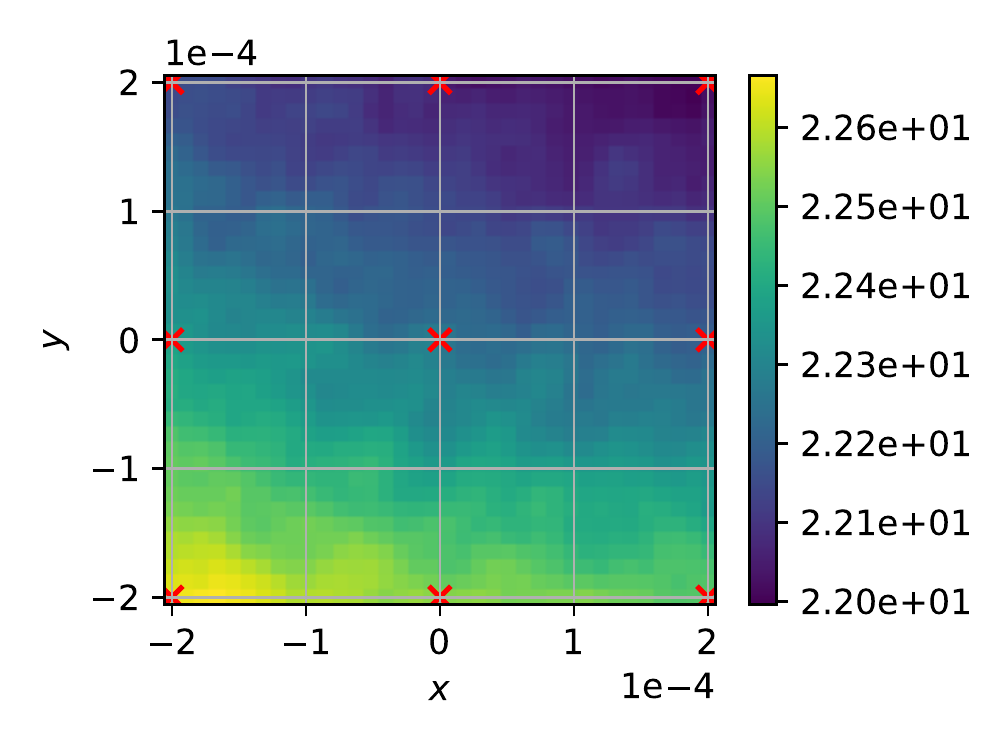}};
                \node[align=center] at (0,3.5){\includegraphics[height=3.6cm]{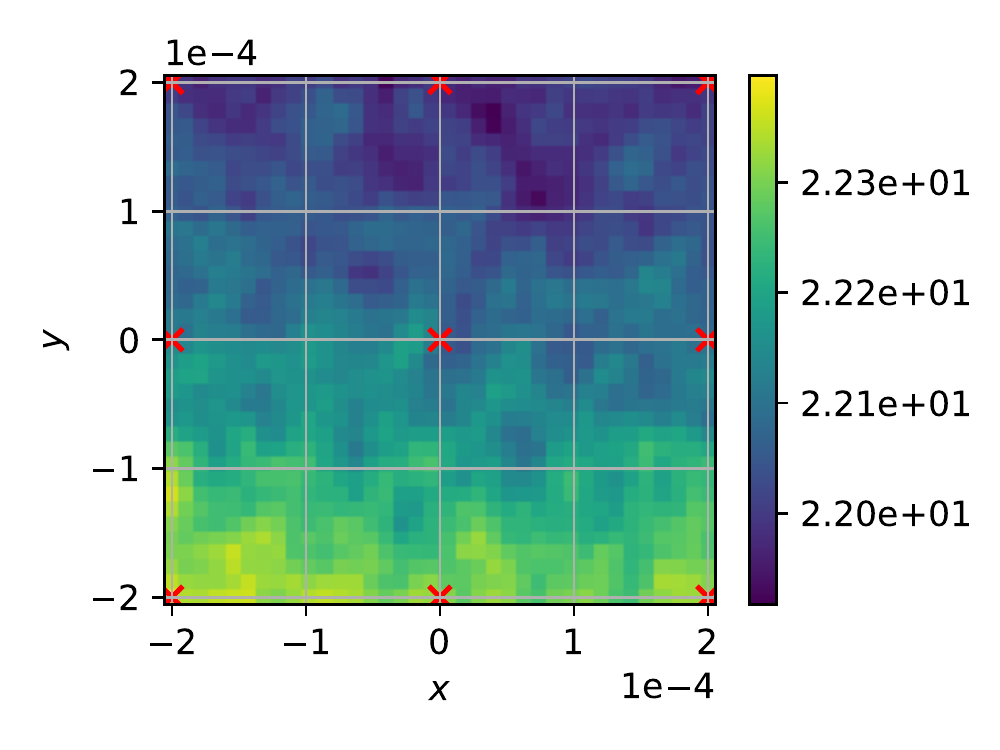}};
                \node[align=center] at (0,7){\includegraphics[height=3.6cm]{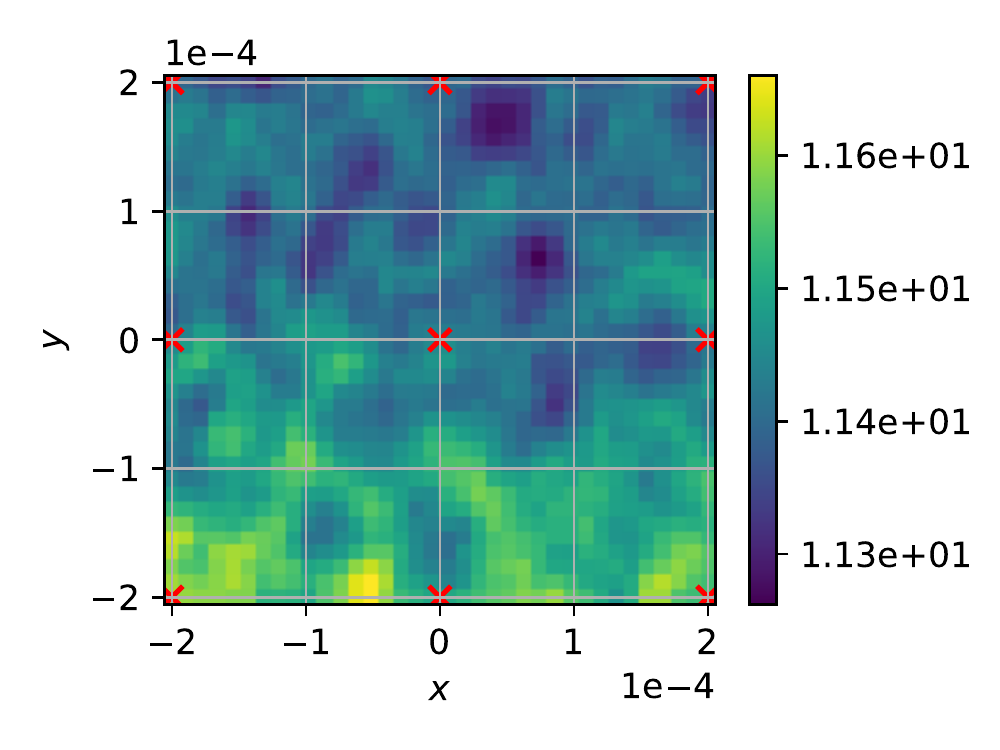}};
            \end{tikzpicture}
        \end{center}
        \vspace{-2em}
        \caption{Energy profiles for \\ $K\in \{1, 10, 100\}$ images. \label{fig:oscillations_TV2:2}}
    \end{subfigure}
    \caption{\rev{Spurious minimizers with variations in the density space. Different sampling schemes are generated with the Sparkling algorithm for various target densities. The densities are linear combinations the two leading principal components, indexed by $x$ and $y$ in Fig.~\ref{fig:oscillations_TV2:1}. The sampling scheme corresponding to the density in the middle ($x=y=0$) is displayed in Fig.~\ref{fig:oscillations_TV_left}. The plots in Fig.~\ref{fig:oscillations_TV2:2} show the energy profile for continuously varying target densities. The number of images $K$ used to evaluated the energy varies from $K=1$ (top) to $K=100$ (bottom). The 3x3 densities in Fig.~\ref{fig:oscillations_TV2:2} are indicated by red crosses in Fig.~\ref{fig:oscillations_TV2:2}.  We consider a total variation reconstructor and $25\%$ undersampling. Similarly to the experiment in Fig.~\ref{fig:energyprofiles}, observe the numerous spurious minimizers occuring whatever the number $K$ of images in the database.\label{fig:oscillations_TV2}}}
\end{figure*}

\rev{A natural approach to solve \eqref{eq:objective_rho} is to optimize the coefficients $z\in\R^{L}$ using a gradient based algorithm. Indeed, one may hope that the reparameterization of the cost function with a density prevents the appearance of spurious minimizers described in \cite{gossard2022spurious}. Unfortunately, this is not the case and gradient based algorithms might be trapped in such local minimizers.
Fig.~\ref{fig:oscillations_TV} and \ref{fig:oscillations_TV2} illustrate this fact.
In Fig.~\ref{fig:oscillations_TV}, the baseline sampling scheme is shifted continuously in the $x$ and $y$ directions. The cost function is evaluated for each shift and displayed in the right. Observe that many local minimizers are present.
Similarly, in Fig.~\ref{fig:oscillations_TV2}, a target density is varied continuously in a subspace consisting of $L=2$ eigen-elements $(\mu_1,\mu_2)$.
Again, the energy profiles on the right are highly oscillatory.}

Overall, this experiment shows that the gradient direction is not meaningful: it oscillates in an erratic way.
This advocates for the use of 0\textsuperscript{th} order optimization methods.
A significant advantage of this observation is that it allows discarding the memory and time issues related to automatic differentiation. 

\subsubsection{Bayesian optimization}

As can be seen from the energy profiles in Fig.~\ref{fig:oscillations_TV}, the cost function seems to be decomposable as a smooth function plus an oscillatory one of low amplitude. 
This calls for the use of algorithms that i) sample the function at a few scattered points, ii) construct a smooth surrogate approximation, iii) find a minimizer of the surrogate and add it to the explored samples, iv) go back to ii). 

Bayesian Optimization (BO) \cite{frazier2018tutorial} is a principled approach that follows these steps. 
It seems particularly adequate since it models uncertainty on the function evaluations and comes with advanced solvers \cite{balandat2020botorch}. Its application is nonetheless nontrivial and requires some care in our setting. We describe some technical details hereafter.

Consider an objective function of the form
\begin{equation*}
\inf_{z\in \Cc} \E \left( F(z, V)\right),
\end{equation*}
where \rev{the expectation is taken with respect to a random vector $V$}. In our setting, $V$ models both the noise $n$ and the input images $x_k$. We consider $V$ to be a random vector taken uniformly inside a database. In that setting, Bayesian optimization requires the following ingredients:
\begin{enumerate}
    \item An initial sampling set. 
    \item A black-box evaluation routine of $F(z,V)$. 
    \item A family of interpolation functions together with a regression routine. 
    \item A solver that minimizes the regression function. 
\end{enumerate}
Hereafter, each choice made in this work is described.
 
\paragraph{The initial sampling set}

To initialize the algorithm, we need the convex set $\Cc$ to be covered as uniformly as possible in order to achieve a good uniform approximation of the energy profile. 
In this work, we used a maximin space covering design \cite{pronzato2017minimax}. 
The idea is to construct a discrete set $\Zc=\{z_1,\hdots, z_P\}$ that solves approximately
\begin{equation}
    \max_{\Zc \in \Cc^P} \min_{p'\neq p}\|x_p-x_{p'}\|_2.
\end{equation}
In words, we want the minimal distance between pairs of points in $\Zc$ to be as large as possible. 
This problem is known to be hard. In this work we used the recent solver proposed in \cite{debarnot2022deep} together with the Faiss library \cite{johnson2019billion}.

\paragraph{The evaluation routine}

Evaluating the cost function \eqref{eq:objective_rho} is not an easy task. 
For just one realization of the noise $n$ and image $x_k$, we need a fast reconstruction method and a fast way to evaluate the non-uniform Fourier transform. The technical details are provided in the appendix \ref{sec:implementation}.
Second, $K$ might be very large. For instance, the fastMRI knee training database contains more than 30 000 slices of size $320\times 320$.
Hence, it is impossible to compute the complete function and it is necessary to either pick a random, but otherwise fixed subset of the images, or to consider random batches that would vary from one iteration to the next.
A similar comment holds for the noise term $n$.

While Bayesian optimization allows the use of random functions, \rev{it requires evaluating expectations, i.e. integrals. This is typically achieved with Monte-Carlo methods, which is computationally costly}. Hence, in all the forthcoming experiments, we will fix a subset of $K$ images. In practice, we observed that using random batches increases the computational load without offering perceptible advantages.

\paragraph{The interpolation process}

In Bayesian optimization, a Gaussian process is used to model the underlying unknown function.
This random process models both the function and the uncertainty associated with each prediction.
This uncertainty is related to the fact that the function $F$ is evaluated only at a finite number of points hence leading to an unknown behavior when getting distant from the samples.
It is also related to the fact that the function evaluations might be noisy.
Every sampled point has a zero variance when using a fixed realization or a low variance when using random noise and batches.
The variance increases with the distance between the sampled points.

In our experiments, the Gaussian process is constructed using a Matern kernel of parameter $5/2$, which is a popular choice for dimensions in the range $[5,20]$. It is defined as
\begin{equation*}
\Phi(z_1,z_2) = \left( 1 + \frac{\sqrt{5}\|z_1-z_2\|_2}{\nu} + \frac{5\|z_1-z_2\|_2^2}{3\nu^2} \right) \exp\left( \frac{-\sqrt{5}\|z_1-z_2\|_2}{\nu}\right),
\end{equation*}
where $\nu$ is a scaling parameter that controls the smoothness of the interpolant and its point-wise variance. 
In practice, the value of $\nu$ is a parameter that is optimized at each iteration when fitting the Gaussian process to the sampled data.

The interpolant mean and its variance are then evaluated by solving a linear system constructed using the kernel $\Phi$ and the sampled points $z_1,\hdots, z_P$. We refer to \cite{frazier2018tutorial} for more details.

\paragraph{Sampling new points}

Bayesian optimization works by iteratively sampling new points. The point in the sampling set with lowest function value, is an approximation of the minimizer. 
To choose a new point, there is a trade-off between finding a better minimizer in the neighborhood of this point and space exploration. Indeed, big gaps in between the samples could hide a better minimizer. This trade-off is managed through a so-called utility function. In this work, we chose the expected improvement \cite{frazier2018tutorial}, resulting in a new function $\Lc(z)$. The new sampled point is found by solving a constrained non-convex problem:
\begin{equation*}
\inf_{z\in \Cc} \Lc(z)
\end{equation*}
Since the function $\Lc$ is non-convex, we use a multi-start strategy. 
We first sample $1000$ points evenly in $\Cc$ using a maximin design.
Then, we launch many projected gradient descents on $\Lc$ in parallel, starting from those points. 
\rev{Notice that the gradient is evaluated with respect to the surrogate interpolation function, and not with respect to the target density.}
The best critical point is chosen and added as a new sample. 

This process requires projecting $z$ on $\Cc$ defined in \eqref{eq:convex_hull}. To this end, we designed an efficient first order solver.

\section{Numerical experiments and results}
\label{sec:results}

\subsection{The experimental setting}

\paragraph{Database and computing power}

Throughout this section, we used the fastMRI database \cite{zbontar2018fastmri}. It contains MRI images of size $320\times 320$.
We focused on the single coil and fully sampled knee images.
The training set is composed of $973$ 3D volumes, which represents a total of $34~742$ slices. 
The validation set has $199$ volumes and $7135$ slices.

Some images in the dataset have a significant amount of noise. 
This presents three significant drawbacks: \rev{i) the high-frequency contents of the images is increased artificially promoting sampling schemes making it possible to reconstruct noise, ii) the signal-to-noise-ratio of the reconstructed images is decreased artificially and iii) we have shown that noise can dramatically impact the convergence of off-the-grid Fourier sampling optimization \cite{gossard2022spurious}.}
To mitigate these effects, we pre-processed all the slices using a non-local mean denoising algorithm \cite{buades2011non}.

The experiments are conducted on the Jean-Zay HPC facility. 
For each task we use $10$ cores and an Nvidia Tesla V100 with 16GB of memory.

\paragraph{Sampling}

The bounds of the constraint sets in \eqref{eq:constraint_sets} are given by:
\begin{equation}\label{eq:constraints}
    \alpha=\Delta t \gamma \frac{G_{max}}{K_{max}} \quad\text{ and }\quad \beta=\Delta t^2 \gamma \frac{S_{max}}{K_{max}},
\end{equation}
where $\Delta t$ is the sampling step of the scanner. Following \cite{chaithya2020}, we used the following realistic hardware constraints: $G_{max}=40$mT/m, $S_{max}=180$T/m/s, $K_{max}=2\pi$ and $\gamma=42.57$MHz/T. The value of $\Delta t$ is fixed to ensure that at maximal speed, the distance between two consecutive points equals the Shannon-Nyquist rate \cite{lazarus2020correcting}.

We consider two different scenarii: $25\%$ and $10\%$ undersampling.
Each shot consists of $646$ acquisition points and we use $N_s=40$ shots and $N_s=16$ shots respectively for the $25\%$ and the $10\%$ undersampling.
Each shot is constrained to start at the center of the k-space.
The first few points of each trajectory are fixed to be radial, see Appendix \ref{sec:handling_mass_0} for the technical details.

The family of densities is generated using the process described in Section~\ref{subsubsec:dengenerator} with $10^4$ densities generated at random.

\paragraph{Sampling baseline}

All the optimized schemes are compared to a state-of-the-art handcrafted baseline: the Sparkling method described in \cite{lazarus2019sparkling}. 
There, the attraction-repulsion problem \eqref{eq:discrepancy} is solved with a radial density $\rho$ \rev{constructed with a lot of care}.
Its value at the center has been optimized to yield the best possible signal-to-noise ratio on the validation set in a way similar to \cite{chaithya2021learning}.
The corresponding point configuration is given in Fig.~\ref{fig11} and Fig.~\ref{fig12} for the $25\%$ and $10\%$ undersampling rates respectively.
It provides a $7$dB improvement compared to the usual radial lines commonly found in the literature (see the first two rows of Table \ref{table:result_different_optim_TV}).

\paragraph{Image reconstruction}

The experiments are conducted with two reconstruction models:
\begin{itemize}
    \item a total variation reconstruction method with $120$ iterations of Algorithm~\ref{alg:TV_minimization} in Appendix~\ref{sec:reconstruction_algorithm_TV} and with a regularization parameter $\lambda=10^2$ and,
    \item an unrolled network (NN) with $6$ iterations of ADMM and a DruNet as the denoising step \cite{zhang2021plug}, $30$ iterations of the CG algorithm that initializes the ADMM and $10$ iterations of CG to solve the data-consistency equations at each iteration.
\end{itemize}
\rev{The parameter $\lambda$ was optimized to get the best mean square error for the baseline scheme and the number of iterations in the algorithms was chosen optimaly with respect to the reconstruction quality and the available memory on a GPU.}

\begin{figure*}[h!]
    \centering
    \plottraj{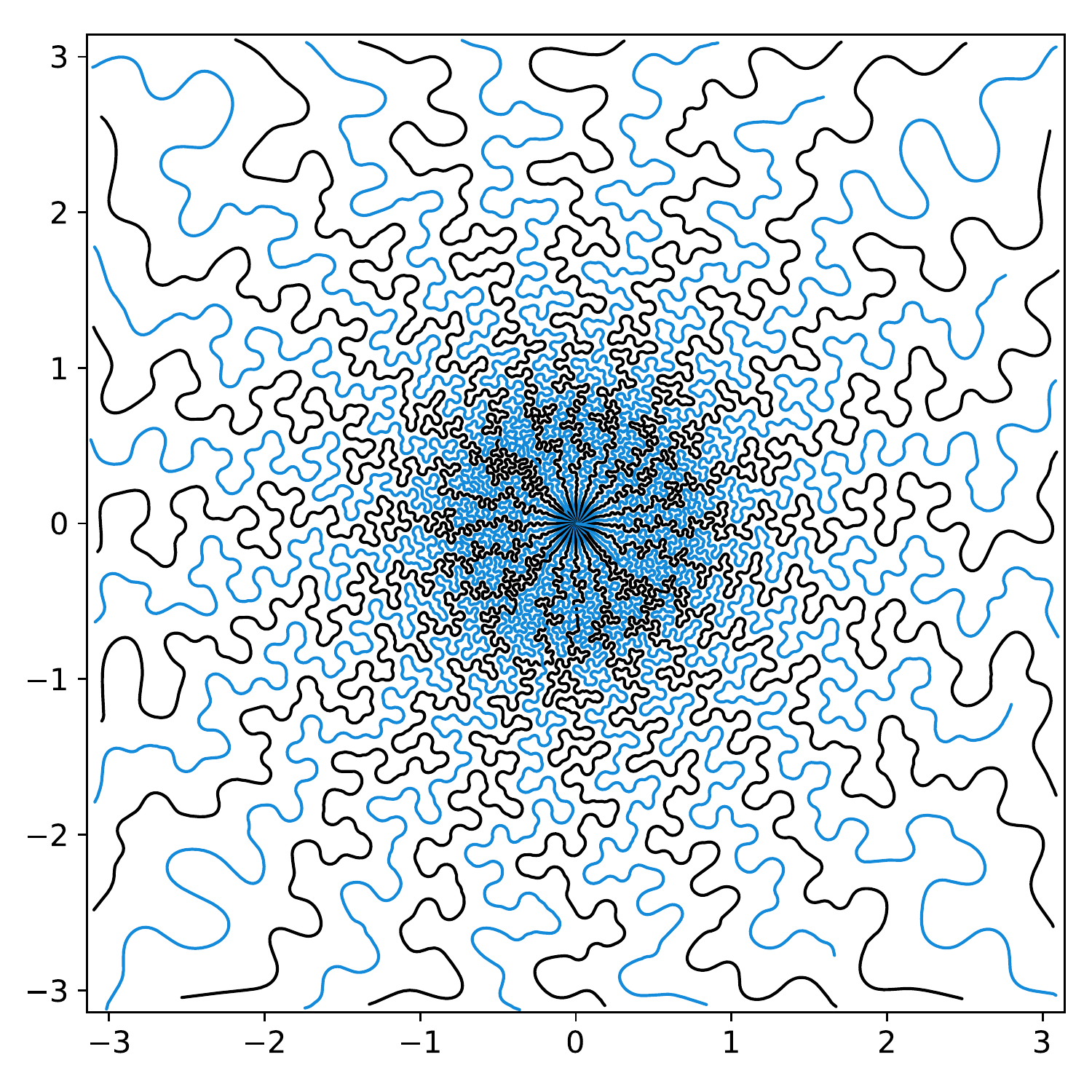}
        {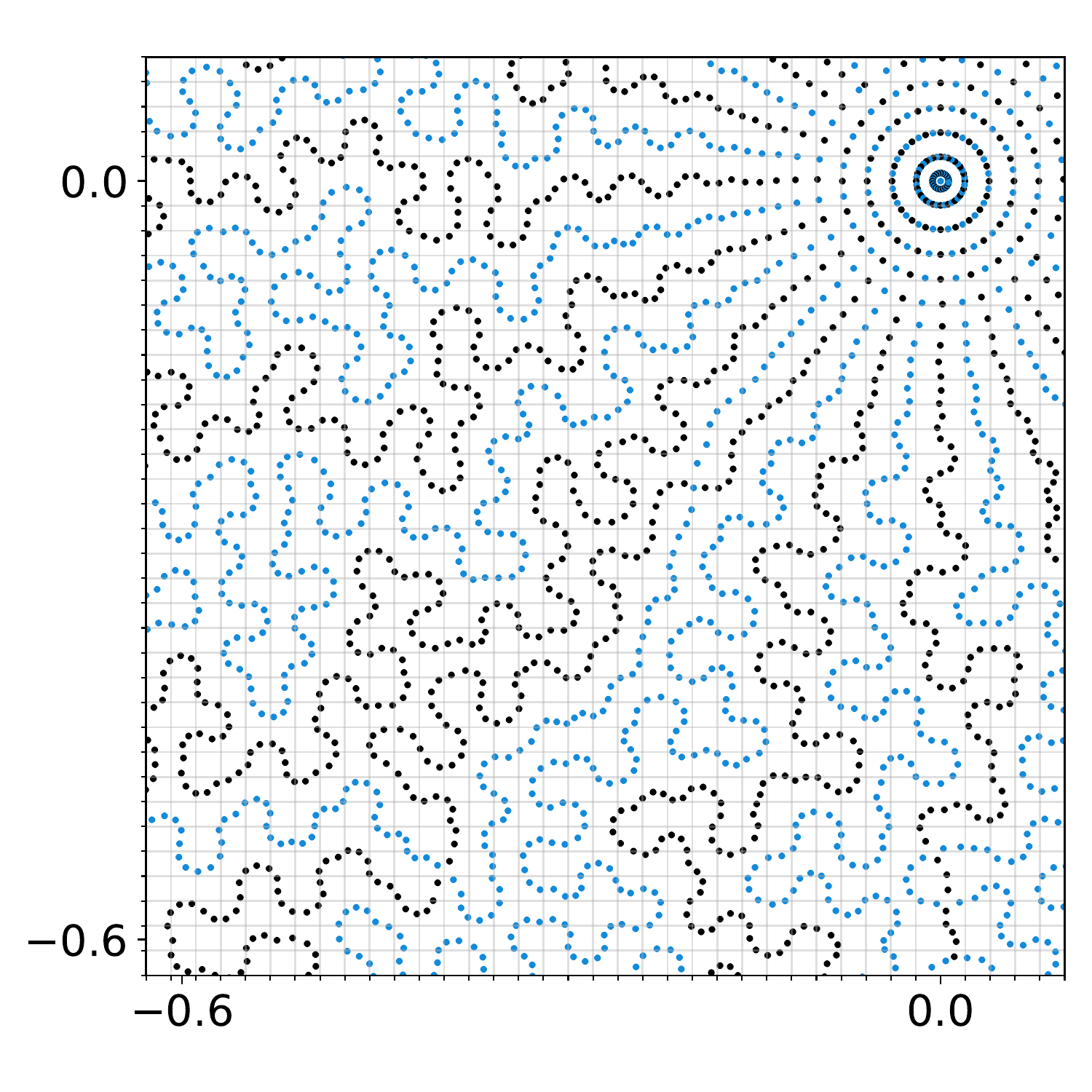}
        {$h(x) = \|x\|_2$ --- $35.10$dB\\ $d(\xi)=0.52$}
        {0}
        {0}
        {0}{fig:dist:linear}{0}
    \hfill
    \begin{subfigure}[t]{.36\linewidth}
        \centering
        \includegraphics[width=\textwidth]{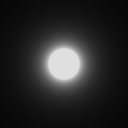}
        \caption{Target radial density}
    \end{subfigure}
    \plottraj{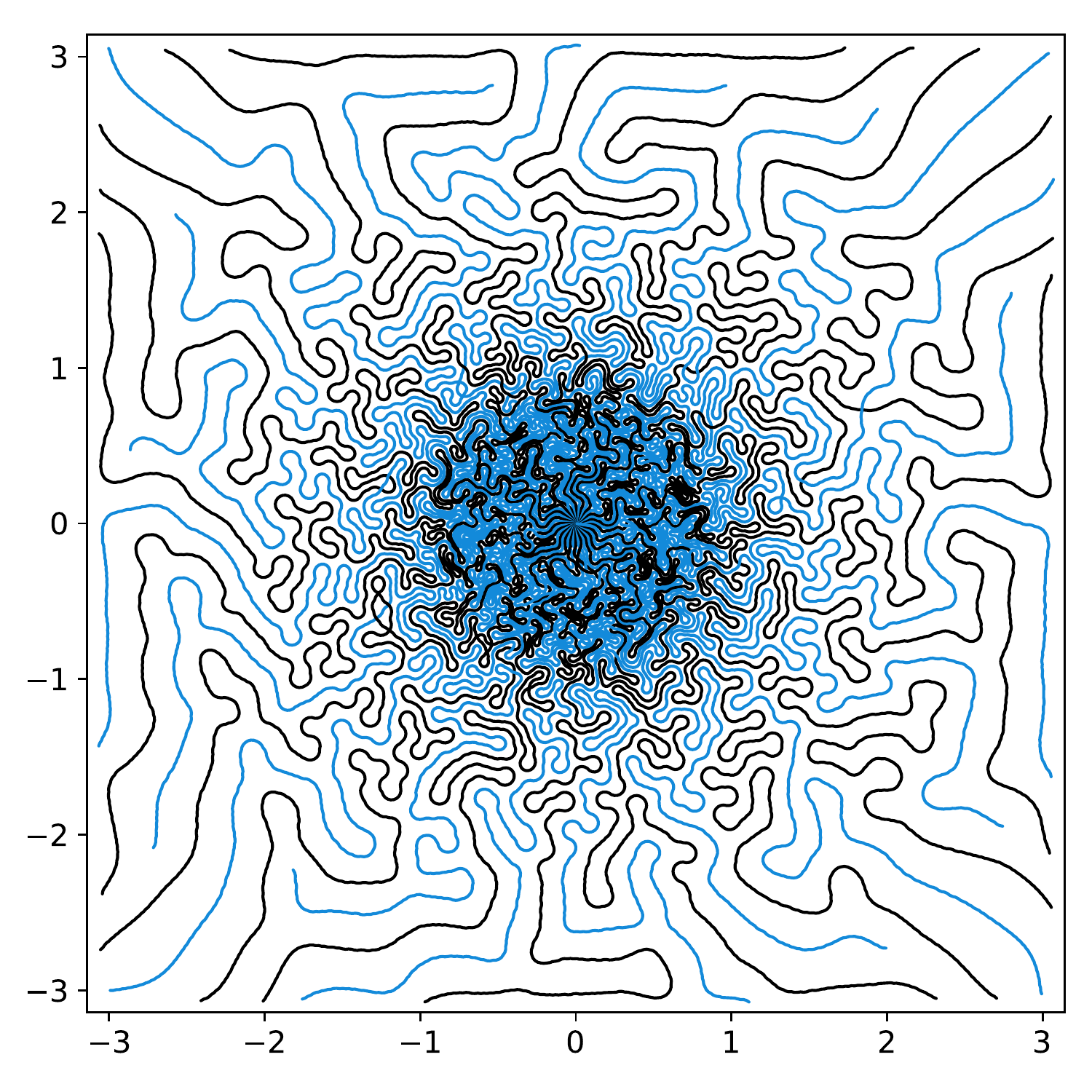}
        {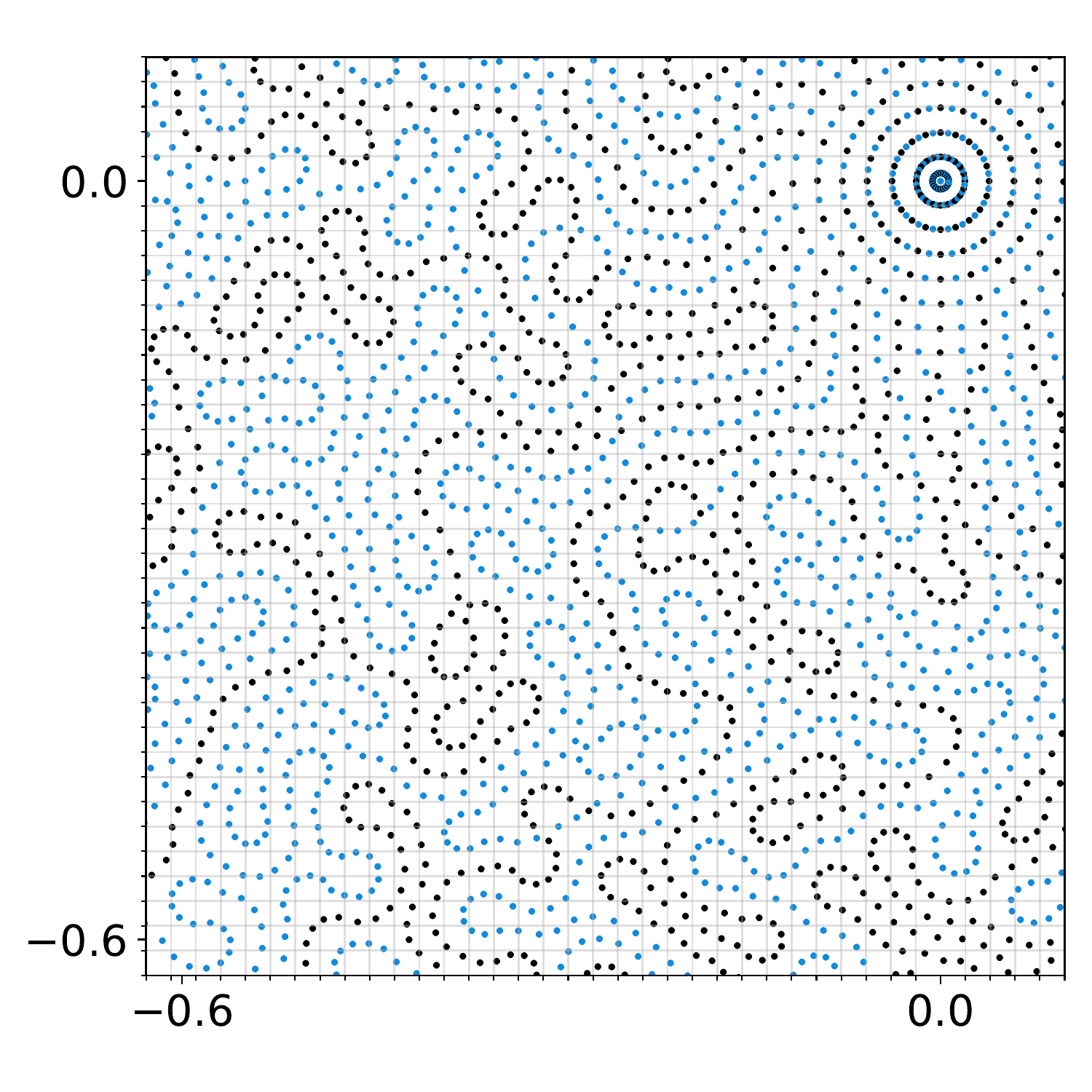}
        {$h(x)=\log(\|x\|_2)$ --- $35.35$dB\\ $d(\xi)=0.72$}
        {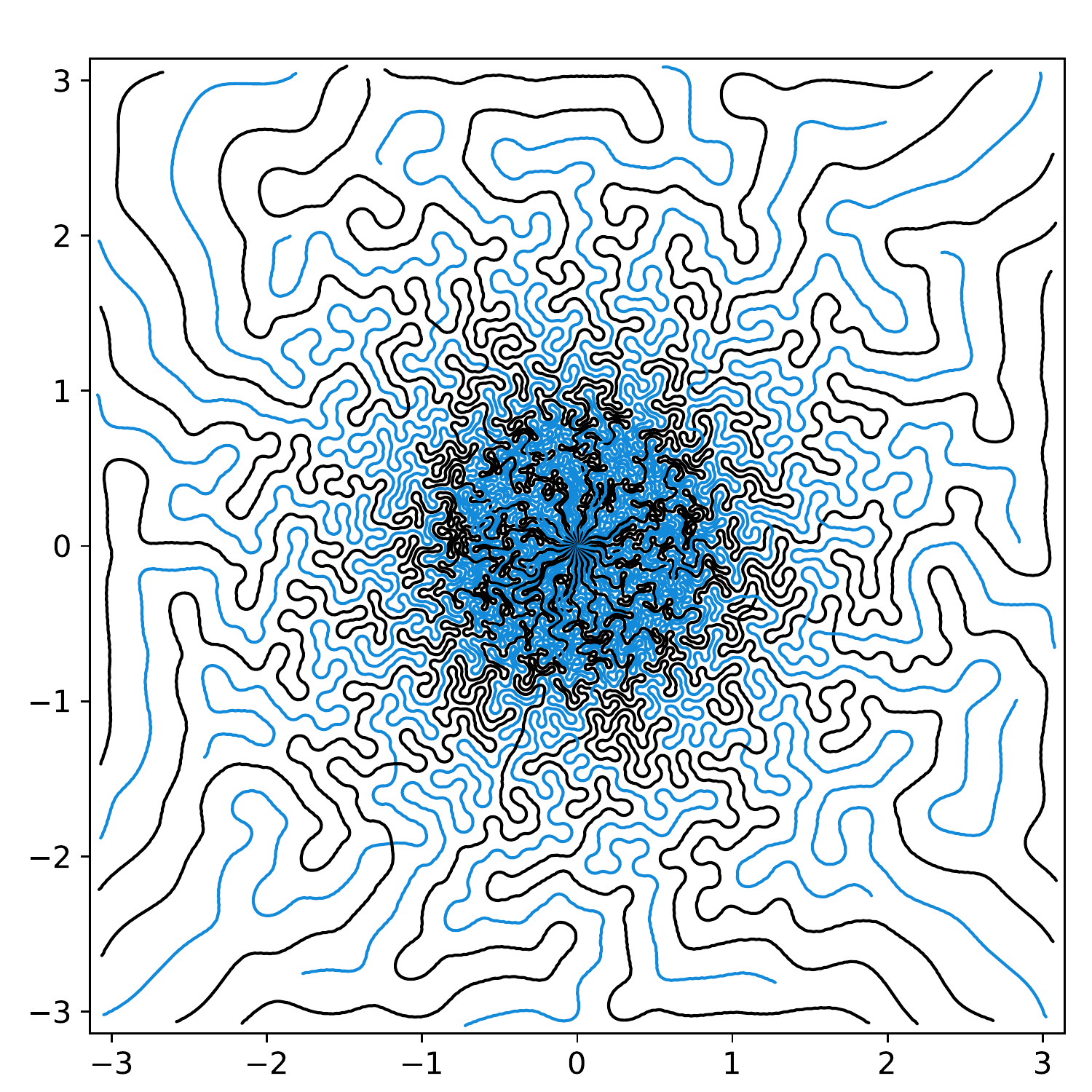}
        {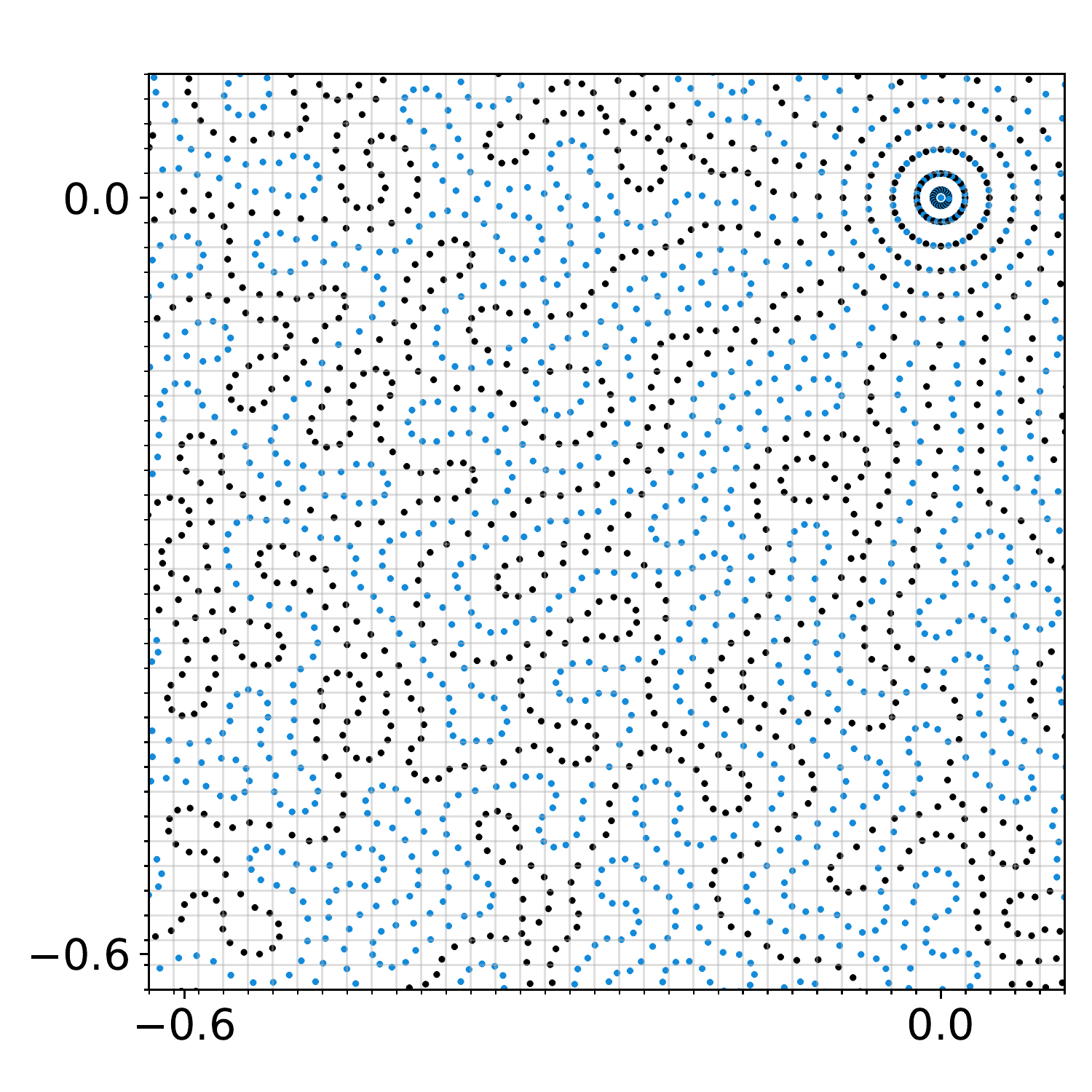}
        {$h(x) = \sqrt{\|x\|_2}$ --- $35.32$dB\\ $d(\xi)=0.69$}{fig:dist:log1}{fig:dist:sqrt1}
    \caption{On the importance of the discrepancy's kernel $h(x)$. The same density is sampled with different kernels. The average PSNR of the reconstructed images on the validation set is displayed with its standard deviation. The average distance between contiguous points on the trajectories is displayed as $d(\xi)$.\label{fig:dist}}
\end{figure*}

\subsection{Choosing a kernel for the discrepancy}\label{subsec:kernel_discrepancy}

In all the previous ``Sparkling'' papers \cite{lazarus2019sparkling,chauffert2017projection}, the kernel function $h(x) = \|x\|_2$ was used. 
This choice seems like the most natural alternative since it is the only one which is scale invariant in the unconstrained setting. 
This means that if $\Xi = (\R^D)^M$ and if a density $\rho$ is dilated by a certain factor, then so is the optimal sampling scheme. 
However, this property is not true anymore when constraints are added. In that case, the choice of kernel turns out to be of importance. 

To illustrate this fact, we considered the three different radial kernels $h(x) = \|x\|_2$, $h(x) = \sqrt{\|x\|_2}$ and $h(x) = \log(\|x\|_2)$.
As can be seen on Fig. \ref{fig:dist}, performance variations of more than $0.2$dB in average  are obtained depending on the kernel. 
The reason is that contiguous points on the trajectories are spaced more or less depending on this choice. 
For instance, observe that the points on the zoom of Fig. \ref{fig:dist:linear} are more packed along the trajectories that on Fig. \ref{fig:dist:log1}.
To compensate for this higher longitudinal density, the sampler then increases the distance between adjacent trajectories, thereby creating holes in the sampling set. 
This is detrimental, since low frequency information is lost in the process.
This effect can be quantified by evaluating the distances between contiguous points in the k-space center. 
As can be seen, it goes from $0.52$ for the usual kernel $h(x)=\|x\|_2$ to a significantly higher value $0.72$ for the logarithmic kernel. The latter kernel creates a higher repulsion for neighboring points.

\begin{figure*}
    \centering
    \plottraj{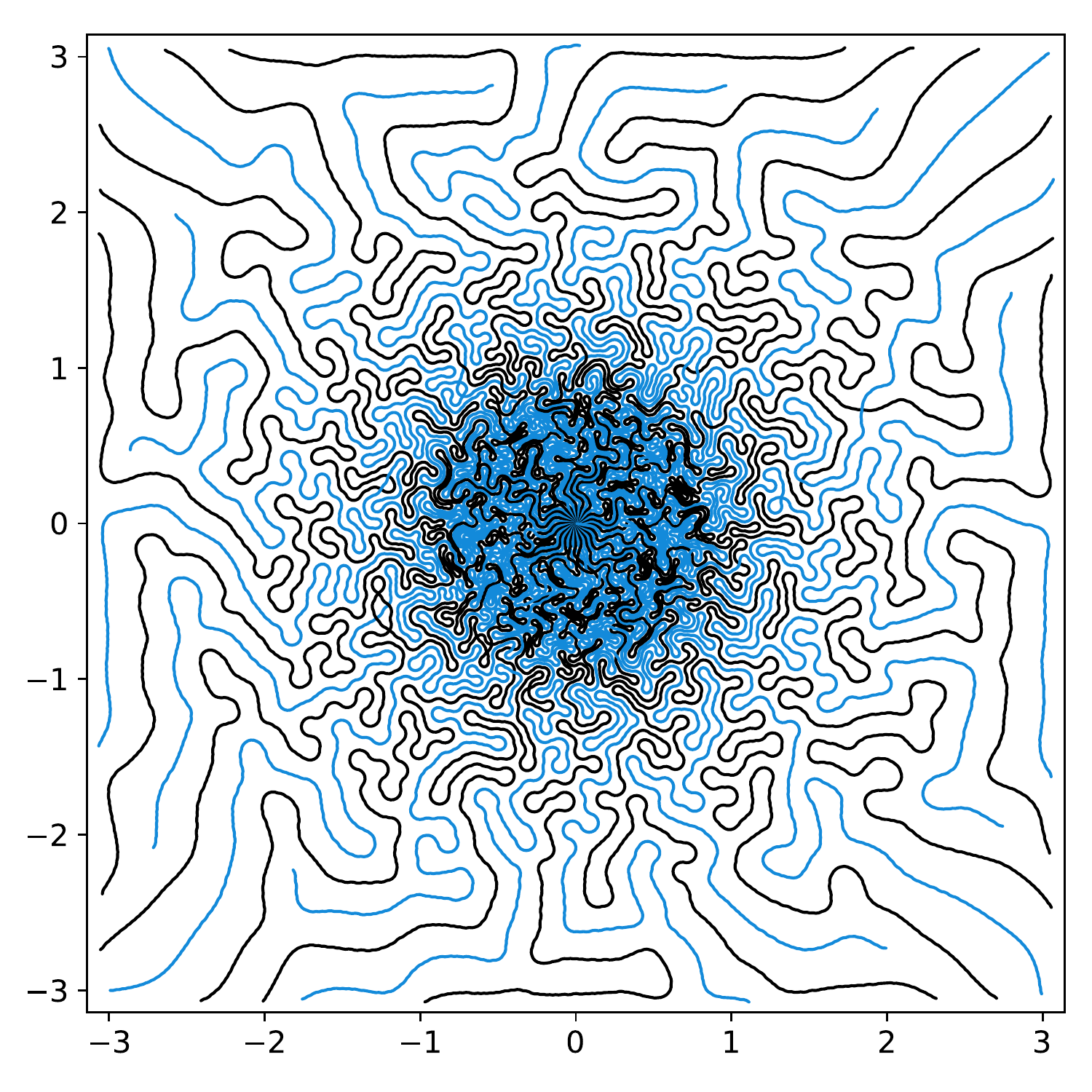}
        {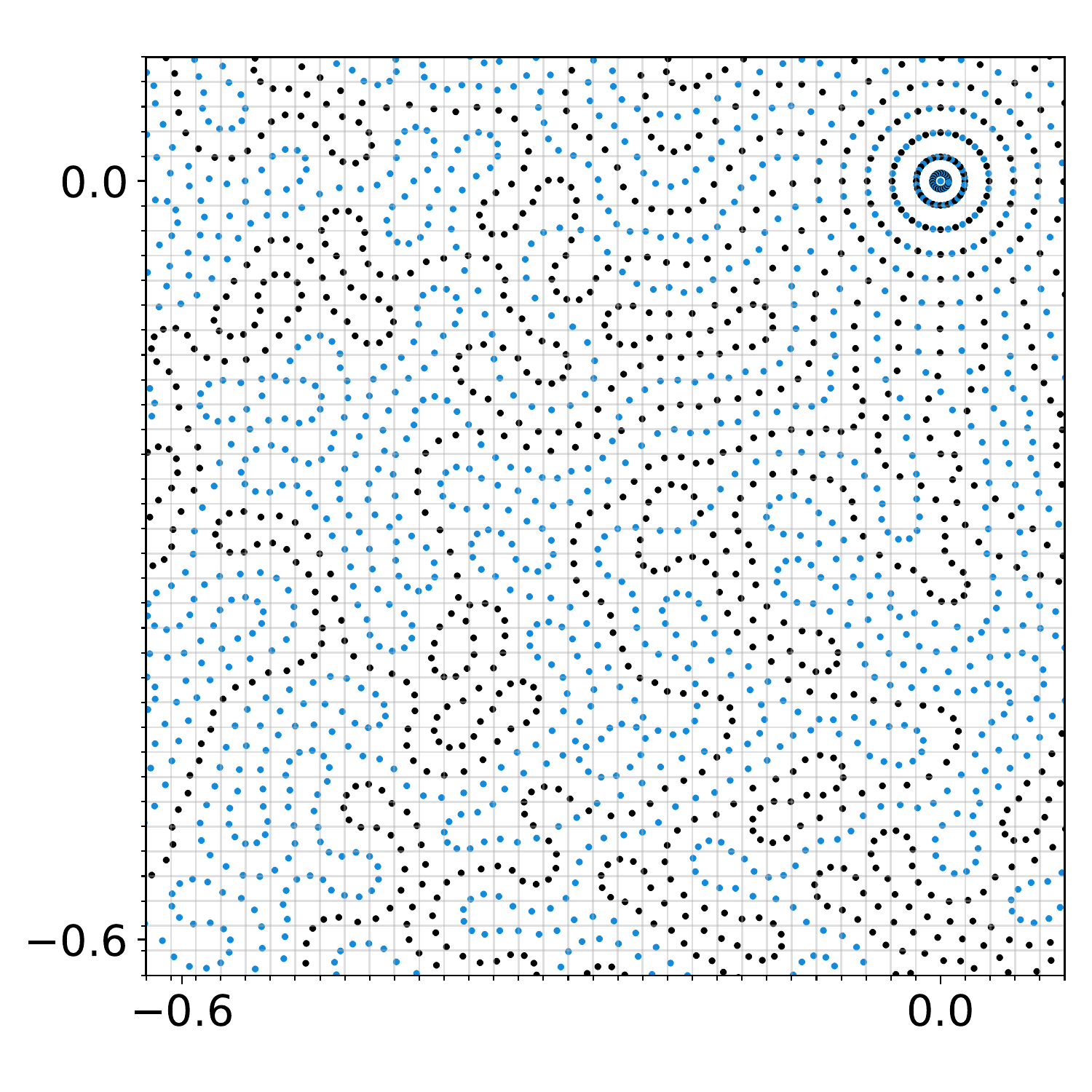}
        {Baseline radial 25\%}
        {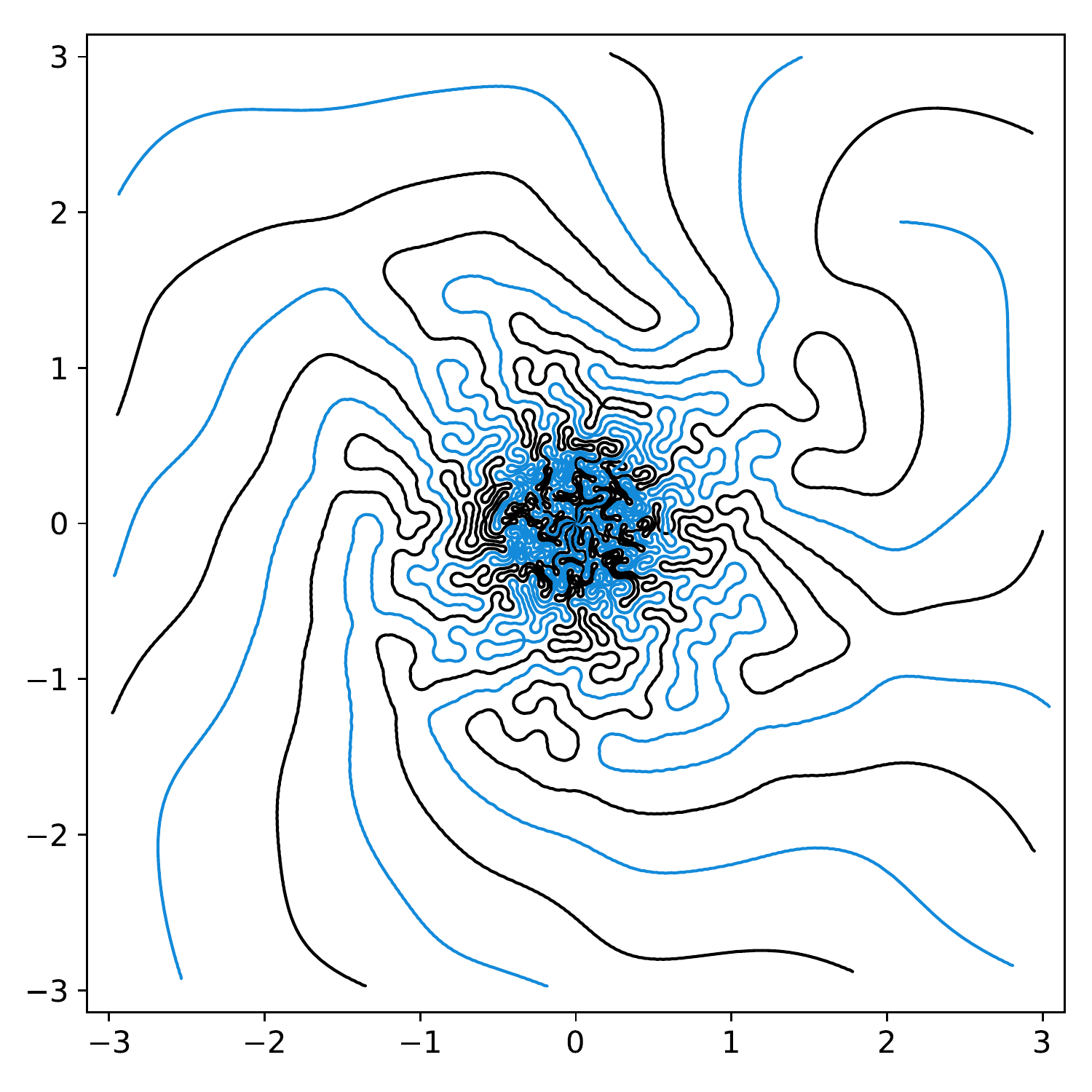}
        {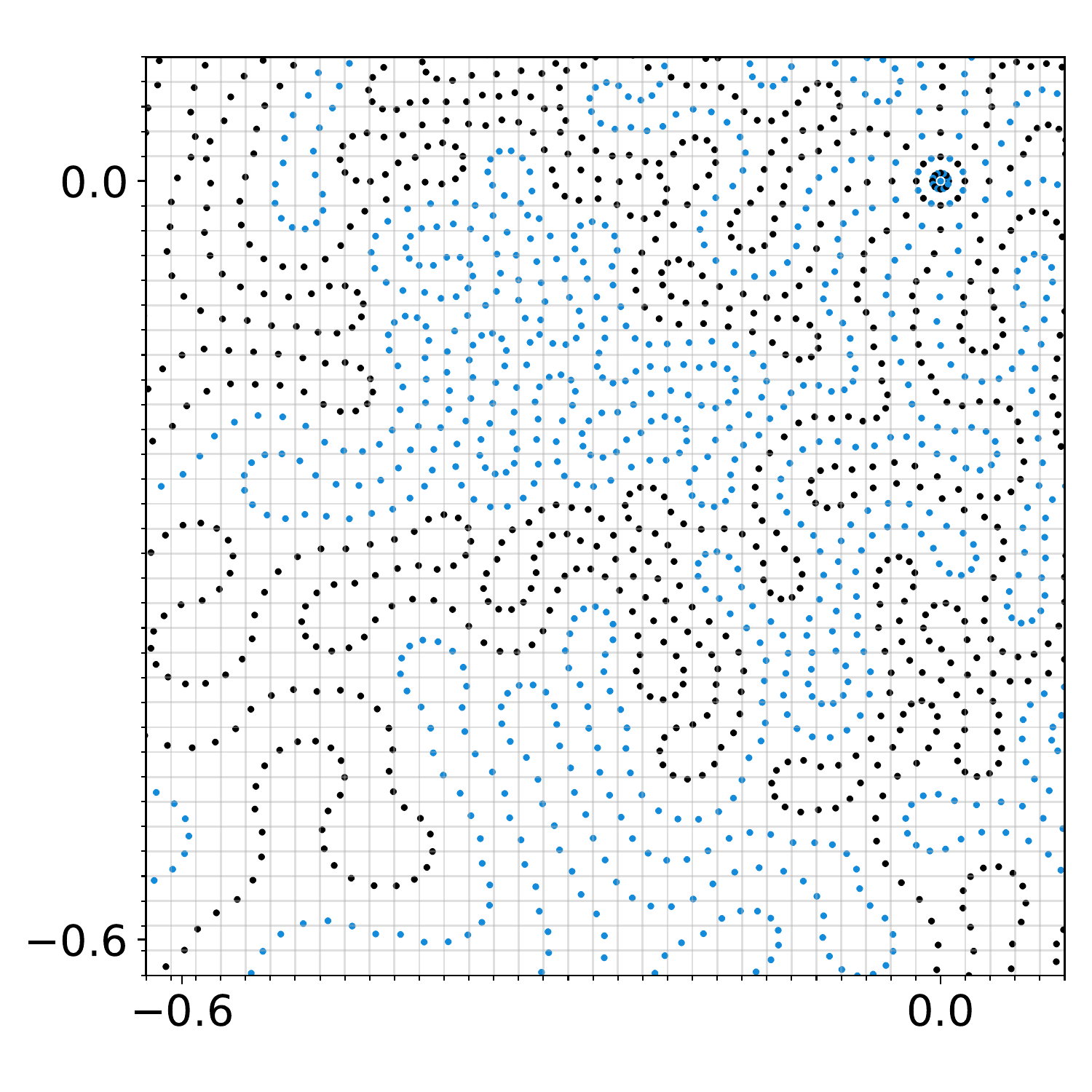}
        {Baseline radial 10\%}{fig11}{fig12}
    \plottraj{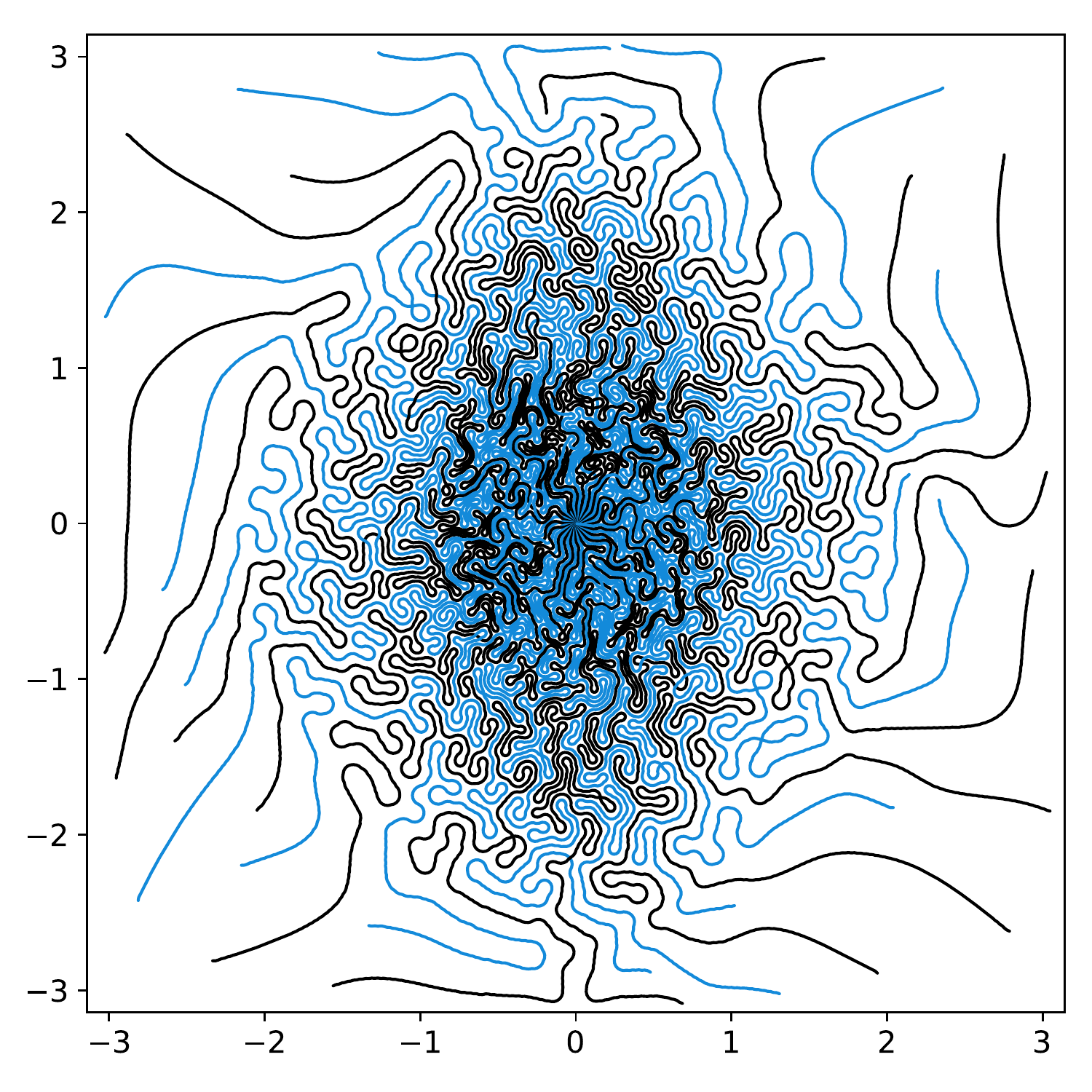}
        {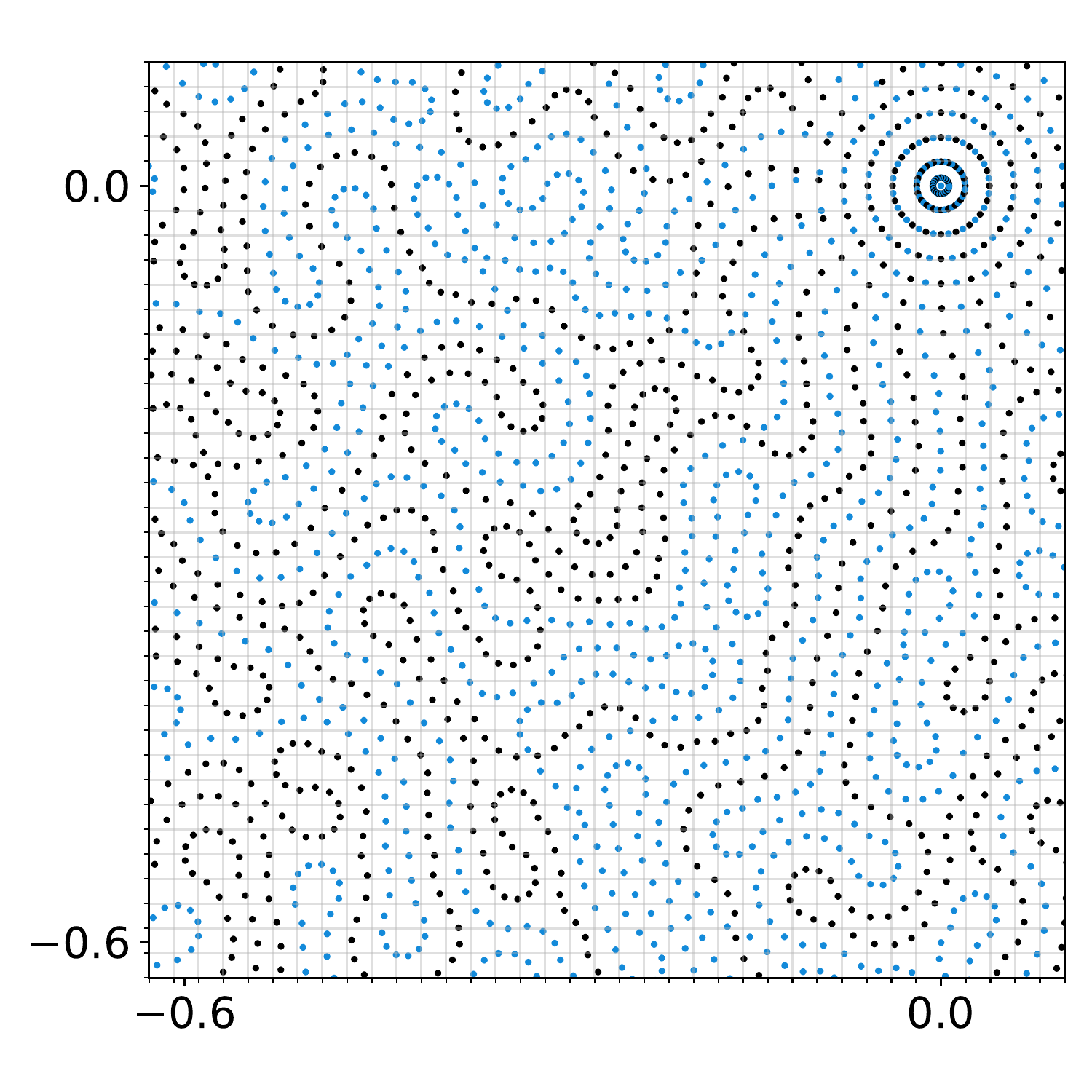}
        {Bayesian density optimization 25\%}
        {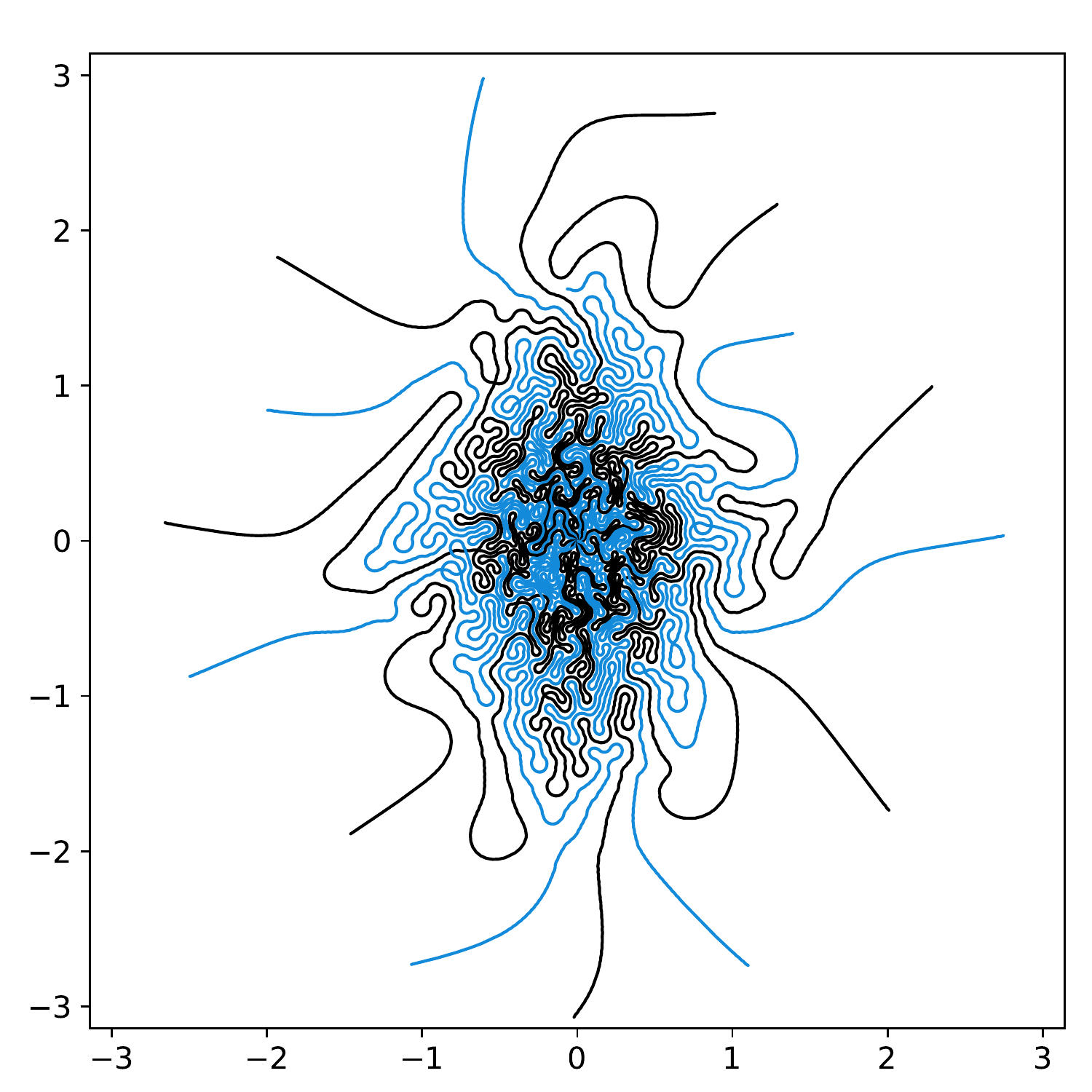}
        {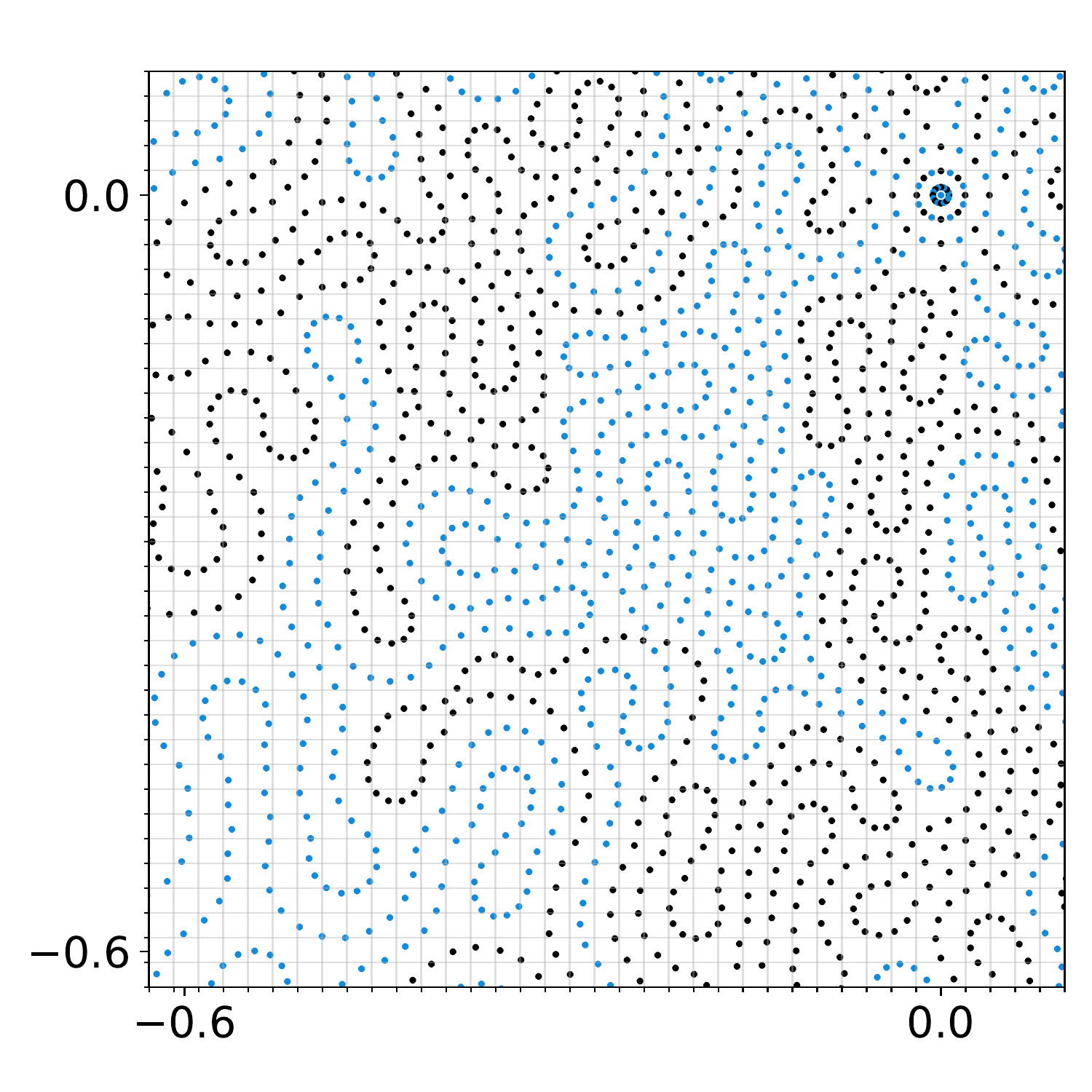}
        {Bayesian density optimization 10\%}{fig21}{fig22}
    \plottraj{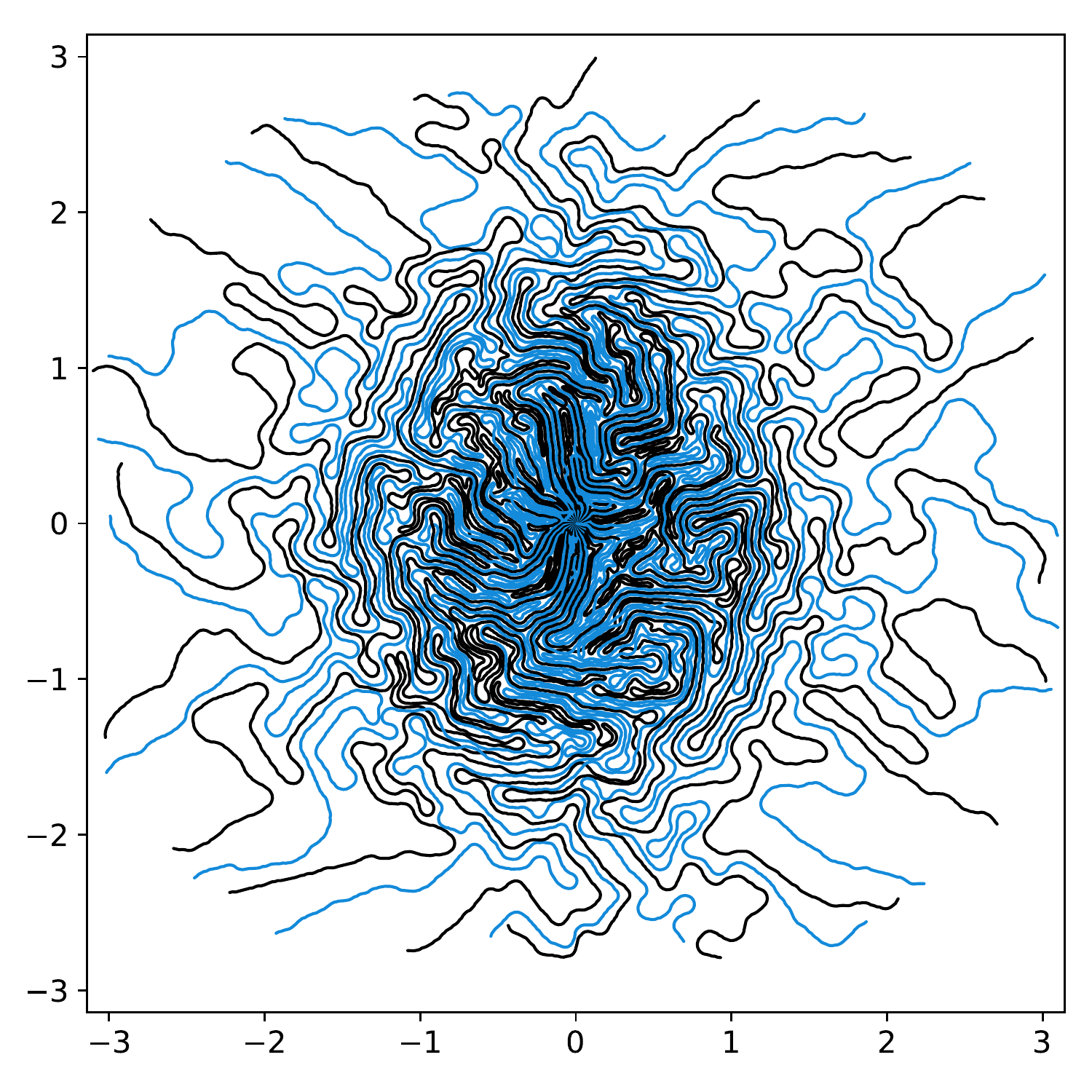}
        {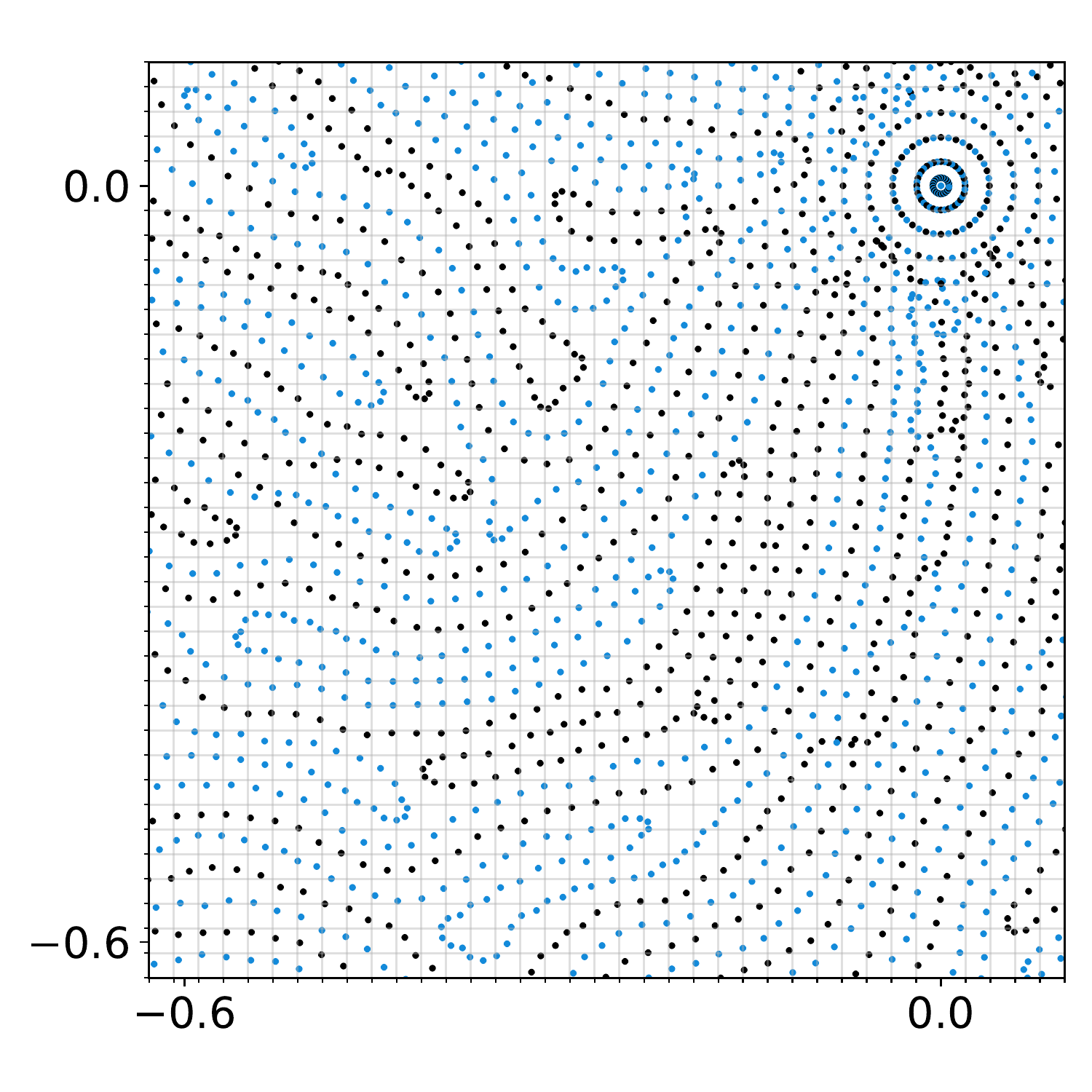}
        {Trajectory optimization 25\%}
        {TV_multiscale_10.pdf}
        {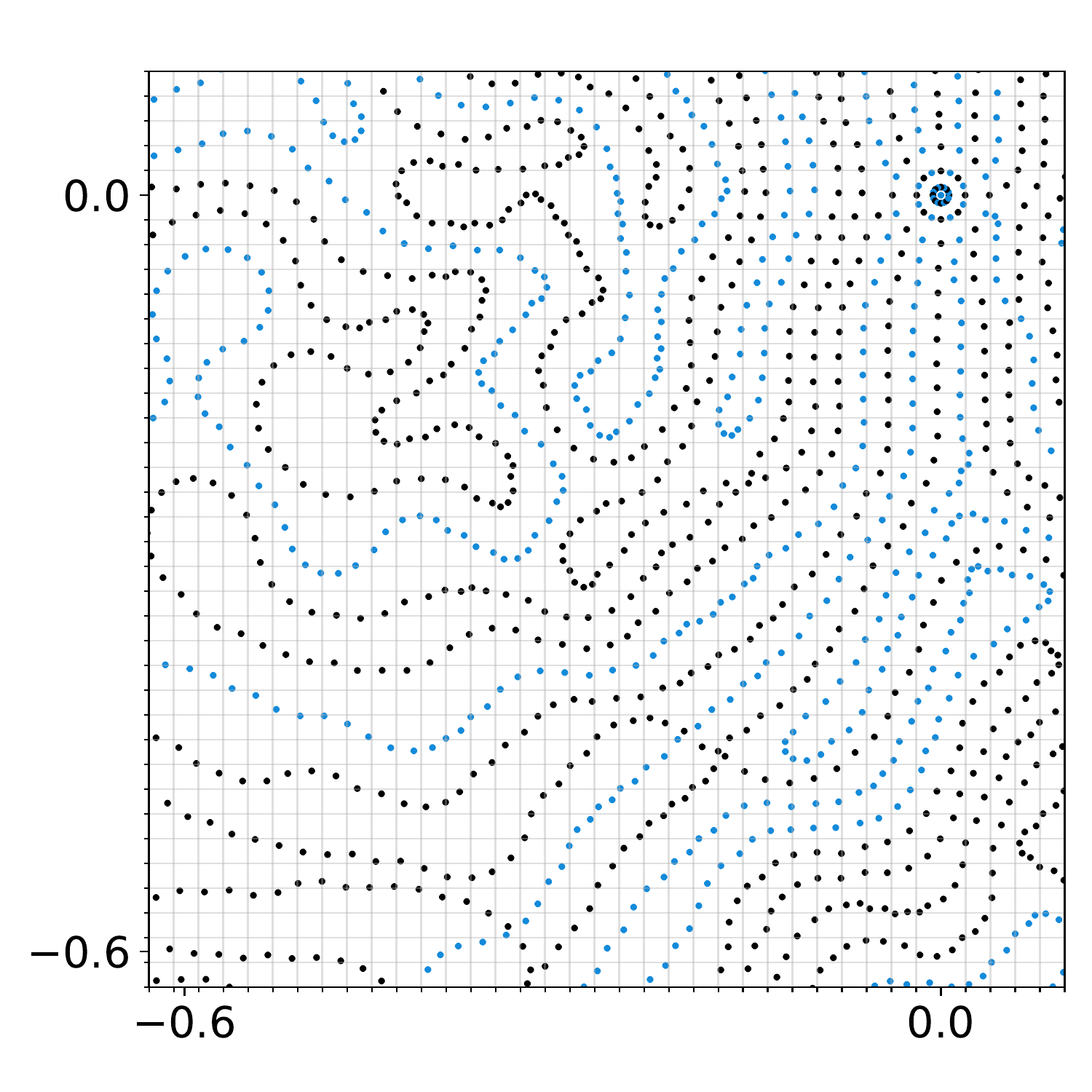}
        {Trajectory optimization 10\%}{fig31}{fig32}
    \caption{Optimized sampling schemes with various optimization approaches. A total variation reconstruction algorithm is used.\label{fig:trajectory_optim_TV}}
\end{figure*}

\subsection{Bayesian optimization: database size and numerical complexity}
\label{subsec:bayesian_optim_numerical}

In this section, we aim at evaluating the computational complexity of the Bayesian optimization routine. 
To this end, we study the impact of the number of images $K$ in the dataset, the size of the initial sampling set and the number of iterations, which are governing the algorithm's complexity.
\rev{Table~\ref{table:bo_results} and \ref{table:computational_time} summarize our main findings for the total variation reconstruction and unrolled neural network.}

\rev{In Table~\ref{table:bo_results}, we see that the number of images $K$ in the dataset has nearly no influence on the quality of the final sampling density.} 
Taking $K=32$ or $K=512$ images yields an identical PSNR on the validation set. This holds both for the 25\% and 10\% undersampling rates.
As can be seen in the Tables, reconstructing as little as $200\times 32$ images is enough to reach the best possible density in the family.
The same conclusion holds for the $10\%$ undersampling rate. This represents $18\%$ of a single epoch.

We also see that the initial sampling set of the convex $\Cc$ plays a marginal role on the quality of the final result. 
In addition, taking a small number of initial points allows \rev{us} to reduce the overall complexity of the algorithm to reach a given PSNR.

\rev{Finally, Table \ref{table:computational_time} allows us to conclude that taking a database of $32$ or $128$ images has only a marginal influence on the computing times (3 hours instead of 4 hours) of the Bayesian optimization routine. This feature is related to the fact that a significant amount of time is spent in the minimization of $\mathcal{L}$, whose numerical cost is independent of the number of images.
Hence, the fact that the algorithm seems to perform well for $32$ images is advantageous mostly when only small datasets can be generated.}

\begin{table}
    \small
    \centering
    \begin{tabular}{ccc|ccc}
        \toprule
        \rev{Recon.} & \makecell{$\#$ init. points} & \makecell{$\#$ evaluations} & \makecell{average PSNR\\$K=32$ images} & \makecell{average PSNR\\$K=128$ images} & \makecell{average PSNR\\$K=512$ images}  \\
        \midrule
        TV & $20$ & $200$ & \makecell{$35.64$dB} & \makecell{$35.65$dB} & \makecell{$35.65$dB} \\
        \midrule
        TV & $100$ & $300$ & \makecell{$35.63$dB} & \makecell{$35.66$dB} & \makecell{$35.66$dB} \\
        \midrule
        TV & $200$ & $300$ & \makecell{$35.65$dB} & \makecell{$35.66$dB} & \makecell{$35.66$dB} \\
        
        \toprule
        NN & $20$ & $200$ & \makecell{$38.14$dB} & \makecell{$38.10$dB} & \makecell{$38.09$dB} \\
        \midrule
        NN & $100$ & $300$ & \makecell{$38.17$dB} & \makecell{$38.05$dB} & \makecell{$38.08$dB} \\
        \midrule
        NN & $200$ & $300$ & \makecell{$38.20$dB} & \makecell{$38.08$dB} & \makecell{$38.10$dB} \\
        \bottomrule
    \end{tabular}
    \caption{Bayesian optimization on a convex set $\Cc$ of dimension $L=20$ using a total variation reconstruction algorithm and an unrolled network for $25\%$ undersampling. The PSNR is evaluated for the optimized density on the validation dataset containing $7135$ images. The total number of cost function evaluations is given in the second column. \label{table:bo_results}}
\end{table}

\subsection{Comparing optimization routines for the total variation reconstructor}\label{subsec:comparison_TV}

In what follows, we aim at comparing two different sampling optimization approaches:
\begin{description}
    \item[Trajectory optimization] The minimization of \eqref{eq:objective_xi} in the space of trajectories. We use a modified version of the multi-scale approach in \cite{wang2022b}, see Appendix \ref{sec:solving_xi}. 
    \item[BO density] The Bayesian approach to minimize \eqref{eq:objective_rho} globally. 
\end{description}
To compare these approaches, we conduct various experiments. The corresponding results are shown in Table~\ref{table:computational_time}, Table~\ref{table:result_different_optim_TV} and Fig~\ref{fig:trajectory_optim_TV}.
Below, we summarize our main findings.

\paragraph{Qualitative comparison of the sampling schemes}

In this paragraph, we compare our method with existing works \cite{wang2022b,weiss2021pilot}. 
The optimized sampling schemes are shown in Fig.~\ref{fig:trajectory_optim_TV} for the TV reconstructor.
In Fig.~\ref{fig:trajectory_optim_TV}, we see the results of the different optimization routines. 

The two optimization methods yield anisotropic sampling schemes with a higher density along the vertical axis.
However the trajectories present significant differences.

The Bayesian optimization yields a sampling scheme which covers the space more uniformly. 
The trajectories have a significantly higher curvature at the k-space center. 
These features are somehow hard-coded within the sampling generator $\Sc_M$ described in Section \ref{sec:sampler}.

The trajectory optimization yields trajectories which are locally linear and aligned at a distance of about a pixel.
This suggests that the trajectory optimization favors Shannon's sampling rate at the center of the k-space.
A potential explanation is as follows. When the sampling points are close to a subgrid \cite{gossard2022spurious}, the adjoint of the forward operator $A(\xi)^*$ is roughly the pseudo-inverse. 
Using a points configuration close to a subgrid therefore helps iterative reconstruction algorithms to converge.

Finally, at the bottom-left of the zoomed region on the 25\% undersampling rate, it seems that Bayesian optimization (Fig.~\ref{fig21}) yields a density slightly higher  than trajectory optimization (Fig.~\ref{fig31}). 
This density is critical for the reconstruction quality and might explain a part of the quantitative differences observed in the next section.

\begin{table}[h!]
    \centering
    \small
    \begin{tabular}{lcc}
    \toprule
    Method & Computational time TV & \rev{Computational time NN}  \\ [0.5ex]
    \midrule
    Trajectory optimization & $85$h & \rev{$97$h} \\
    \rev{14 epochs} & \\
    \hline
    Bayesian optimization &  \\
    \hspace{0.4cm}Optimization $K=32$ & $3$h  & \rev{$4$h} \\
    \hspace{0.4cm}Optimization $K=128$ & $4$h & \rev{$5$h} \\
    \bottomrule
    \end{tabular}
    \caption{Computational time of the different optimization procedures ($25\%$ undersampling) with an NVIDIA Quadro RTX 5000 GPU.}
    \label{table:computational_time}
\end{table}

\paragraph{Performance comparison}

Table~\ref{table:result_different_optim_TV} reveals that the trajectory optimization yields better performance \rev{in average} than the Bayesian optimization approach both for the 25\% ($+0.26$dB) and 10\% ($+0.07$dB) undersampling rates.
This was to be expected since the density optimization is much more constrained.
The difference is however mild.

\begin{figure*}[htbp]
    \centering
    \begin{tabular}{c ccc c ccc}
        & \multicolumn{7}{c}{
            \begin{tikzpicture}
                \node[rounded corners=4pt,fill=black!20] (W) at (0,0) {Worst};
                \node[rounded corners=4pt,fill=black!20] (B) at (13,0) {Best};
                \draw[->,>=latex,thick,shorten >=5pt,shorten <=5pt] (W) -- (B) node[midway,above,rounded corners=4pt]{PSNR difference with baseline} ;
            \end{tikzpicture} 
        } \\

        \rotatebox[origin=l]{90}{BO TV} &
        \begin{subfigure}[t]{0.12\textwidth}
            \centering
            \includegraphics[width=\textwidth]{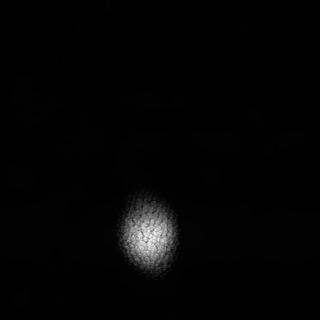}
            \caption*{$-1.3$dB}
        \end{subfigure} &
        \begin{subfigure}[t]{0.12\textwidth}
            \centering
            \includegraphics[width=\textwidth]{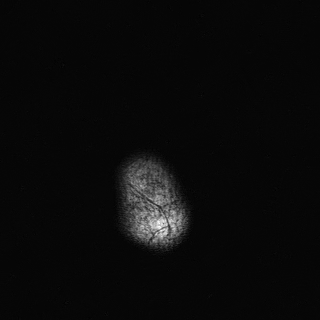}
            \caption*{$-1.1$dB}
        \end{subfigure} &
        \begin{subfigure}[t]{0.12\textwidth}
            \centering
            \includegraphics[width=\textwidth]{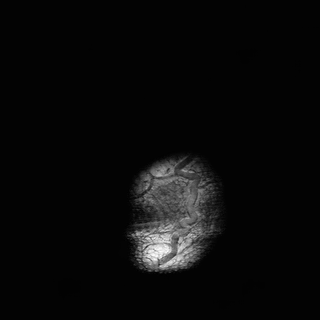}
            \caption*{$-1.1$dB}
        \end{subfigure} &
        \begin{tikzpicture}
            \draw[color=white](0,0) rectangle (0.2,1.5);
            \draw[color=black](0,0.8) node{$\hdots$};
        \end{tikzpicture}   &
        \begin{subfigure}[t]{0.12\textwidth}
            \centering
            \includegraphics[width=\textwidth]{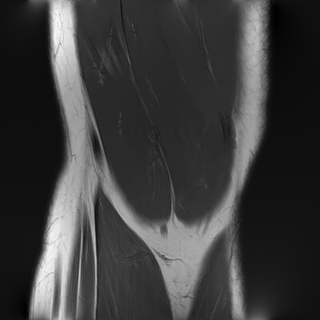}
            \caption*{$+1.4$dB}
        \end{subfigure} &
        \begin{subfigure}[t]{0.12\textwidth}
            \centering
            \includegraphics[width=\textwidth]{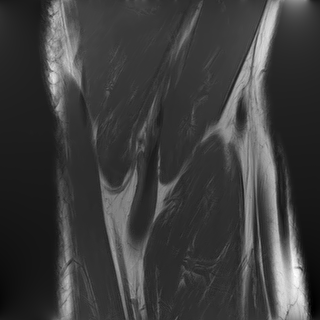}
            \caption*{$+1.4$dB}
        \end{subfigure} &
        \begin{subfigure}[t]{0.12\textwidth}
            \centering
            \includegraphics[width=\textwidth]{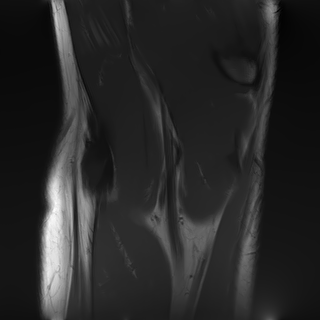}
            \caption*{$+1.4$dB}
        \end{subfigure} \\

        \rotatebox[origin=l]{90}{TO TV} &
        \begin{subfigure}[t]{0.12\textwidth}
            \centering
            \includegraphics[width=\textwidth]{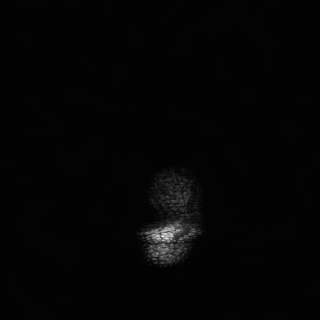}
            \caption*{$-1.4$dB}
        \end{subfigure} &
        \begin{subfigure}[t]{0.12\textwidth}
            \centering
            \includegraphics[width=\textwidth]{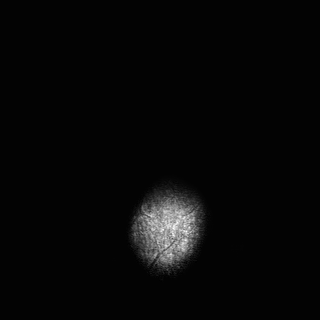}
            \caption*{$-0.9$dB}
        \end{subfigure} &
        \begin{subfigure}[t]{0.12\textwidth}
            \centering
            \includegraphics[width=\textwidth]{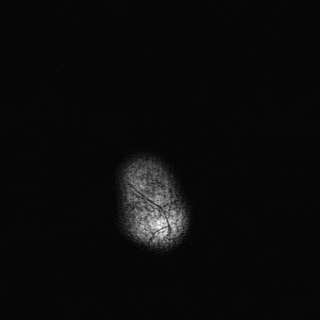}
            \caption*{$-0.7$dB}
        \end{subfigure} &
        \begin{tikzpicture}
            \draw[color=white](0,0) rectangle (0.2,1.5);
            \draw[color=black](0,0.8) node{$\hdots$};
        \end{tikzpicture}   &
        \begin{subfigure}[t]{0.12\textwidth}
            \centering
            \includegraphics[width=\textwidth]{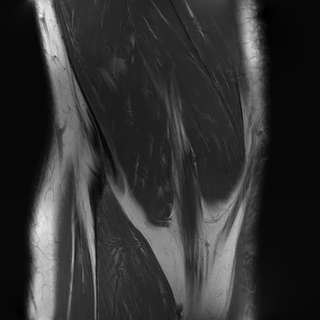}
            \caption*{$+1.6$dB}
        \end{subfigure} & 
        \begin{subfigure}[t]{0.12\textwidth}
            \centering
            \includegraphics[width=\textwidth]{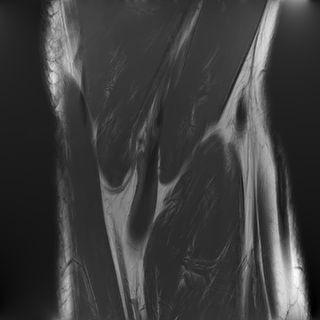}
            \caption*{$+1.7$dB}
        \end{subfigure} &
        \begin{subfigure}[t]{0.12\textwidth}
            \centering
            \includegraphics[width=\textwidth]{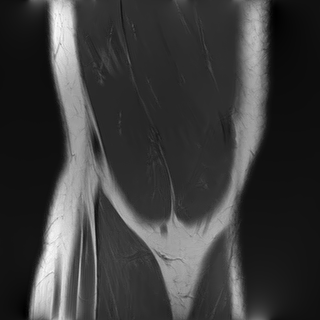}
            \caption*{$+1.8$dB}
        \end{subfigure} \\

        \rotatebox[origin=l]{90}{BO NN} &
        \begin{subfigure}[t]{0.12\textwidth}
            \centering
            \includegraphics[width=\textwidth]{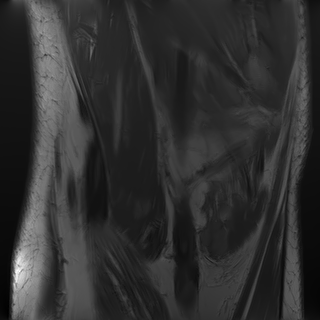}
            \caption*{$-1.3$dB}
        \end{subfigure} &
        \begin{subfigure}[t]{0.12\textwidth}
            \centering
            \includegraphics[width=\textwidth]{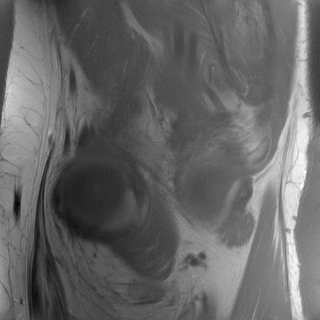}
            \caption*{$-1.1$dB}
        \end{subfigure} &
        \begin{subfigure}[t]{0.12\textwidth}
            \centering
            \includegraphics[width=\textwidth]{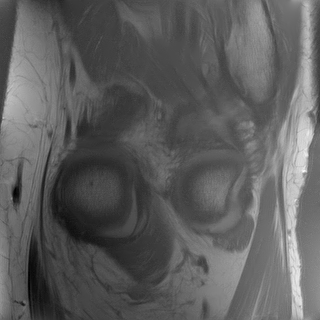}
            \caption*{$-1.1$dB}
        \end{subfigure} &
        \begin{tikzpicture}
            \draw[color=white](0,0) rectangle (0.2,1.5);
            \draw[color=black](0,0.8) node{$\hdots$};
        \end{tikzpicture}   &
        \begin{subfigure}[t]{0.12\textwidth}
            \centering
            \includegraphics[width=\textwidth]{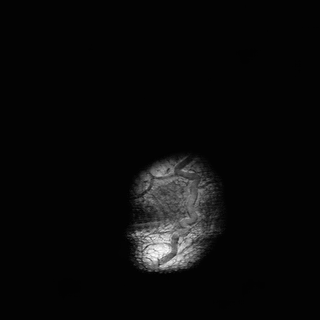}
            \caption*{$+4.8$dB}
        \end{subfigure} & 
        \begin{subfigure}[t]{0.12\textwidth}
            \centering
            \includegraphics[width=\textwidth]{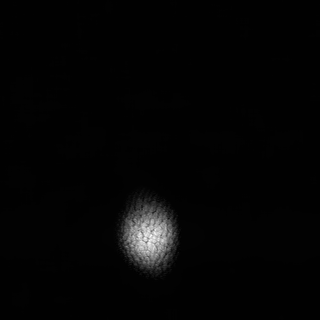}
            \caption*{$+4.9$dB}
        \end{subfigure} &
        \begin{subfigure}[t]{0.12\textwidth}
            \centering
            \includegraphics[width=\textwidth]{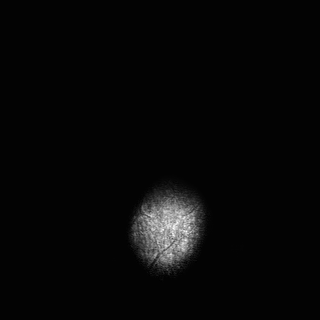}
            \caption*{$+5.3$dB}
        \end{subfigure} \\

        \rotatebox[origin=l]{90}{TO NN} &
        \begin{subfigure}[t]{0.12\textwidth}
            \centering
            \includegraphics[width=\textwidth]{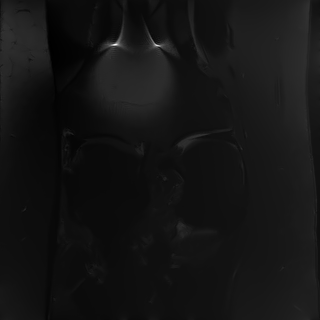}
            \caption*{$0.1$dB}
        \end{subfigure} &
        \begin{subfigure}[t]{0.12\textwidth}
            \centering
            \includegraphics[width=\textwidth]{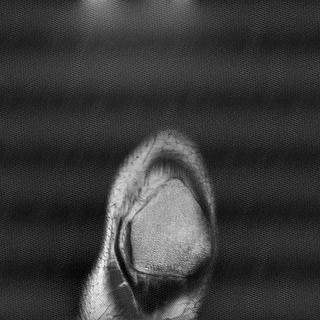}
            \caption*{$0.3$dB}
        \end{subfigure} &
        \begin{subfigure}[t]{0.12\textwidth}
            \centering
            \includegraphics[width=\textwidth]{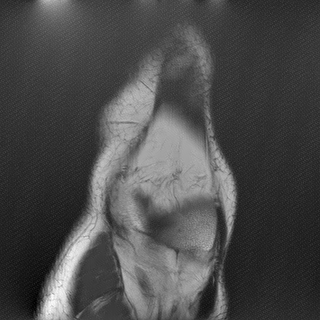}
            \caption*{$0.3$dB}
        \end{subfigure} &
        \begin{tikzpicture}
            \draw[color=white](0,0) rectangle (0.2,1.5);
            \draw[color=black](0,0.8) node{$\hdots$};
        \end{tikzpicture}   &
        \begin{subfigure}[t]{0.12\textwidth}
            \centering
            \includegraphics[width=\textwidth]{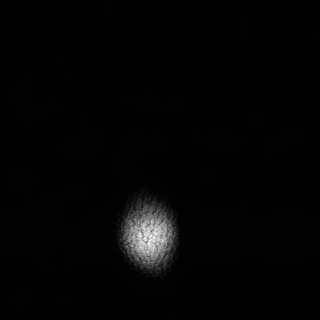}
            \caption*{$+5.1$dB}
        \end{subfigure} & 
        \begin{subfigure}[t]{0.12\textwidth}
            \centering
            \includegraphics[width=\textwidth]{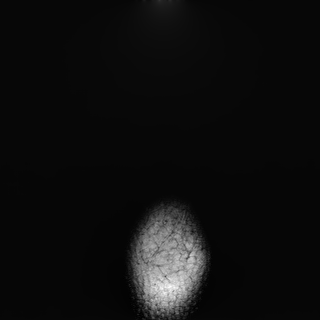}
            \caption*{$+5.2$dB}
        \end{subfigure} &
        \begin{subfigure}[t]{0.12\textwidth}
            \centering
            \includegraphics[width=\textwidth]{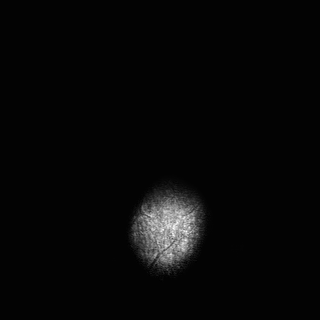}
            \caption*{$+5.8$dB}
        \end{subfigure}
    \end{tabular}
    \caption{\rev{PSNR differences between the optimized sampling schemes and the baseline for different images. The images were sorted with increasing PSNR differences from the left to the right. BO corresponds to the result with Bayesian optimization and TO corresponds to trajectory optimization. TV correponds to a total variation reconstruction and NN to an unrolled ADMM reconstruction. In this experiment, we used sampling schemes with $25\%$ undersampling. The numbers below the images are the PSNR differences between the baseline sampling scheme and the optimized trajectories.}}
    \label{fig:sample_best_worst_TV}
\end{figure*}
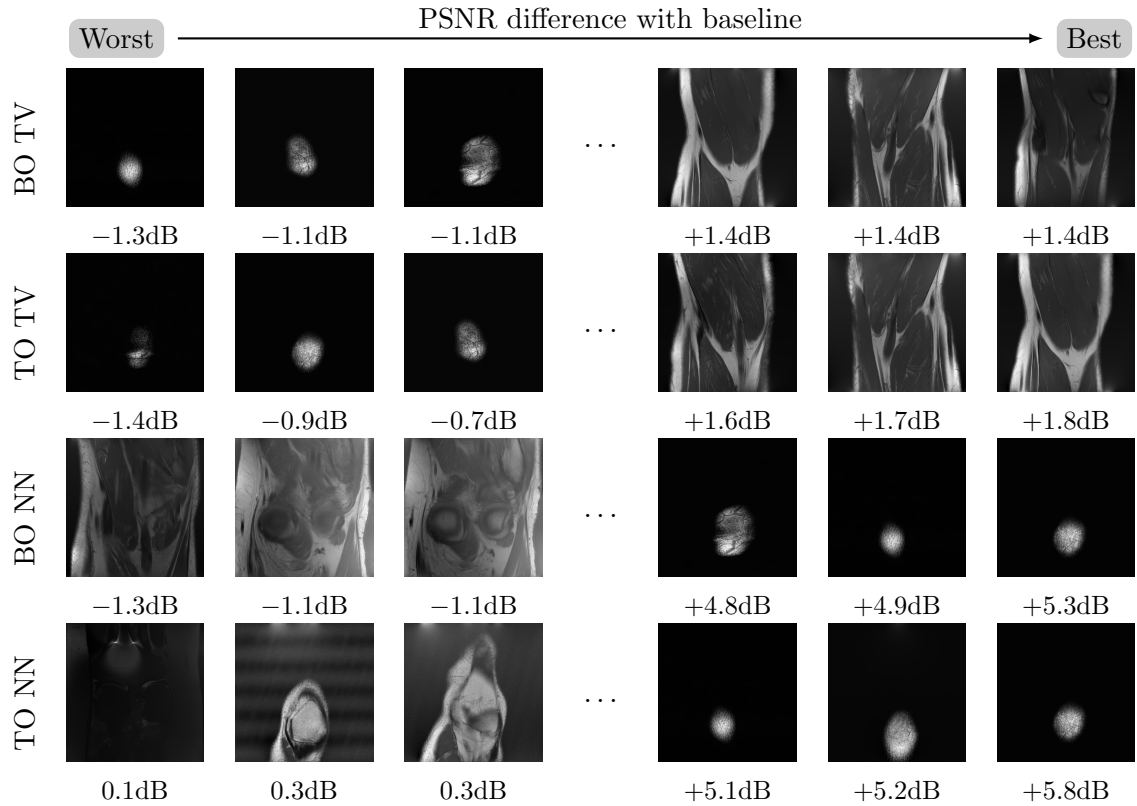

\rev{In Fig.~\ref{fig:sample_best_worst_TV}, we display some images which benefited the least (resp. the most) from the sampling scheme optimization, with respect to the baseline. For the total variation reconstructor, some images actually suffer from the sampling scheme optimization as can be seen on the left of the picture. However, they correspond to slices which are not dominant in the dataset. On the other hand, the images that benefit the most from trajectory optimization correspond to central slices, prevalent in the set. As for the neural network, the best increases are obtained with the outer slices, which could be due to the large constant areas, the network is able to reproduce easily. Finally, the largest decreases for the neural network as a reconstruction mapping are not clearly associated to a specific type of slice.}

\begin{table}
    \centering
    \small
    \begin{tabular}{>{\raggedright\arraybackslash}p{9em} >{\centering\arraybackslash}p{12em} >{\centering\arraybackslash}p{12em}}
    \toprule
    Method & $25\%$ & $10\%$ \\ [0.5ex]
    \midrule
    Radial scheme & \makecell{$27.87$dB \\ $0.66$} & \makecell{$24.28$dB \\ $0.57$} \\
    \hline
    Sparkling radial (baseline) & \makecell{$35.35$dB \\ $0.85$} & \makecell{$32.94$dB \\ $0.79$} \\
    \hline
    Bayesian optim. $K=128$ & \makecell{$35.66$dB ($+0.31$) \\ $0.86$} & \makecell{$33.41$dB ($+0.47$) \\ $0.80$} \\
    \hline
    Trajectory optim. $K=34742$ & \makecell{$35.92$dB ($+0.57$) \\ $0.87$} & \makecell{$33.48$dB ($+0.54$) \\ $0.80$} \\
    \hline
    Trajectory optim. $K=32$ & \makecell{$35.67$dB ($+0.32$) \\ $0.86$} & \makecell{$32.84$dB ($-0.10$) \\ $0.79$} \\
    \hline
    Trajectory optim. $K=128$ & \makecell{$35.67$dB ($+0.32$) \\  $0.86$} & \makecell{$32.89$dB ($-0.05$) \\ $0.79$} \\
    \bottomrule
    \end{tabular}
    \caption{Comparison of different optimization procedures for the TV reconstructor with different numbers of images $K$ in the training set. For each test case, the first line is the PSNR and the second line is the SSIM. \rev{We added the gain in PSNR with respect to the baseline in parentheses.} \label{table:result_different_optim_TV}}
\end{table}

\paragraph{Computing times}

Table~\ref{table:computational_time} gives the computation times for each method with the total variation reconstruction method.
The proposed approach has the significant advantage of giving an optimized sampling scheme with guarantees on the underlying density with a reduced computational budget and with a reduced number of images.
As can be seen, our approach requires only $32$ images and 3 hours.
This has to be compared to the 85 hours (3 days and a half) needed by the trajectory optimization routine.

This feature is a significant advantage of our approach. 
It could be key element when targeting high resolution images or 3D data.

\paragraph{Size of the training set}

As advertised, the Bayesian optimization approach works even for small datasets.
The trajectory optimization routine also  provides competitive results with only $32$ images in the training set. 
However, the performance collapses for the $10\%$ undersampling rate. 
Increasing the size of the training set to $K=128$ does not improve the situation.
This feature is in strong favor of our approach, when having access to a limited dataset.

\subsection{Comparing optimization routines for a neural network reconstructor}\label{subsec:comparison_NN}

The aim of this section, is to compare three different sampling optimizers:
\begin{itemize}
    \item The Bayesian density optimization solver proposed in this paper. 
    \item The trajectory optimization solver with a fixed unrolled neural network trained on a family of sampling schemes, see Appendix \ref{sec:training_on_family}. 
    This is a novelty of this paper. 
    \item An optimization routine minimizing the trajectories and the unrolled network weights simultaneously, as proposed in \cite{wang2022b,weiss2021pilot}.
\end{itemize}

\paragraph{Qualitative comparisons}

\begin{figure*}
    \centering
    \plottraj{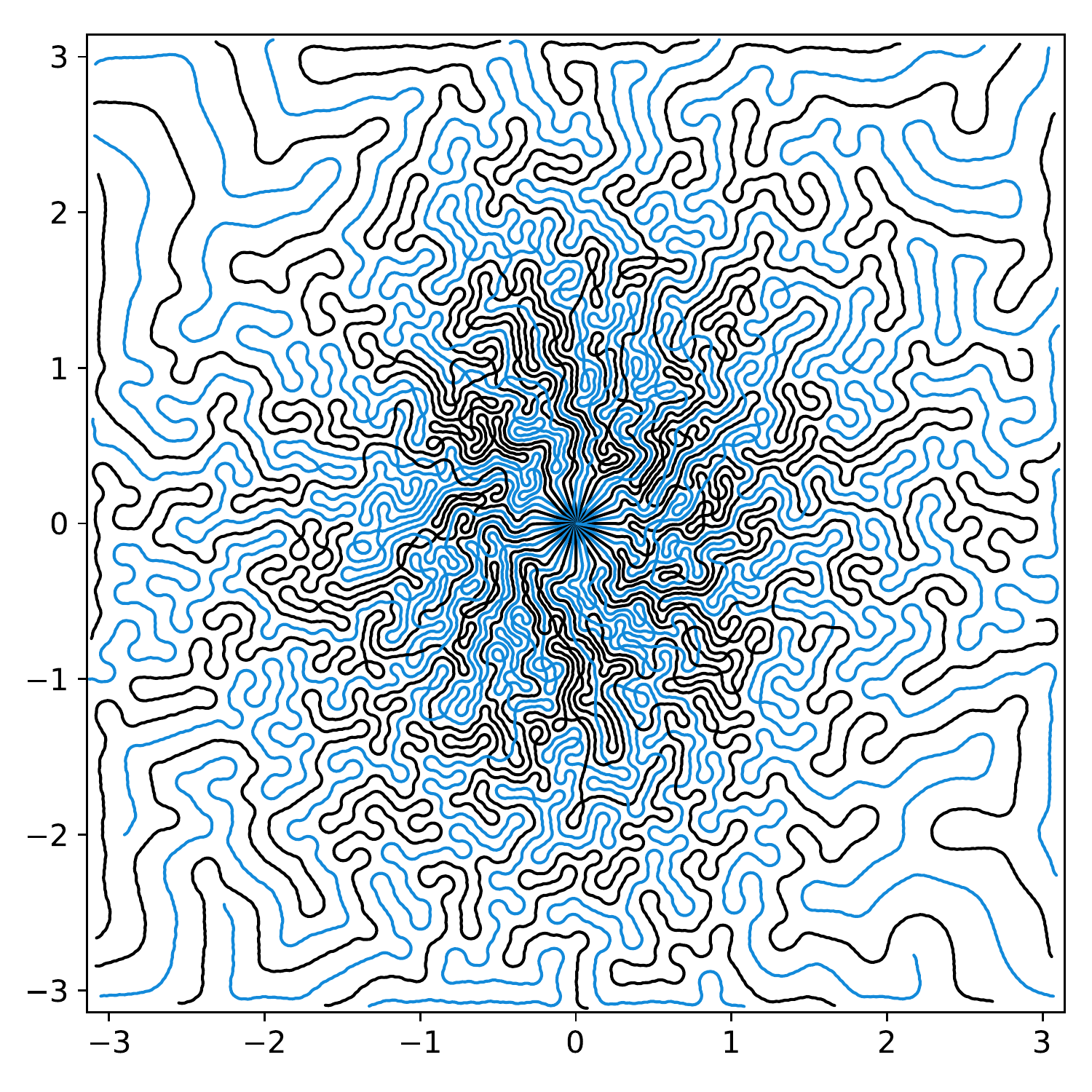}
        {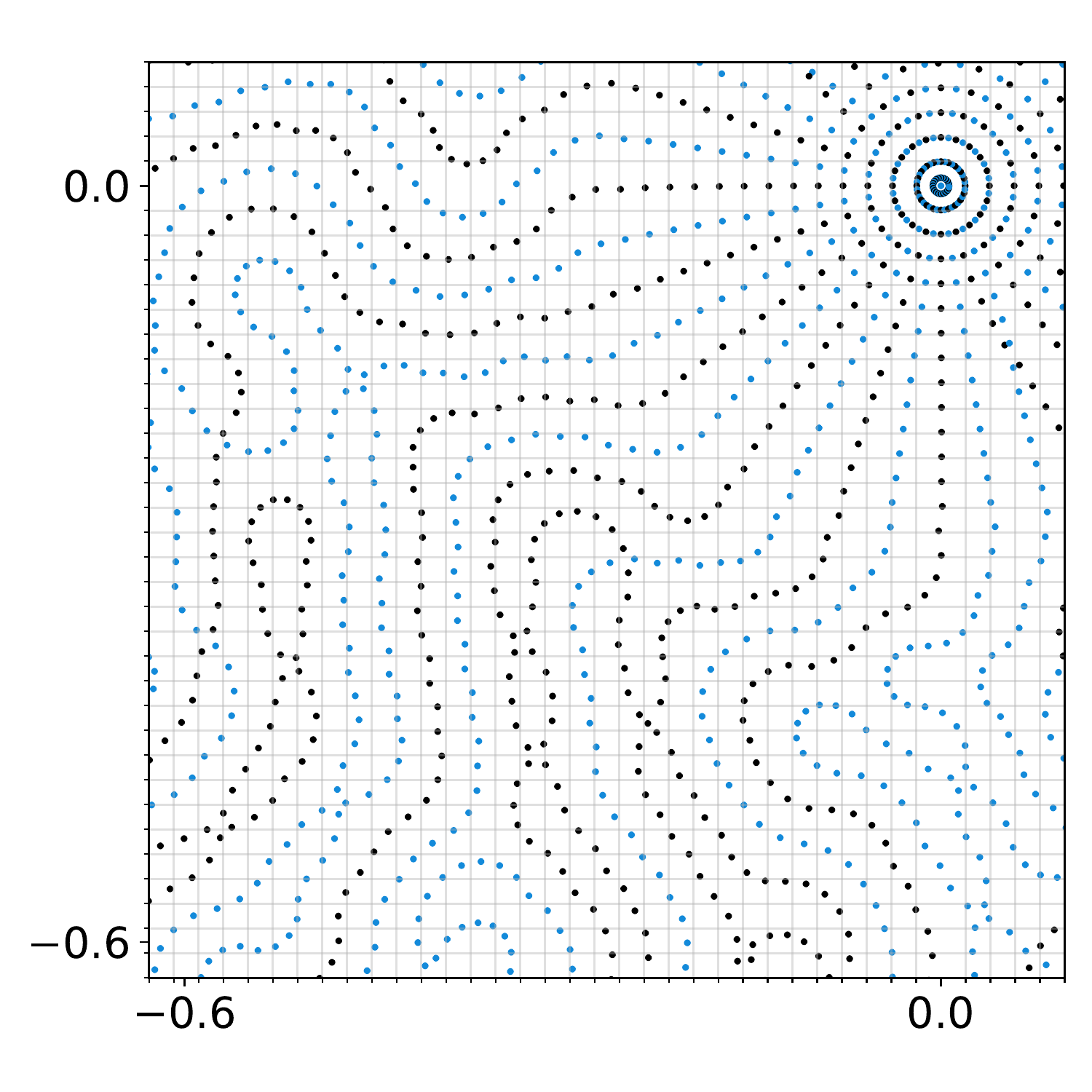}
        {Bayesian density optim. $25\%$}
        {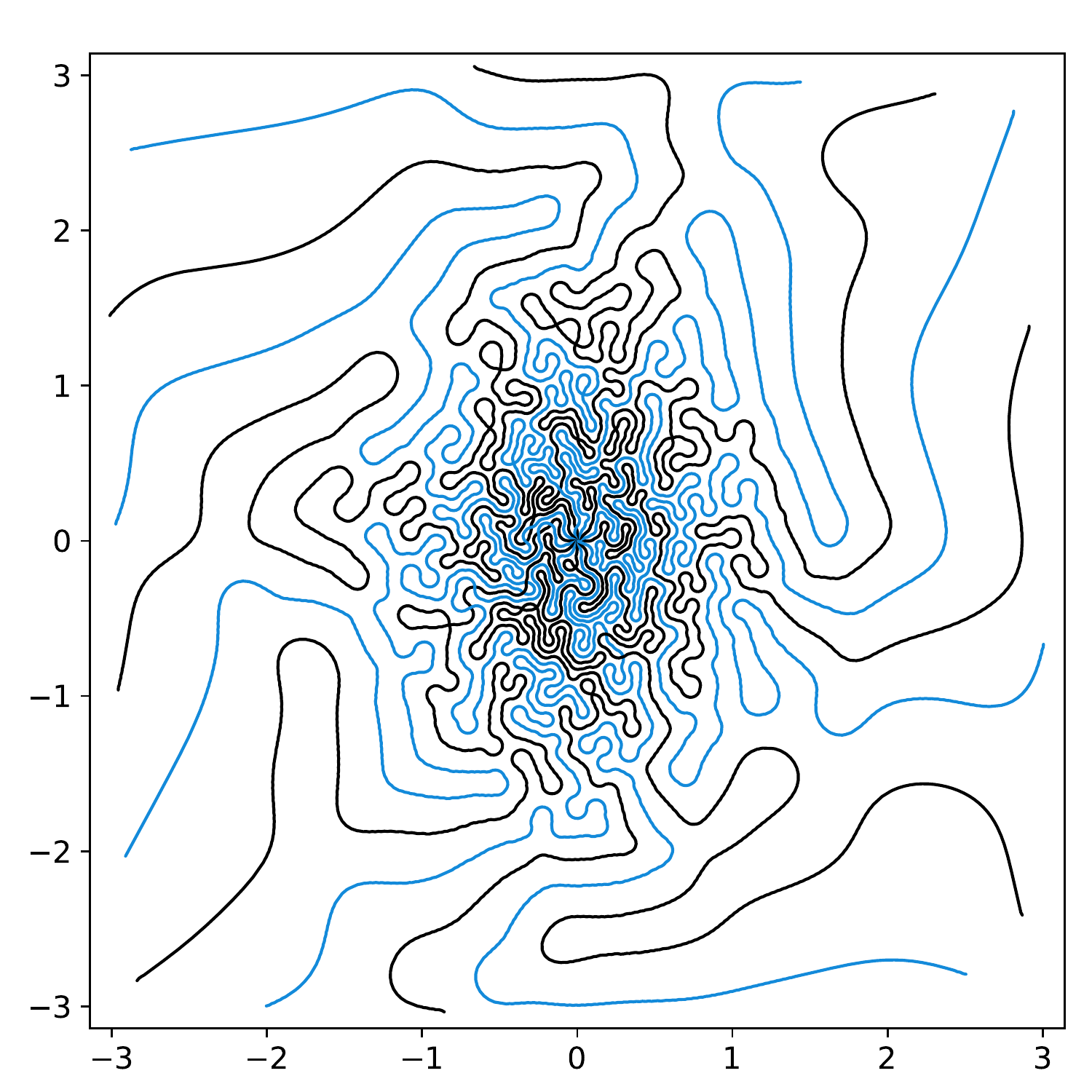}
        {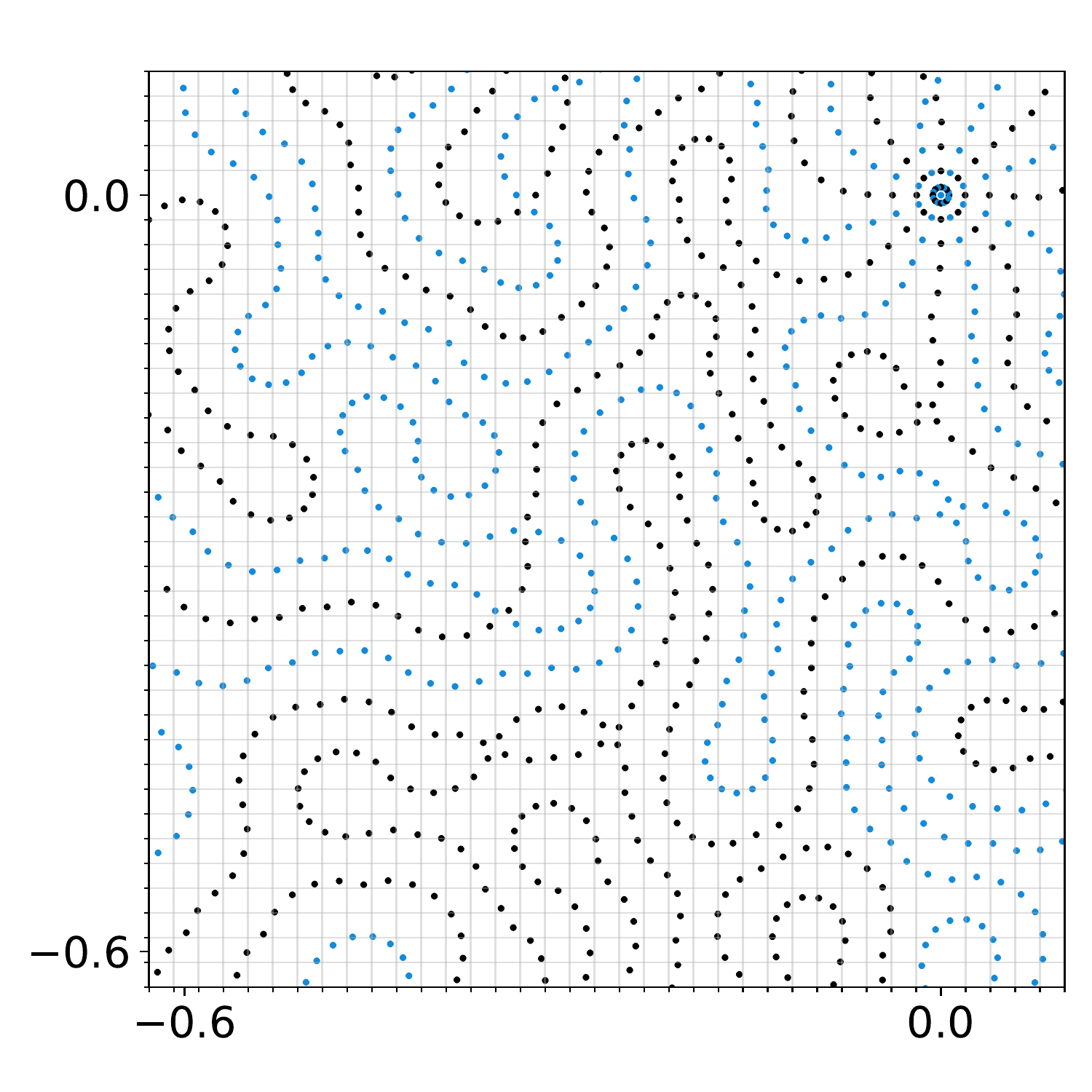}
        {Bayesian density optim. $10\%$}{}{}
    \plottraj{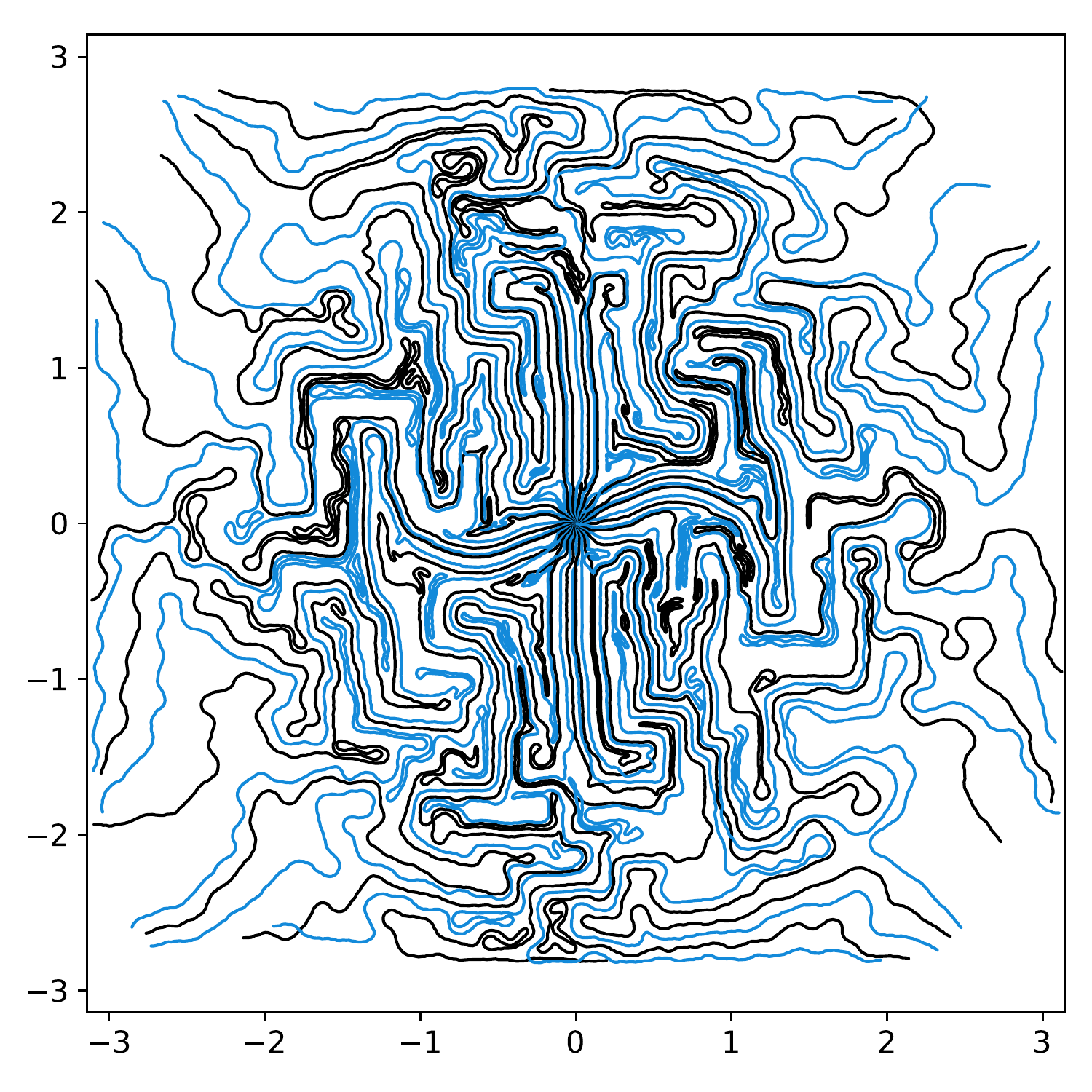}
        {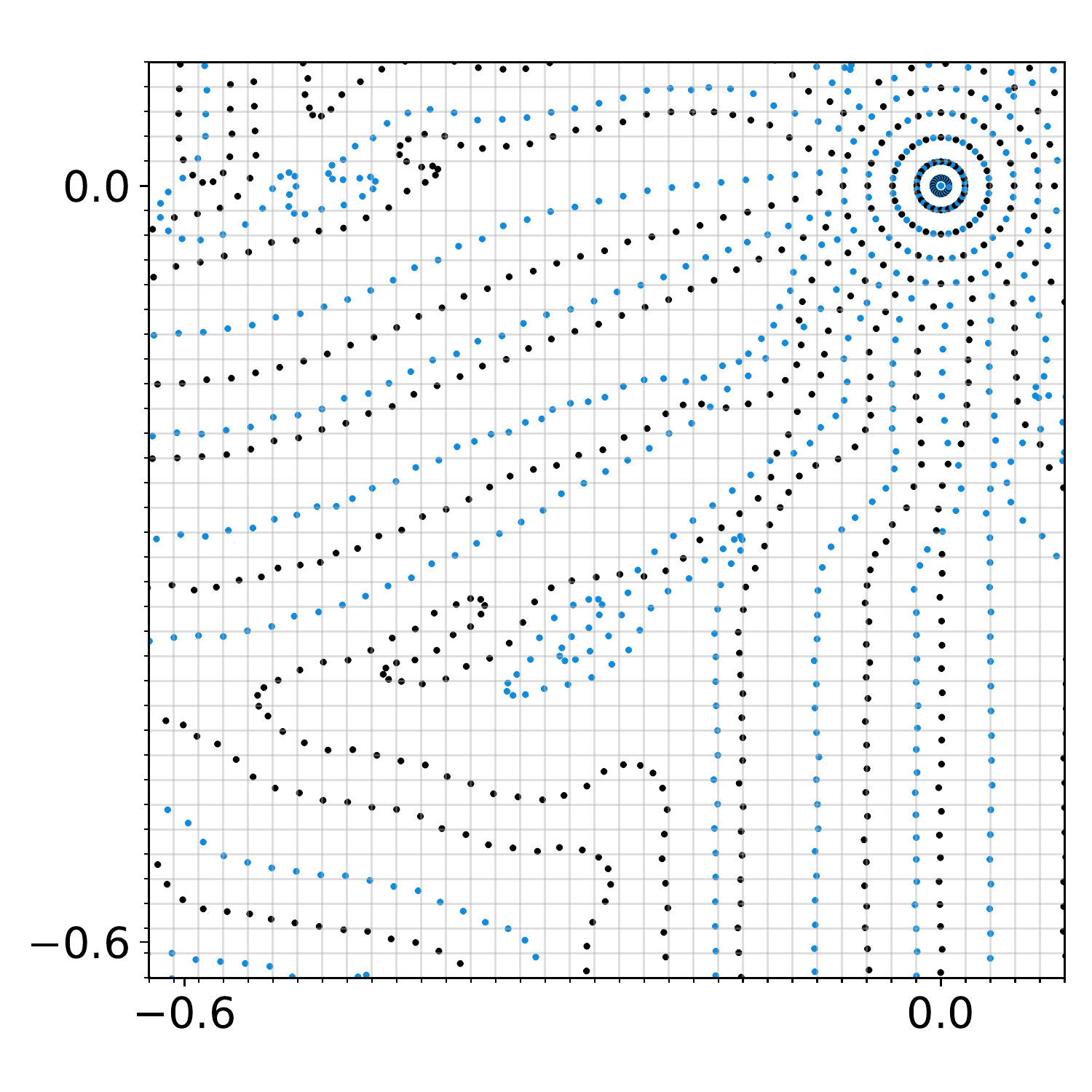}
        {Joint traj. optim. $25\%$}
        {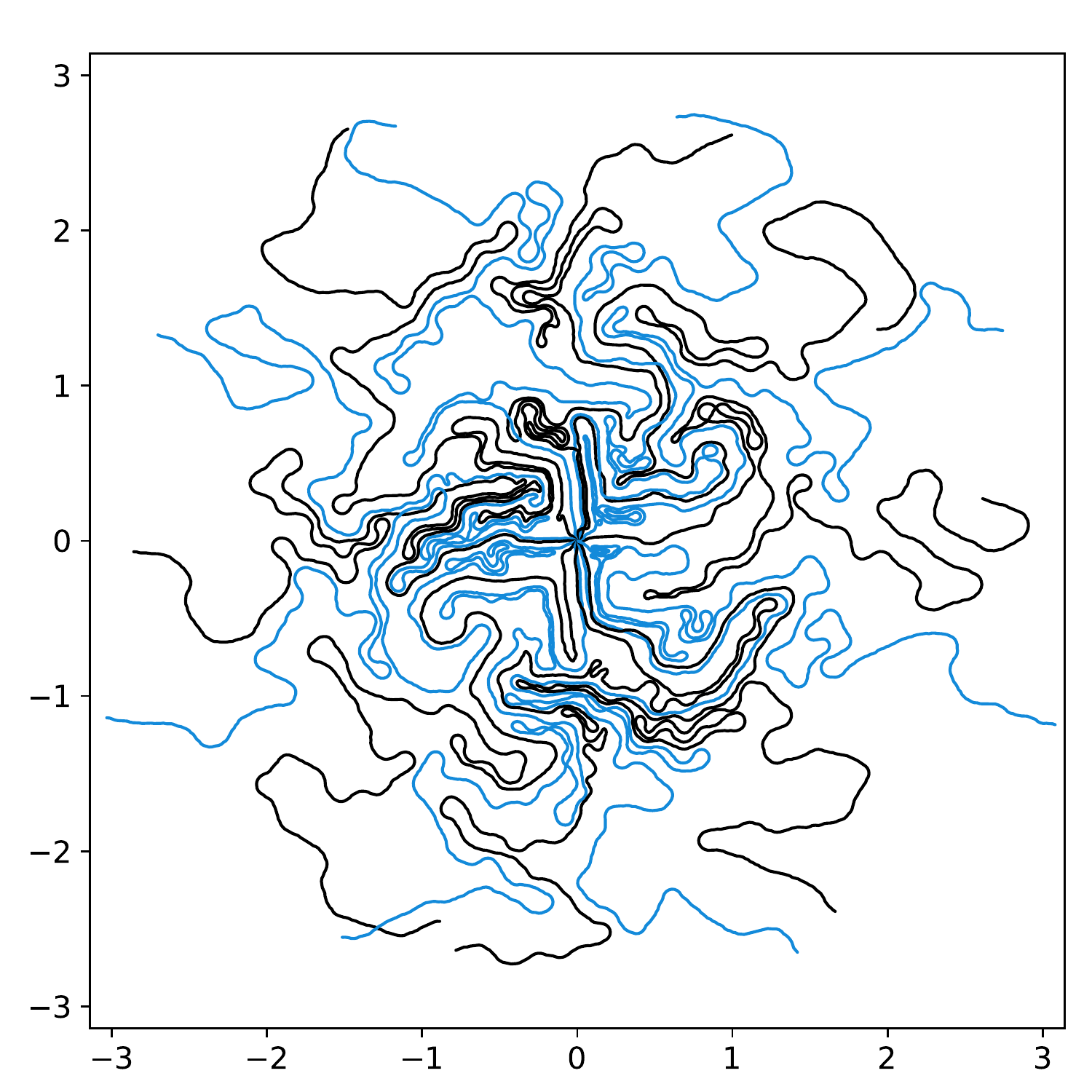}
        {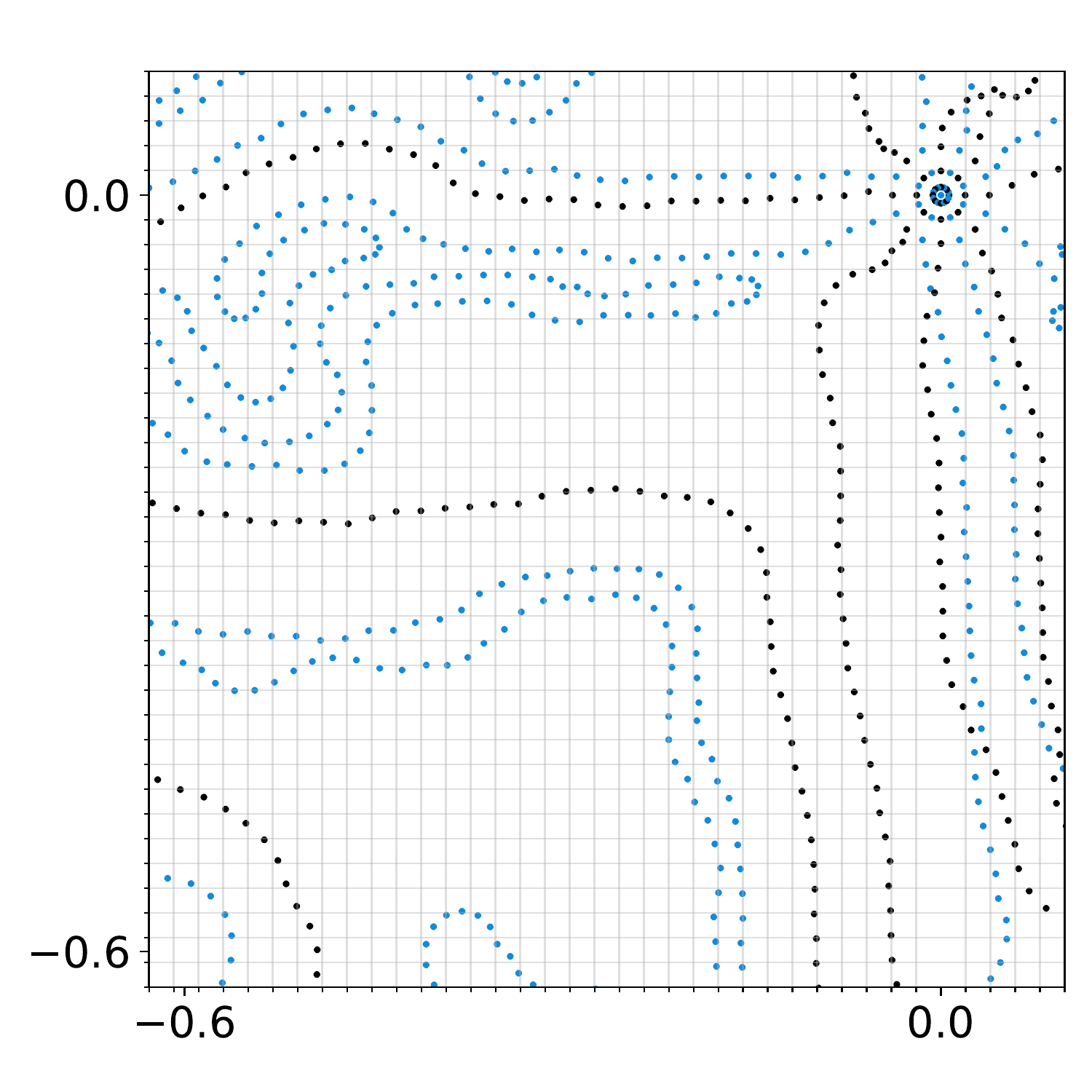}
        {Joint traj. optim. $10\%$}{}{}
    \plottraj{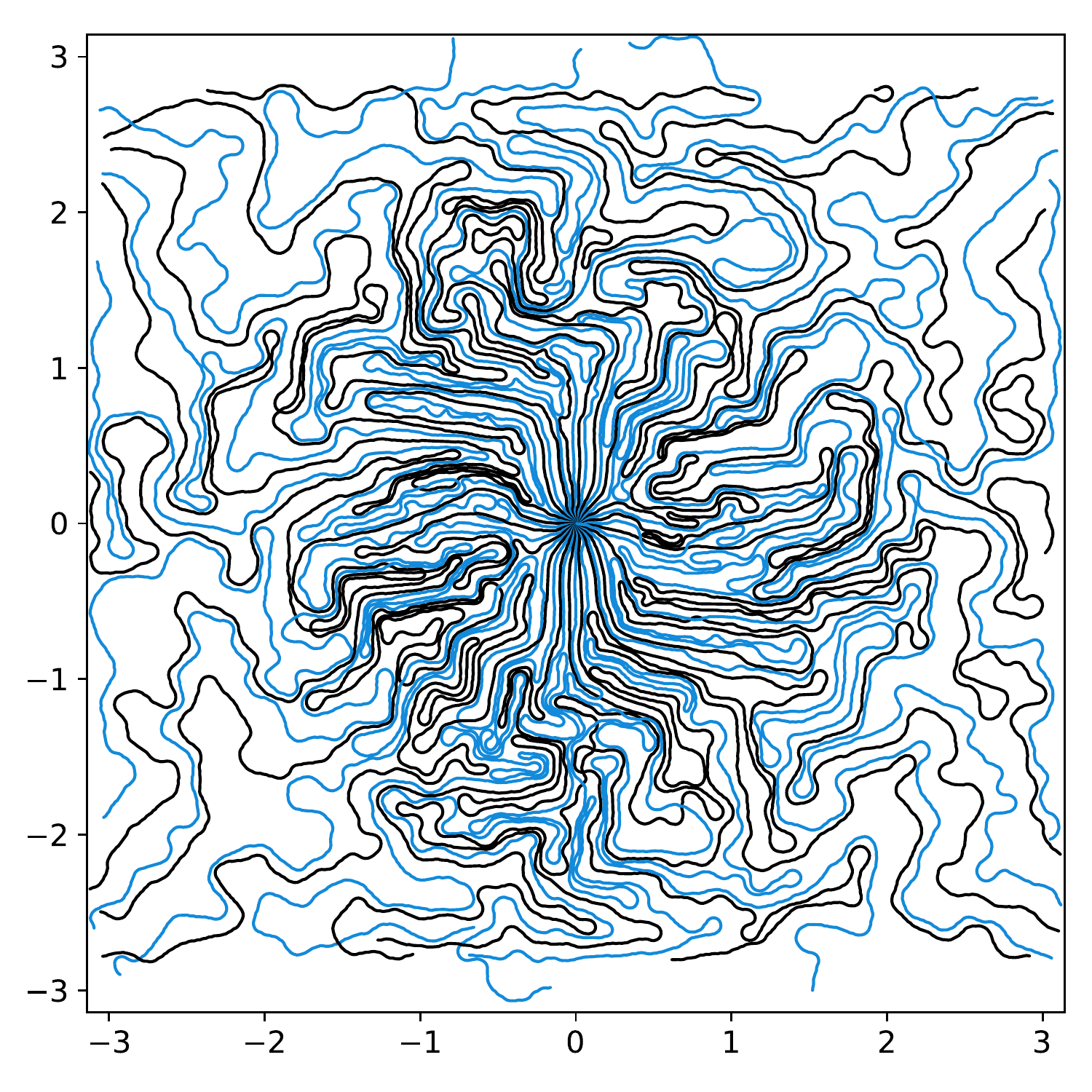}
        {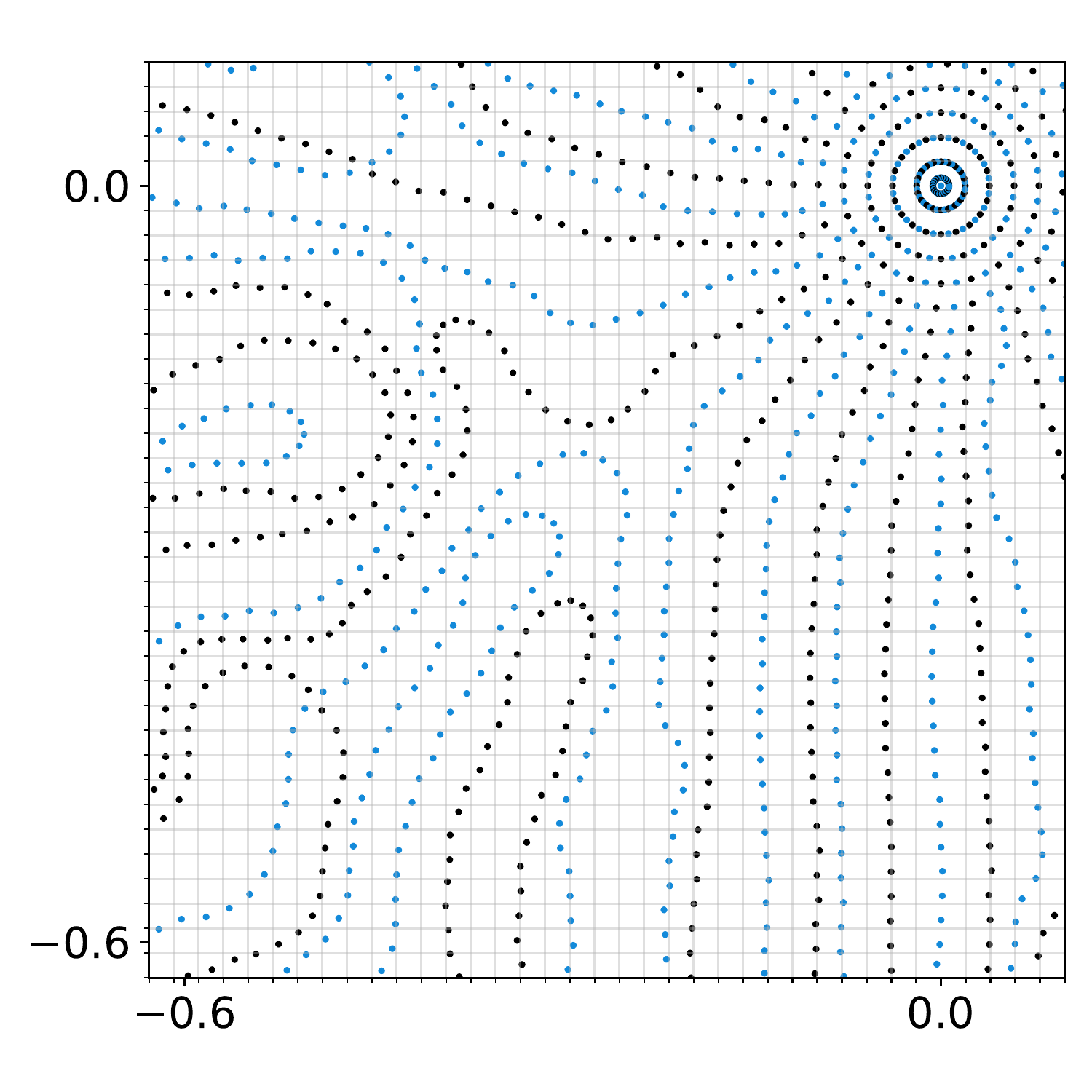}
        {Traj. optim. with fixed net $25\%$}
        {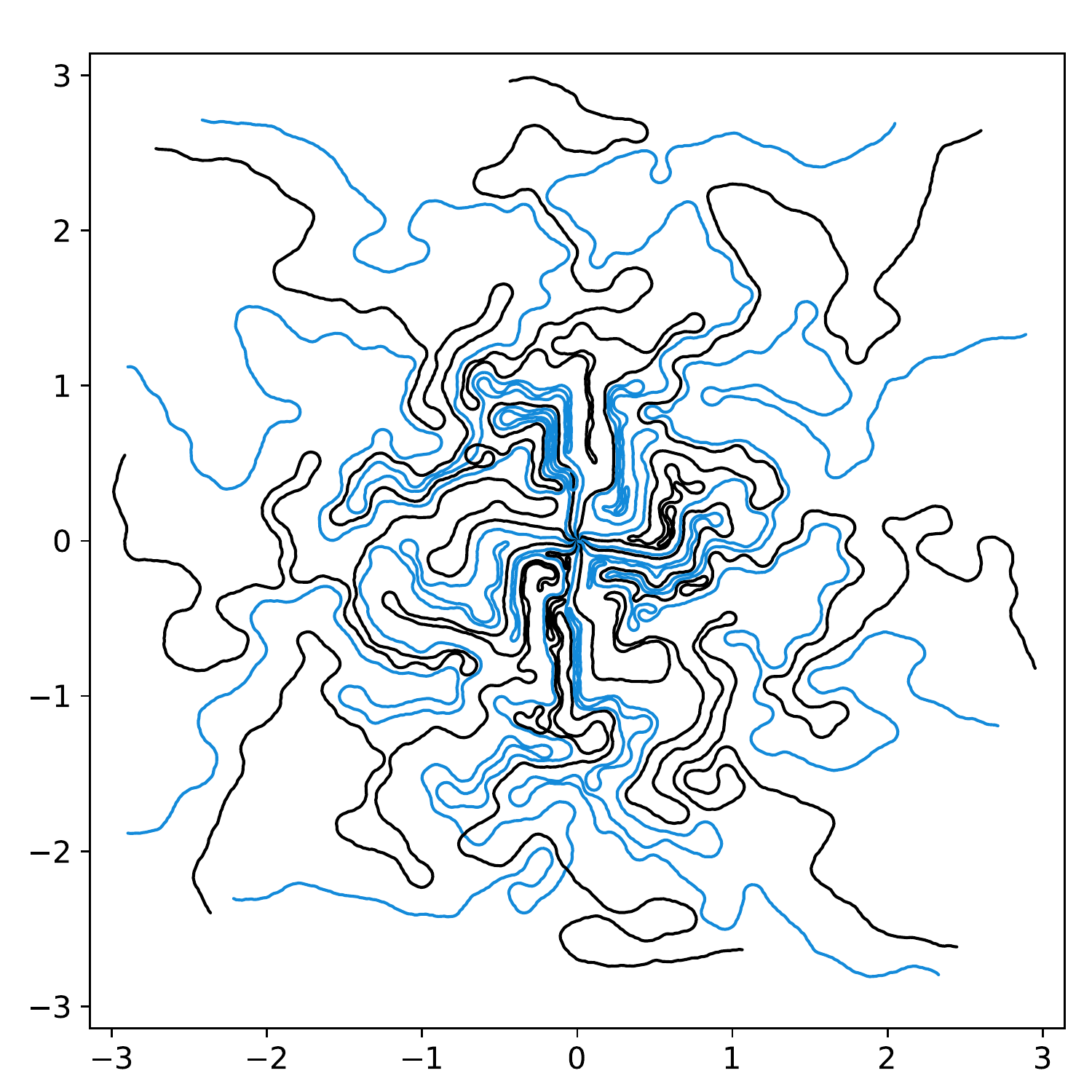}
        {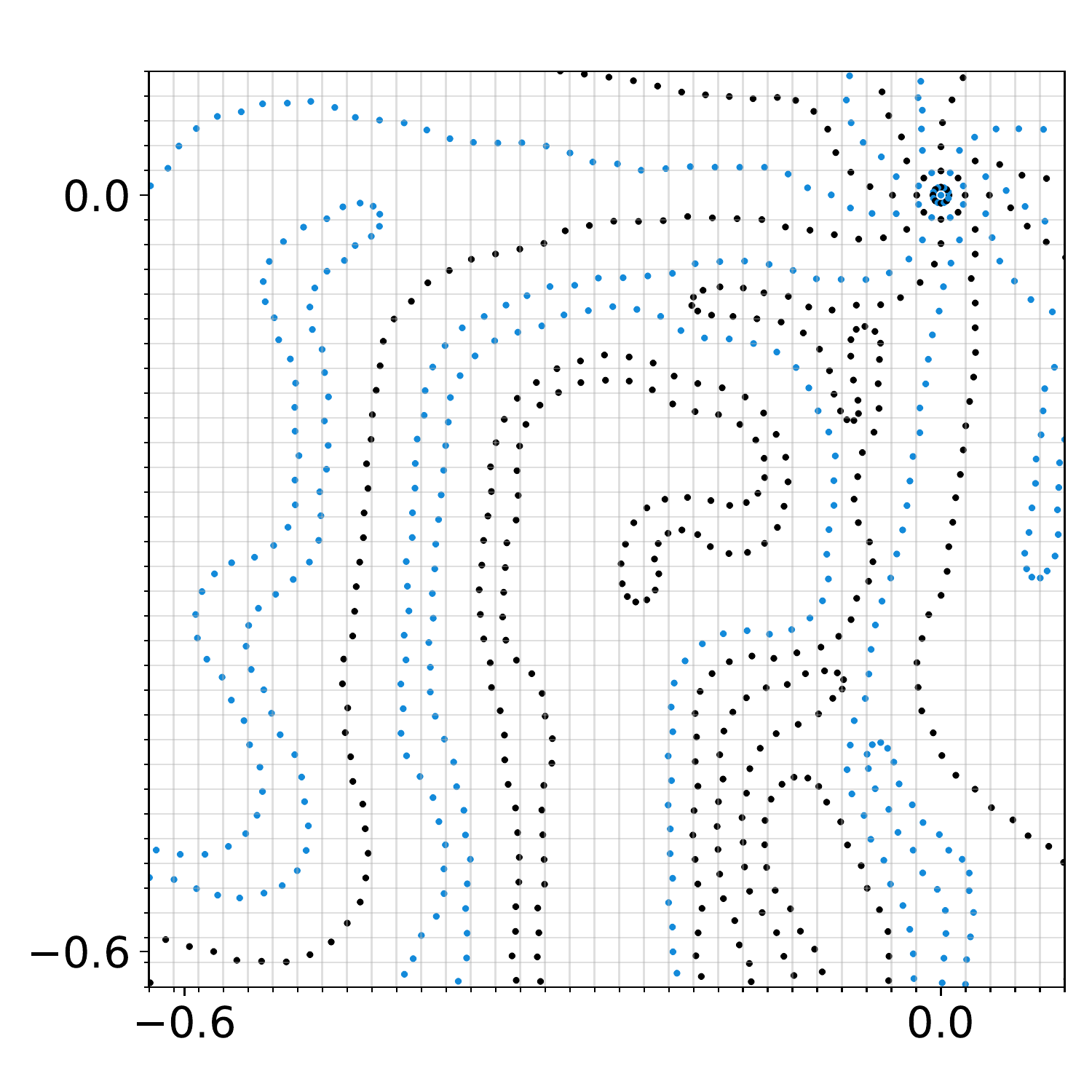}
        {Traj. optim. with fixed net $10\%$}{}{}
    \caption{Optimized sampling schemes with the various optimization approaches for a neural network reconstruction.}
    \label{fig:optimmultiNN}
\end{figure*}

The differences between the density optimization and the trajectory optimization can be observed on Fig. \ref{fig:optimmultiNN}. 
They are much more pronounced than for the total variation reconstructor. 
Surprisingly, the trajectory optimized sampling schemes leave large portions of the low frequencies unexplored.
Hence, it seems that the unrolled network is able to infer low frequency information better than the traditional total variation prior.
This suggests that the existing compressed sampling theories designed for the Fourier-Wavelet system have to be revised significantly to account for the progress in neural network reconstructions. 
The optimization of a trajectory for a fixed sampling scheme or the joint optimization yield qualitatively similar trajectories, with perhaps larger unexplored parts of the k-space for the fixed reconstruction method. 

\paragraph{Quantitative comparisons}

\begin{table}
    \centering
    \small
    \begin{tabular}{>{\raggedright\arraybackslash}p{12em} >{\centering\arraybackslash}p{12em} >{\centering\arraybackslash}p{12em}}
    \toprule
    Method & $25\%$ & $10\%$ \\ [0.5ex]
    \midrule
    Baseline with unrolled net & \makecell{$37.26$dB \\ $0.89$} & \makecell{$34.49$dB \\ $0.83$} \\
    \hline
    BO scheme $K=128$ with unrolled net & \makecell{$38.20$dB ($+0.94$) \\ $0.91$} & \makecell{$35.13$dB ($+0.64$) \\ $0.86$} \\
    \hline
    Traj. optim. with fixed unrolled net \rev{$K=34742$} & \makecell{$39.09$dB ($+1.83$) \\ $0.92$} & \makecell{$35.65$dB ($+1.16$) \\ $0.85$} \\
    \hline
    Joint optim. multi-scale \rev{$K=34742$} & \makecell{$39.03$dB ($+1.77$) \\ $0.92$} & \makecell{$35.53$dB ($+1.04$) \\ $0.85$} \\
    \bottomrule
    \end{tabular}
    \caption{Comparison of different optimization procedures for the unrolled ADMM reconstructor. For each test case, the first line is
the PSNR and the second line is the SSIM. \rev{The increase in PSNR compared to the baseline scheme is shown in parentheses}.\label{table:result_different_optim_NN}}
\end{table}
Table~\ref{table:result_different_optim_NN} allows  comparing the different methods quantitatively.
BO yields a PSNR increase almost twice lower than the multi-scale optimization (+0.94dB VS +1.83dB for 25\% and +0.64dB VS +1.16dB for 10\%).
This can likely be explained by the fact that the chosen family of densities (dimension 20) is unable to reproduce the complexity of the optimized trajectories.
It is possible that richer sampling densities could reduce the gap between both approaches.
However, Bayesian optimization is known to work only in small dimension and it is currently unclear how to extend the method to this setting. 

Interestingly, the trajectory optimized with a fixed unrolled neural network trained on a family provides slightly better results ($\approx +0.1$dB) than the joint optimization.
This suggests that the joint optimization gets trapped in a local minimizer since it can only be better if optimized jointly with the reconstructor.

\section{Conclusion}
\label{sec:conclu}

In this work, we designed efficient optimization algorithms that either optimize trajectories directly or learn a sampling density and an associated sampling pattern in MRI. Overall, the main highlights of this work are:
\begin{itemize}
 	\item The compressed sensing theories designed for the Fourier-Wavelet system with $\ell^1$ reconstruction (e.g. \cite{adcock2021compressive}) seem nearly optimal from an experimental point of view. Sampling schemes can be designed based on a density that is close to Shannon's rate at the k-space center and that decays towards the high frequencies. The precise shape of the density depends on the images structure. 
 	\item In that context, the Bayesian optimization of densities is an attractive method to design sampling schemes. It works with small datasets, ensures the convergence to a global minimizer. Its performance is close to much heavier trajectory optimizers and is from one to two orders of magnitude faster. 
 	\item In the case of unrolled neural network reconstructions, the proposed Bayesian optimization framework is still interesting with gains of up to $1$dB in average on the fastMRI knee validation set. However, the gain can be nearly doubled with a direct optimization of the trajectories. A possible explanation for this fact is that the family of densities is too poor to describe the best convoluted trajectories. \rev{Another one is that the theoretical bases of the compressed sensing theory break down when using neural network reconstruction methods. This calls for a renewed theory for this fast expanding field}.
 	\item As points of minor importance, we improved the Sparkling trajectories \cite{lazarus2019sparkling}, by changing the discrepancies and provided various improvements to the direct optimization of trajectories by using the Extra-Adam algorithm to handle hard constraints and by training reconstruction networks on families of operators.
    \item \rev{As a prospect, the extensions of these ideas to 3D multi-coil imaging is particularly relevant. There is no a priori obstacle to apply this rather lightweight formalism and to obtain dataset tailored sampling distributions.}
 \end{itemize} 

\acks{P. Weiss and F. de Gournay were supported by the ANR JCJC Optimization on Measures Spaces ANR-17-CE23-0013-01 and the ANR-P3IA-0004 Artificial and Natural Intelligence Toulouse Institute.
This work was granted access to the HPC resources of IDRIS under the allocation AD011012210 made by GENCI.}

%
\ethics{The work follows appropriate ethical standards in conducting research and writing the manuscript, following all applicable laws and regulations regarding treatment of animals or human subjects.}

\coi{We declare we don't have conflicts of interest.}

{\small
\bibliography{biblio}
}

\clearpage
\appendix

\section{Implementation details}
\label{sec:implementation}

\subsection{TV reconstruction algorithm\label{sec:reconstruction_algorithm_TV}}

In this part we detail the TV iterative reconstruction algorithm that is used in this paper.
We consider a regularized version of the total variation of the form 
\begin{equation*}
TV_\epsilon(x) = \sum_{n=1}^N \sqrt{\| (\nabla x)[n] \|_2^2 + \epsilon^2}.
\end{equation*}
Given $y\in\C^M$, the solver of problem \eqref{eq:reconstructor_TV} is given in Algorithm~\ref{alg:TV_minimization}.
The parameter $\alpha$ drives the acceleration and $D$ is the dimension, here $D=2$.
It corresponds to a Nesterov accelerated gradient descent \cite{nesterov1983method} with a regularized version of the $\ell^1$ norm.
\begin{algorithm}
	\begin{algorithmic}
	\Require Number of iterations $Q$.
	\State Set $z^{(0)}=x^{(0)}=0$, $\tau = \frac{1}{\|A(\xi)\|_{2\to 2}^2 + 4D\lambda/\epsilon}$.
	\ForAll{$q = 0$ to $Q-1$}
		\State $r^{(q)} = A(\xi)^*(A(\xi)z^{(q)}-y)$
		\State $\displaystyle x^{(q+1)}=z^{(q)}- \tau \left[ r^{(q)}+ \lambda \nabla TV_{\epsilon}(z^{(q)}) \right]$
		\State $\displaystyle z^{(q+1)}=x^{(q+1)}+\alpha(x^{(q+1)}-x^{(q)})$
	\EndFor
	\State \Return $x^{(Q)}$.
	\end{algorithmic}
\caption{A TV minimization algorithm} \label{alg:TV_minimization}
\end{algorithm}
A critical point is the choice of the step $\tau$ in Algorithm~\ref{alg:TV_minimization}.
This step is computed using the spectral norm of the data fidelity term which can be computed using a power iteration method for each point configuration $\xi$.
The resulting step is taken into account in the computation of the gradient with respect to the locations $\xi$ of the cost function in~\eqref{eq:objective_xi}.

\subsection{The unrolled neural network \label{sec:appendix_neural_net_definition}}

The neural network based reconstruction is an unrolled network. 
The one used in this work is based on the ADMM (Alternative Descent Method of Multipliers) \cite{ng2010solving}.
It consists in alternating a regularized inverse followed by a denoising step with a neural network.
If $\Dc_{\lambda^{(p)}}$ denotes the denoiser used at iteration $p$, the unrolled ADMM can be expressed through the sequence:
\begin{equation*}
    \begin{dcases}
        x^{(p+1)} = \left( A(\xi)^*A(\xi)+\beta\Id \right)^{-1}\left( A(\xi)^*y+\beta z^{(p)}-\mu^{(p)} \right) \\
        z^{(p+1)} = \Dc_{\lambda^{(p)}}\left( x^{(p+1)}+\frac{\mu^{(p)}}{\beta} \right) \\
        \mu^{(p+1)} = \mu^{(p)}+\beta\left( x^{(p+1)}-z^{(p+1)} \right)
    \end{dcases}
\end{equation*}
with a pseudo-inverse initialization $z^{(0)} = A(\xi)^\dag y$.

In this work, we use the DruNet network \cite{zhang2021plug} to define the denoising mappings $\Dc_{\lambda^{(p)}}$.
We choose an ADMM algorithm for the following reasons:
\begin{enumerate}
\item for well-spread sampling schemes, the matrix $A(\xi)^*A(\xi)$ has a good conditioning and the linear system that has to be inverted can be solved in less than a dozen iterations,
\item it has demonstrated great performance to solve linear inverse problems in imaging, including image reconstruction from Fourier samples \cite{wang2022b}.
\end{enumerate}

We opted for a different network at each iteration instead of a network that shares its weights accross all iterations.
This leads to slightly higher performance at the price of a slightly harder to interpret architecture (see e.g.  \cite{genzel2022near} for a similar discussion in CT reconstruction).

\subsection{Training the reconstruction network for a family of operators \label{sec:appendix_neural_net_training}}
\label{sec:training_on_family}

Following \cite{gossard2022training}, we trained our network in a non usual way. 
Instead of training the denoising networks $\Dc_{\lambda^{(p)}}$ for a single operator $A(\xi_0)$, we actually trained it for a whole family of operators $\mathcal{A}=\{A(\xi), \xi \in \mathcal{F}\}$, where $\mathcal{F}$ is a large family of sampling schemes. 
We showed in \cite{gossard2022training}, that this simple approach yields a much more robust network, which is adaptive to the forward operator.

In our experiments, the network is trained on a family of $10^3$ sampling schemes that are generated using the attraction-repulsion minimization problem \eqref{eq:discrepancy}. These schemes are parameterized by densities that are within $\Cc$.
This pretraining step consists of $32$ epochs with a batch of $8$ images using the Adam optimizer with default parameters ($\beta_1=0.9$ and $\beta_2=0.999$).
The step for the CNN weights is set to $10^{-4}$ with a multiplicative update of $0.95$ after each epoch.
The measurements are perturbed by an additive white noise (see $n$ in \eqref{eq:objective_xi}).

\subsection{Joint optimization}

Instead of optimizing the sampling scheme for a fixed network, we can also optimize jointly the sampling locations together with the network weights. 
This approach was proposed in \cite{wang2022b,weiss2021pilot}.
Due to memory requirements, we set the batch size to $7$ for the unrolled network in our training procedure.
The step size for the CNN weights in this experiment is also set to $10^{-4}$ with the default Adam parameters.

\subsection{Computational details}\label{subsec:computational_details}
In this paragraph, we describe the main technical tools used to optimize the reconstruction process. 

\subsubsection{Solving the particle problem \eqref{eq:objective_xi}}
\label{sec:solving_xi}

Problem \eqref{eq:objective_xi} is a highly non-trivial problem. Two different computational solutions were proposed in \cite{weiss2021pilot,wang2022b}. 
In this work, we re-implemented a solver with some differences outlined below.

First, the optimization problem \eqref{eq:objective_xi} involves a nontrivial constraint set $\Xi$.
While the mentioned works use a penalization over the constraints, we enforce the constraints by using a projection at each iteration.
Handling constraints in stochastic optimization was first dealt with stochastic mirror-prox algorithms \cite{juditsky2011solving}. 
This approach turned out to be inefficient in practice. We therefore resorted to an extension of Adam in the constrained case called Extra-Adam \cite{gidel2018variational}.
The step size was set to $10^{-3}$ and the default Adam parameters $\beta_1=0.9$ and $\beta_2=0.999$.
We observed no significant difference by tuning these last two parameters.
We also use a step decay of $0.9$ each fourth of epoch and batch size of $13$, which is the largest achievable by our GPU.

Similarly to \cite{weiss2021pilot,wang2022b}, we use a multi-scale strategy. The trajectories are defined through a small number of control points, that progressively increases across iterations. We simply use a piecewise linear discretization (contrarily to higher order splines in \cite{wang2022b}). The initial decimation factor is $2^7$ and is divided by two every two epochs. This results in a total number of epochs equal to $14$ and takes about 86 hours for a total variation solver. In comparison \cite{wang2022b}, reports a total of $40$ epochs.

\subsubsection{Implementing the Non-uniform Fourier Transform (NUFT)}

Various fast implementations of the Non-uniform Fourier Transform \eqref{eq:def_NUFFT} are now available \cite{keiner2009using,fessler2003nonuniform,shih2021cufinufft,muckley2020torchkbnufft}.
In this work, we need a pyTorch library capable of backward differentiation.
Evaluating the gradient of the cost function in \eqref{eq:objective_xi} or in \eqref{eq:objective_rho} indeed requires computing the differential of the forward operator $A(\xi)$ with respect to $\xi$.
This can be done by computing $D$ non-uniform Fourier transforms (see \cite{wang2022b,gossard2022spurious,wang2021efficient}).
Different packages were tested and we finally opted for the cuFINUFFT implementation \cite{shih2021cufinufft}.
The bindings for different kind of NUFT are available at \href{https://github.com/albangossard/Bindings-NUFFT-pytorch/}{https://github.com/albangossard/Bindings-NUFFT-pytorch/}.

\subsubsection{Minimizing the discrepancy}

The minimization of the discrepancy \eqref{eq:discrepancy} is achieved with a gradient descent, as was proposed in the original paper \cite{schmaltz2010electrostatic}, see Algorithm~\ref{alg:discrepancy_minimization}. The input parameters are the initial sampling set $\xi^{\textrm{ini}}$, the target density $\rho$ and a step-size $\tau>0$.
The step size needs to be carefully chosen to ensure a fast convergence. The optimal choice can be shown to be related to the minimal distance between adjacent points. In our experiments, it was tuned by hand and fixed respectively to $2\times 10^4$ and $5\times 10^{3}$ for the $25\%$ and $10\%$ undersampling schemes.

Computing the gradient requires to compute pairwise interactions between all particles: in our codes, it is achieved using \href{https://www.kernel-operations.io/}{PyKeOps} \cite{charlier2021kernel}.
This approach presents the advantages of being fast, adapting to arbitrary kernels $h$  and to natively allow backward differentiation within PyTorch. 
For a number of particles $M$ above $10^6$, fast multipole methods might become preferable \cite{chaithya2020}.

\begin{algorithm}
    \begin{algorithmic}
        \State Set $\zeta^{(0)}=\xi^{\textrm{ini}}$
        \For{$j=1\ldots J$}
        \State $\zeta^{(j)} = \Pi_\Xi\left(\zeta^{(j-1)} - \tau \nabla_1\dist(\zeta^{(j-1)},\rho)\right)$
        \EndFor
        \State Set $\xi^{(n)}=\zeta^{(J)}$
    \end{algorithmic}
    \caption{Gradient descent to minimize \eqref{eq:discrepancy}. \label{alg:discrepancy_minimization}}
\end{algorithm}

\subsubsection{Handling the mass at 0}
\label{sec:handling_mass_0}

An important issue is related to the fact that all trajectories start at the k-space origin. This creates a large mass for the sampling scheme at 0. When minimizing a discrepancy between the sampling scheme and a target density, the sampling points are therefore repulsed from the origin, creating large holes at the center. To avoid this detrimental effect, we fix rectilinear radial trajectories at the origin at maximal acceleration until a distance of 0.5 pixel between adjacent trajectories samples is reached. This creates a fixed pattern in the k-space center, which can easily be seen in the zoom of Fig.~\ref{fig11} and Fig.~\ref{fig12}. We compute the discrepancy only at the exterior of a disk centered at the origin containing this fixed pattern.

\subsubsection{Projection onto the constraint set}

The projector onto the constraint set is used twice in this work.
First the Extra-Adam algorithm requires a Euclidean projector on the constraint set $\Xi$ to solve \eqref{eq:objective_xi}.
This projector is also needed to compute one evaluation of the sampler in \eqref{eq:discrepancy}.
In this project, we used the dual approach proposed in \cite{chauffert2016projection}, implemented on a GPU.
This algorithm can be implemented in PyTorch, and can be differentiated. This allows computing the gradient of the overall function in \eqref{eq:objective_rho}.

\end{document}